\documentclass[12pt,a4paper]{article}
\pdfoutput=1
\usepackage{jheppub}
\usepackage{graphicx}
\usepackage{amssymb}
\usepackage{amsmath}
\usepackage{topcapt}
\usepackage{epstopdf}
\preprint{DIAS-STP-14-03\\}

\title{On the Phase Structure of Commuting Matrix Models}

\author[a]{Veselin G. Filev}
\author{and}
\author[b]{Denjoe O'Connor}

\affiliation[a,b]{School of Theoretical Physics, Dublin Institute for Advanced Studies\\
10 Burlington Road, Dublin 4, Ireland.}
\emailAdd{denjoe@stp.dias.ie}
\emailAdd{vfilev@stp.dias.ie}

\abstract{We perform a systematic study of commutative $SO(p)$
  invariant matrix models with quadratic and quartic potentials in the
  large $N$ limit. We find that the physics of these systems depends
  crucially on the number of matrices with a critical r\^ole played by
  $p=4$. For $p\leq4$ the system undergoes a phase transition
  accompanied by a topology change transition. For $p> 4$ the system
  is always in the topologically trivial phase and the eigenvalue
  distribution is a Dirac delta function spherical shell. We verify
  our analytic work with Monte Carlo simulations.}


\subheader{}

\begin{document}
\maketitle

\section{Introduction}
Multi-matrix models arise in a wide variety of settings and are
believed to play a fundamental r\^ole in string theory.  One such
model, the IKKT matrix model has been proposed as a non-perturbative
definition of string theory \cite{Ishibashi:1996xs} and its quantum
mechanical relatives are fundamental to current understanding of
M-theory. There are even recent indications that four of its 
space-time dimensions may be dynamically large in a cosmological scenario
\cite{Kim:2011cr}.

Although a non-perturbative formulation of M-theory in terms of its
fundamental degrees of freedom is still lacking, the best candidate
for such a formulation appears to be the infinite matrix size limit of
a matrix model of some kind. The leading candidate for such a
formulation is the BFFS model \cite{Banks:1996vh,Townsend:1995kk}
which was conjectured to capture the entire dynamics of M-theory and
shown to contain perturbative string states
\cite{Banks:1996my,Dijkgraaf:1997vv}. Relatives of this model such as
the BMN model \cite{Berenstein:2003gb} or models derived from the ABJM
model\footnote{Kovacs et al \cite{Kovacs:2013una} establish a natural
  and direct connection between a certain sector of the ABJM theory
  and the BMN model \cite{Berenstein:2003gb}.}
\cite{Aharony:2008gk,Kovacs:2013una} are also considered possible
viable candidates for such a non-perturbative formulation.  

All of these conjectured formulations of M-theory are regularised
versions of the supermembrane. They are based on the matrix
regularisation of membranes introduced by Hoppe
\cite{Hoppe:PhDThesis1982} and extended to the supermembrane in
\cite{Townsend:1995kk} and \cite{de Wit:1988ig}. They also arise as
dimensionally reduced 4-dimensional or 3-dimensional supersymmetric
field theories.

Multi-matrix models further arise in lower dimensional variants of the
IKKT model \cite{Connes:1997cr}, in the low energy dynamics of
$D$-branes \cite{Kazakov:1998ji} and simple models of emergent
geometry \cite{DelgadilloBlando:2007vx,DelgadilloBlando:2012xg} and
emergent gravity \cite{Steinacker:2012ra,Blaschke:2010ye} and
dimensionally reduced Yang-Mills models
\cite{Krauth:1998xh,Krauth:1998yu,Hotta:1998en,Ambjorn:2000bf,Ambjorn:2000dx,Azuma:2004zq}.
Many of these models will have regimes where commuting matrices play a
r\^ole.

In \cite{Filev:2013pza,O'Connor:2012vr} it was established that the
unique rotationally invariant three dimensional joint eigenvalue 
distribution that corresponds to a parabolic one dimensional
distribution is the uniform distribution within a ball of radius
$R$. It was also established that the strong coupling limit of Hoppe's
two matrix model \cite{Hoppe:PhDThesis1982} which describes the low
energy dynamics of $D0$-branes \cite{Kazakov:1998ji} in ${\cal N}=1$
supersymmetric Yang-Mills in four dimensions was captured by commuting
matrices.  In part the motivation for the current paper
arose from this earlier work coupled with a desire to understand
commuting matrices in and of themselves.

To our knowledge no systematic study of commutative matrix
models, with general potential, has been undertaken prior to the current
work. An understanding of commutative matrix models fills a gap in the
literature and because of the simplicity of these systems the results
may prove useful in a wider context. Such models, of course, also have
an intrinsic interest in their own right.

In this paper we show that due to rotational invariance we can recover
the full joint eigenvalue distribution from that of the one
matrix distribution, but only when the eigenvalue distribution of the
full system is topologically trivial. We begin by studying
Gaussian distributions (considered previously  in 
refs.~\cite{Berenstein:2005aa,Berenstein:2005jq, Aharony:2007rj,Berenstein:2008eg}) 
and find that the generalisation of the Wigner
distribution for $p=1$ becomes the uniform distribution within a disk
for $p=2$, but for $p=3$ the distribution is
$\rho_{(3)}(\vec{x})=\frac{3}{4\pi^3}\frac{1}{\sqrt{\frac{4}{3}-\vec{x}^2}}$,
which is divergent at the boundary but still integrable. We find a
special r\^ole is played by $p=4$ as it is the critical dimension
where the distribution is a Dirac delta function on the unit sphere:
$\rho_{(4)}(\vec{x})=\frac{1}{\pi^2}\delta(1-\vec{x}^2)$. For all
$p>4$ only spherical shells occur and the Gaussian distributions of
commuting matrices have eigenvalue distributions $\rho_p(\vec
x)=\frac{2}{\Omega_{p-1}}\,\delta(1-\vec x^2)$ where $\Omega_{p-1}$ is
the volume of the unit $p-1$-sphere.

When considering models with the quartic potential, $a\vert
\vec{x}\vert^2+b\vert \vec{x}\vert^4$, we find that for $p=1,2$ and
$3$ the system has a phase transition at the critical values
$a_c=-2\sqrt{b}$ for $p=1$; $a_c=0$ for $p=2$ and the surprising
positive value $a_c=\frac{\sqrt{20 b}}{3}$ for $p=3$.  There is no
transition for $p > 4$, rather the distribution is concentrated on
the sphere irrespective of the potential.

The principal results of this paper are:
\begin{itemize}
\item We find that there is a special r\^ole played by $p=4$, it is
  the critical dimension where shell solutions become the
  energetically preferred eigenvalue configurations.
\item The eigenvalue distributions for Gaussian ensembles of $p$
  rotationally invariant commuting matrices with $p=2,3$ and $4$ can
  be obtained by lifting the Wigner semi-circle distribution. The
  distribution for $p=4$ is a spherical $\delta$-function shell. The
  distributions for $p>4$ are $\delta$-function shells but cannot be
  obtained by lifting the Wigner distribution. We derive an analytic
  technique for the reduction (or lifting) of commuting models with
  arbitrary rotationally invariant potentials.
\item Commuting matrices with quartic potential $V(\vec{x})=a\vert
  \vec{x}\vert^2+b\vert \vec{x}\vert^4$ have phase transitions of 3rd
  order for $p=1$, 6th order for $p=2$ and 4th order for $p=3$. In
  these transitions the eigenvalue density undergoes a one-cut to
  two-cut transition for $p=1$, a disk to annulus transition for $p=2$
  and a ball to shell transition for $p=3$. For $p\ge 4$ there is a
  phase transition from a spherical shell to a metastable phase
  comprising a mixture of shell and uniform distributions. The
  metastable phase exists only for negative $b$ and sufficiently large
  $a$.
\item The critical transitions occur at $a_c=-2\sqrt{b}$ for $p=1$,
  $a_c=0$ for $p=2$ and $a_c=\sqrt{20b}/3$ for $p=3$.
  For $p=4$ the metastable shell-mixture transition occurs at $b_c=0$
  with $a^2 >|6b|$. There is also an instability transition at
  $a^2=-6b$.  For all $p >4$ and $b>0$ the strong eigenvalue repulsion
  forces all of the eigenvalues onto a shell and there is no
  transition.
\end{itemize}

The structure of the paper is as follows: 

In Section \ref{TheModel} we
describe the family of commuting matrix models we consider and obtain
the integral equation satisfied by the joint eigenvalue distribution
for these systems in the large matrix size limit. We further show how
the eigenvalue density, integral kernel and effective action can be
reduced to a lower dimensional system and lifted back to the original
dimension due to rotational invariance. 

In section \ref{Gaussian model} we study Gaussian systems in different
dimensions. We show that for $p=2,3$ and $4$ the eigenvalue
distribution is simply the rotationally invariant lift of the Wigner
semicircle.  We further show that a further lift to $p =5$ does not
yield a normalisable positive distribution, however we establish by
studying the effective action that the least action is given by
spherical shells. Spherical shells are the preferred distributions for
all $p>4$. We finish the section by confirming this conclusion with
Monte Carlo simulations.

In Section \ref{NonGaussianPotentials} we study the quartic potential
$V(\vec{x})=a\vert\vec{x}\vert^2+b\vert \vec{x}\vert^4$ and study the
phase structure of these systems. We find that the well known 3rd
order transition at $a_c=-2\sqrt{b}$ of the $p=1$ model becomes a 6-th
order transition for $p=2$ and occurs at $a_c=0$ while for $p=3$ the
transition occurs at the positive value $a_c=\sqrt{20b}/3$ and is
fourth order. We conclude the section by showing that for $p=4$ there is 
a phase transition from a spherical shell eigenvalue distribution  to a 
metastable phase comprising a mixture of shell and uniform distributions. 
The metastable phase exists only for negative $b$ and sufficiently large $a$.

The paper finishes with our conclusions and discussion in Section
\ref{Conclusions}.

The results of this paper should have applications wherever an ensemble of 
commuting matrices form a good approximation.

\section{Commuting matrix model}\label{TheModel}
\subsection{The model}
We consider a commuting SO({p}) invariant $p$-matrix model with partition function:
\begin{equation}
{\cal Z}=\int\hat{\cal D}\vec X\, e^{-N\,{\rm tr}\, V_p[\vec X] }    \ ,\label{partition1}
\end{equation}
where $\vec X$ is an array of $p$, $N\times N$ commuting hermitian
matrices, $\hat{\cal D}\vec X$ is the corresponding invariant measure
and $V(\vec X)$ is an SO({p}) invariant potential. The set of
commuting hermitian matrices $\vec X$, can be parameterised by a set
of real diagonal matrices $\vec\Lambda$ and an unitary matrix $U$:
\begin{equation}
\vec X =U^{\dagger}\,\vec\Lambda\,U\ .
\end{equation}
The corresponding Jacobian is given by:
\begin{equation}
J=\left(\prod_{i\neq j}|\vec\lambda_i-\vec\lambda_j|\right)\,{\rm det}\Big|\Big|\frac{\delta\theta_{rs}}{\delta u_{lm}}\Big|\Big|\ ,\label{Jacob}
\end{equation}
where $\theta =U^{\dagger}\,dU$ and $u_{lm}$ are coordinates on $SU(N)$. The partition function (\ref{partition1}) can be written as:
\begin{equation}\label{partition2}
\frac{{\cal Z}}{{\rm vol}\,SU(N)}=\,\int \prod_{i} d^p\lambda_i \,\, e^{-N^2\,\left[\frac{1}{N}\sum\limits_{i}\,V_p(|\vec \lambda_i|)-\frac{1}{2\,N^2}\sum\limits_{i\neq j}\,\log (\vec\lambda_i-\vec\lambda_j)^2 \right]}    \ .\end{equation}
The resulting effective action (we divide by $N^2$) for the eigenvalues $\vec\lambda$ is:
\begin{equation}
S_{\rm{eff}}[\vec \lambda]=\frac{1}{N}\sum_{i}\,V_p(|\vec \lambda_i|)-\frac{1}{2N^2}\sum_{i\neq j}\log(\vec\lambda_i-\vec\lambda_j)^2\ .\label{eff-action}
\end{equation}
At large $N$ the dynamics is dominated by the saddle point. Varying with respect to $\lambda_i$ we obtain:
\begin{equation}\label{saddle}
\frac{V_p'(|\vec\lambda_i|)}{2|\vec\lambda_i|}\vec\lambda_i = \frac{1}{N}\sum_{j}\,\frac{\vec\lambda_i-\vec\lambda_j}{(\vec\lambda_i-\vec\lambda_j)^2}
\end{equation}
Equation (\ref{saddle}) determines the eigenvalue distribution in the large $N$ limit and admits rotationally invariant shell solutions. The only shell solution consistent with SO(p) invariance is a $p-1$ dimensional spherical shell. These solutions have been considered in refs.~\cite{Berenstein:2005aa,Berenstein:2005jq, Aharony:2007rj,Berenstein:2008eg} for gaussian potential, where the authors argued that the radius of the spherical shell is independent on the number of the commuting matrices. One can show that the same holds for any potential. Indeed, it is straightforward to verify that the vector equation (\ref{saddle}) is satisfied by a homogeneous spherical eigenvalue  distribution of radius $R$, provided the radius satisfies:
\begin{equation}\label{shell}
R\,V_p'(R)=1
\end{equation}
Equation (\ref{saddle}) admits also $p$-dimensional (``fat'') rotationally invariant solutions, which may or may not be energetically favoured relative to the shell solution. To explore these solutions we consider a course grained approximation: 
\begin{equation}
\vec\Lambda_i\to \vec x\ ,\quad\quad\frac{1}{N}\sum_i\to \int d^p x\,\rho_{p}(\vec x)\ 
\end{equation}
and extremize the following functional:
\begin{eqnarray}\label{effectiveS}
S_{p}[\rho_p]&=&\int d^p x\,\rho_{p}(\vec x)V_p(|\vec x|)-\frac{1}{2}\int\int d^p x \, d^p x'\rho_{p}(\vec x)\,\rho_{p}(\vec x')\log(\vec x -\vec x')^2\nonumber\\
&+&\mu_p\left(\int d^p x\rho_{p}(\vec x)-1\right)
\end{eqnarray}
Upon variation with respect to $\rho$ we obtain the integral equation:
\begin{equation}\label{int-eq-V}
\mu_p+V_p(|\vec x|)=\int d^p x'\,\rho_{p}(\vec x')\,\log (\vec x  -\vec x')^2\ ,
\end{equation}
differentiating equation (\ref{int-eq-V}) with resect to $\vec x$ we obtain:
\begin{equation}\label{diff-int-eq-V}
\frac{V_p'(|\vec x|)}{2\,|\vec x|}\,\vec x=\int d^p x'\,\rho_{p}(\vec x')\,\frac{\vec x-\vec x'}{(\vec x-\vec x')^2}\ ,
\end{equation}
which we recognise as the continuous  limit of equation (\ref{saddle}). This of course is not surprising, since as long as we are dealing with $p$-dimensional distributions it shouldn't matter when we take the continuous limit. It turns out that instead of directly solving equation (\ref{diff-int-eq-V}) in $p$ dimensions one can use the properties of the logarithmic kernel in equation (\ref{effectiveS}) to reduce the problem to a lower dimensional one. Let us study this in more details. 

\subsection{Reducing the effective action}
The rotational invariance of the potential $V_p(\vec x)$ suggests that the system settles in a rotationally invariant eigenvalue distribution, which is fully characterised by its radial distribution. In such cases the distribution can be reduced to lower dimensions without any loss of information. Furthermore, the reduced distribution can also be lifted back to higher dimensions. This opens up the possibility to reduce a higher dimensional problem down to one or two dimensions where it can be analysed more easily, the obtained one-dimensional distribution can then be lifted back to higher dimensions. What makes this approach valuable is that the logarithmic kernel in the effective action (\ref{effectiveS}) is preserved under such dimensional reduction. Furthermore, for polynomial potential the reduction just alters the coefficients of the polynomial. This suggests (naively) that the saddle point equation of a given problem can be analysed only in one dimension and the solution to the analogous problem in higher dimensions 
can be obtained by simply lifting the one dimensional distribution. It turns out that this is the case only for distributions with simple topology and the description breaks down if the distribution undergoes a topology change transition (look at section \ref{NonGaussianPotentials}). This still leaves a large class of problems for which reducing the distribution can be useful. To describe how this works let us first focus on the reduction from $p$ to $p-1$ and $p-2$ dimensions, we have:
\begin{eqnarray}\label{ptop-1}
\rho_{p-1}(\vec x)&=&\int\limits_{-\sqrt{R^2-\vec x^2}}^{\sqrt{R^2-\vec x^2}}\,dy\,\rho_{p}(\sqrt{\vec x^2+y^2})\ ,\\
\rho_{p-2}(\vec x)&=&2\pi\,\int\limits_{0}^{\sqrt{R^2-\vec x^2}}\,dr\, r\, \rho_p(\sqrt{\vec x^2+r^2}) \ .\label{ptop-2}
\end{eqnarray}
These relations can be inverted by solving the integral equations (\ref{ptop-1}) and (\ref{ptop-2}). The result is \cite{Filev:2013pza}:
\begin{eqnarray}\label{p-1top}
\rho_p(x)&=&\frac{1}{\pi\,x}\frac{d}{dx}\int\limits_R^x dr\,\frac{\rho_{p-1}(r)\,r}{\sqrt{r^2-x^2}}\\
\rho_p(x)&=&-\frac{{\rho_{p-2}\,'(x)}}{2\pi\,x}\label{p-2top}
\end{eqnarray}
Our strategy is to describe how the $p$ dimensional action (\ref{effectiveS}) reduces to $p-2$ dimensions and then to show how a two dimensional action reduces to one dimension. In this way we can reduce both odd and even dimensional actions down to one dimension. Let us begin by reducing the potential term in (\ref{effectiveS}). Using equation (\ref{p-2top}) we obtain:
\begin{equation}
\int d^px\, V_p(x)\rho_p(x)=-\frac{1}{2\pi}\Omega_{p-1}\int dx\,x^{p-2}V_p(x)\rho_{p-2}\,'(x)=\int d^{p-2}x V_{p-2}(x)\rho_{p-2}(x)\ ,
\end{equation}
where:
\begin{equation}\label{Vptop-2}
V_{p-2}(x)=\left(\,1+\frac{1}{p-2}\,x\,\frac{d}{dx}\,\right)V_p(x)\ 
\end{equation}
and $\Omega_{p-1}$ is the volume of the $p-1$ dimensional sphere. Note that if $V_p$ is a polynomial of $x$ of a certain degree, $V_{p-2}$ is also a polynomial of the same degree, just the coefficients change according to (\ref{Vptop-2}). 

Next we focus on the reduction of the logarithmic kernel in (\ref{effectiveS}). Using the rotational invariance of the distribution $\rho_p(x)$, we can write:
\begin{equation}\label{reduction1}
\int\int d^p x \, d^p x'\rho_{p}(\vec x)\,\rho_{p}(\vec x')\log(\vec x -\vec x')^2=\int\int d x \, d x' \,\rho_{p}(x)\,K_p(x,x')\,\rho_{p}(x'),
\end{equation}
where the kernel $K_p(x,x')$ is given by:
\begin{equation}\label{Kp}
K_p(x,x')=\frac{4\pi^p}{\Gamma(\frac{p}{2})^2}\,x^{p-1}\,x'^{p-1}\,\left(\log\left(x^2+x'^2\right)-\frac{a}{2p}\,_3F_2(1,1,3/2;2,1+p/2;a)   \right)\ ,
\end{equation}
where $a=(4x^2x'^2)/(x^2+x'^2)^2$. Substituting equation (\ref{p-2top}) for $\rho_p(x)$ into equation (\ref{reduction1}) and integrating by parts for $p>2$ we obtain:
\begin{equation}
\int\int d^p x \, d^p x'\rho_{p}(\vec x)\,\rho_{p}(\vec x')\log(\vec x -\vec x')^2=\int\int d x \, d x' \,\rho_{p-2}(x),\rho_{p-2}(x')\,\frac{\partial^2}{\partial x\partial x'}\left(\frac{K_p(x,x')}{4\pi^2 x x'}\right)\ .
\end{equation}
One can show that:
\begin{equation}
\frac{\partial^2}{\partial x\partial x'}\left(\frac{K_p(x,x')}{4\pi^2 x x'}\right)=K_{p-2}(x,x')+\frac{2}{p-2}\Omega_{p-3}^2\,x^{p-3}x'^{p-3}\ ,
\end{equation}
where $\Omega_{p-3}$ is the volume of the unite $p-3$ sphere. For the reduced logarithmic term we obtain:
\begin{eqnarray}\label{reduced2}
\int\int d^p x \, d^p x'\rho_{p}(\vec x)\,\rho_{p}(\vec x')\log(\vec x -\vec x')^2&=&\int\int d^{p-2} x \, d^{p-2} x'\rho_{p-2}(\vec x)\,\rho_{p-2}(\vec x')\log(\vec x -\vec x')^2\nonumber\\
&&+\frac{2}{p-2}\left(\int d^{p-2} x\, \rho_{p-2}(\vec x)\right)^2\ .
\end{eqnarray}
Using (\ref{p-2top}) one can also show that:
\begin{equation}\label{reduced3}
\int d^p x \,\rho_p(\vec x)=\int d^{p-2} x\, \rho_{p-2}(\vec x)\ .
\end{equation}
Equation (\ref{reduced3}) implies that if $\rho_p$ is normalised to one so is $\rho_{p-2}$, which suggests that the last term on the right-hand site of equation (\ref{reduced2}) is just the constant $2/(p-2)$. Finally defining $\mu_{p-2}=\mu_p$ we can write:
\begin{equation}\label{SptoSp-2}
S_p[\rho_p]=S_{p-2}[\rho_{p-2}]-\frac{1}{p-2}\left(\int d^{p-2} x\, \rho_{p-2}(\vec x)\right)^2
\end{equation}
and because the equations of motion for $\mu_p$ and $\mu_{p-2}$ imply that both $\rho_p$ and $\rho_{p-2}$ are normalised to one, the effective actions $S_p$ and $S_{p-2}$ differ only by a constant and describe equivalent physics.

If $p$ is odd and $p>2$ one can repeat this procedure until one reduces the problem down to one dimension. For the relation between the effective actions one obtains:
\begin{equation}\label{SptoS1}
S_p[\rho_p]=S_1[\rho_1]-(\log2+\frac{1}{2}H_{p/2-1})\left(\int d x\, \rho_{1}(x)\right)^2\ ,
\end{equation}
where $H_n$ is the harmonic number. For the saddle point equation for $\rho_1$ we obtain:
\begin{equation}\label{int-eq-V1}
-2\log2-H_{p/2-1}+\mu_p+V_1(x)=\int dx'\,\rho_{1}(x')\,\log ( x  - x')^2\ ,
\end{equation}
where $V_1$ is reduced using equation (\ref{Vptop-2}) and we have used that $\mu_1=\mu_p$ by definition. One can show that equations (\ref{SptoS1}) and (\ref{int-eq-V1}) are still valid for even $p$. Indeed, for even $p$ one can use equation (\ref{SptoSp-2}) to reduce to two dimensions arriving at:
\begin{equation}\label{SptoS2}
S_p[\rho_p]=S_2[\rho_2]-\frac{1}{2}H_{p/2-1}\,\left(\int d^2 x\, \rho_{2}(\vec x)\right)^2\ ,
\end{equation}
Finally, one can use equation (\ref{p-1top}) (see Appendix \ref{AppendixA}) to show that:
\begin{equation}
\label{S2toS1}
S_2[\rho_p]=S_1[\rho_1]-\log2\,\left(\int d x\, \rho_{1}(x)\right)^2\ ,
\end{equation}
now combining equations (\ref{SptoS2}) and (\ref{S2toS1}) one verifies that equation (\ref{SptoS1}) is valid also for even number of commuting matrices $p$.

\section{Gaussian model}\label{Gaussian model}

In this section we  focus on the properties of commuting matrix models with a quadratic potential: 
\begin{equation}\label{Gaussian-p}
V_p(|\vec x|)=\frac{1}{2}\,\vec x^{\,2}\ .
\end{equation}
We begin by studying the joint eigenvalue distributions for various number of commuting matrices.

\subsection{Gaussian model in various dimensions}\label{gaussian-vardim}

Using equation (\ref{Vptop-2}) one can reduce the potential (\ref{Gaussian-p}) to two or one dimensions depending on whether $p$ is even or odd. In even dimensions one can use equation (\ref{V2toV1}) to reduce the potential further to one dimension. It is easy to verify that the reduced potential is:
\begin{equation}
V_1(x)=\frac{p}{2}\,x^{\,2}\ .
\end{equation}
Substituting $V_1$ into equation (\ref{int-eq-V1}) and differentiating with respect to $x$ we obtain the integral equation:
\begin{equation}
\frac{p}{2}\,x =\int\limits_{-R}^{R} dx'\,\frac{\rho_{(1)}(x)}{x-x'}\ ,
\end{equation}
whose solution is a Wigner semi-circle, which if normalised to one has a radius $R_p^2=4/p$:
\begin{equation}\label{Wigner}
\rho_{(1)}(x)=\frac{p}{2\pi}\sqrt{4/p-x^2}\ .
\end{equation}
Therefore we conclude that for gaussian potential the $p$-dimensional joint eigenvalue distribution is obtained by lifting a Wigner semi-circle distribution using equations (\ref{p-1top}) and (\ref{p-2top}). Let us see how this works in different dimensions.\\
{\bf For $p=1$} we trivially obtain a Wigner semi-circle of radius $R_1=2$. \vspace{0.1cm}\\
{\bf For $p=2$} using equation (\ref{p-1top}) we obtain that the joint eigenvalue distribution is a uniform disk of radius $R_2=\sqrt{2}$:
\begin{equation}\label{WG->2}
\rho_{(2)}(\vec x)=\frac{1}{2\pi}\Theta(2-\vec x^2)\ .
\end{equation}
The distribution (\ref{WG->2}) can easily be obtained directly in two dimensions by using the fact that $\log(\vec x-\vec x')^2$ is proportional to the Green's function of the laplacian in two dimensions (see for example ref. \cite{Berenstein:2005aa}). \vspace{0.1cm} \\
{\bf For $p=3$} we use equation (\ref{p-2top}) to lift the Wigner semi-circle (\ref{Wigner}). We obtain:
\begin{equation}\label{WG->3}
\rho_{(3)}(\vec x)=\frac{3}{4\pi^2}\frac{1}{\sqrt{\frac{4}{3}-\vec x^2}}\ .
\end{equation}
 The distribution in equation (\ref{WG->3}) diverges at the boundary, however it is still integrable. In the next subsection we will compare this distribution to Monte Carlo simulations at large (but finite) $N$ and we will confirm that it is indeed approached by the physical distribution in the large $N$ limit. \vspace{0.1cm}\\
{\bf For $p=4$} it is convenient to first lift the Wigner semi-circle (\ref{Wigner}) to two dimensions using equation (\ref{p-1top}) and then lift from two to four dimensions using equation (\ref{p-2top}). One easily obtains:
\begin{equation}\label{WG->4}
\rho_{(4)}(\vec x)=\frac{1}{\pi^2}\,\delta(1-\vec x^2)\ .
\end{equation}
Note that the distribution $\rho_{(4)}$ is a shell and is thus three (rather than four) dimensional. In fact this is the spherical shell saddle point that we analysed in the previous section. Indeed, if we substitute the potential (\ref{Gaussian-p}) into equation (\ref{shell}) we arrive at unit radius $R_4=1$. It is intriguing that the equation (\ref{WG->4}) which we derived under the assumption of a four dimensional (``fat") distribution agrees with the derivation of the shell saddle point above equation (\ref{shell}). In fact as we are going to see this is no longer the case for dimensions higher than four. \vspace{0.1cm}\\
{\bf For $p>4$} we run into troubles. In even dimensions $p=2n$ ($n>2$) using first equation (\ref{p-1top}) and then equation (\ref{p-2top}) one can show that the distribution is a shell proportional to derivatives of a delta function: $\rho_{(2n)}(\vec x)\propto\delta^{(n-1)}(2/n-\vec x^2)$, which is not a positive function and cannot represent joint eigenvalue distribution. In odd dimensions $p=2n+1$ ($n>1$). The distribution is not integrable. Indeed, using  equation (\ref{p-2top}) one can show that $\rho_{(2n+1)}(\vec x)\propto 1/(4/(2n+1)-\vec x^2)^{(n-1/2)}$, which is not integrable near the boundary for $n>1$.  Therefore we conclude that although the saddle point extremising the effective action (\ref{effectiveS}) can be constructed mathematically by lifting the Wigner semi-circle distribution (\ref{Wigner}), for dimensions higher than four ($p>4$) the mathematical solutions are not physical and cannot be realised as eigenvalue distributions. However the spherical shell saddles derived in the previous section still exist. It is then natural to conclude that for $p>4$ the joint eigenvalue distribution is given by \cite{Berenstein:2005aa}:
\begin{equation}\label{shell-p}
\rho_p(\vec x)=\frac{2}{\Omega_{p-1}}\,\delta(1-\vec x^2)\ ,
\end{equation}
where $\Omega_{p-1}$ is the volume of the unit $p-1$-sphere. 

Overall, we see that the eigenvalue distribution depends crucially on the number of commuting matrices. The different eigenvalue distributions can be split into two classes: The first class is for $p\leq 4$, when the joint eigenvalue distributions are obtained by lifting the Wigner semi-circle distribution (\ref{Wigner}). The second class is for $p\geq4$, when the spherical shell saddles are realised and the radius depends only on the shape of the potential but not on the dimension. Interestingly these two classes overlap at $p=4$ since the three-sphere shell can be obtained in both approaches. In the next subsection we analyse this behaviour and argue that it follows from the principle of least action, which should be valid in the large $N$ limit.

\subsection{Least action analysis}

\begin{figure}[htbp]
  \centering
   \includegraphics[width=10cm]{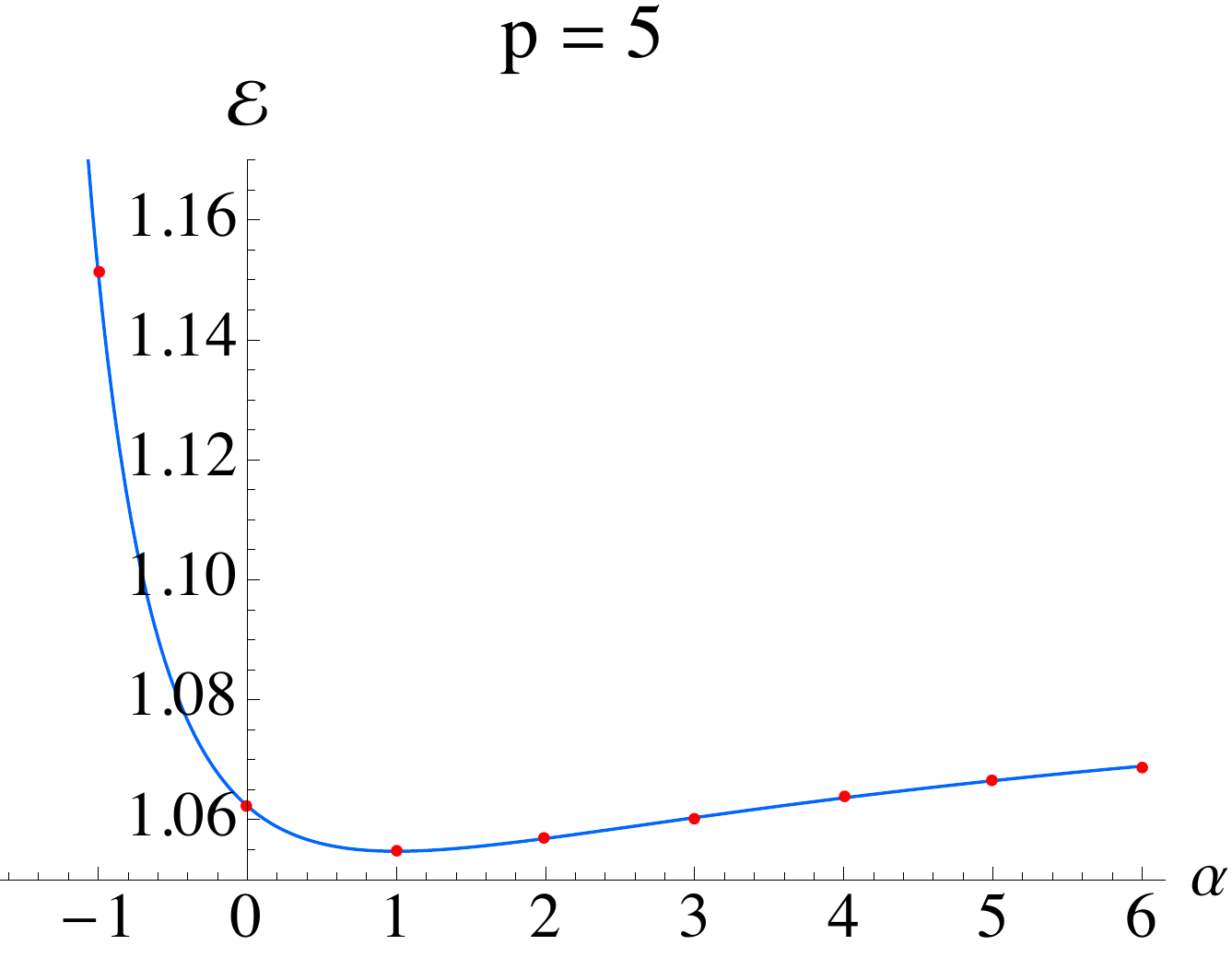}
   \caption{Plot of the $\mathcal{E}$ versus $\alpha$ for $p=5$. One can see that the minimum is realised at $\alpha=1$ and the the function is monotonically increasing for $\alpha>1$.}
   \label{fig:1}
\end{figure}

As we observed above for the gaussian potential (\ref{Gaussian-p}) the possible eigenvalue distributions split into two classes. In particular, we showed that for $p>4$ the joint eigenvalue distribution does not extremise the effective action (\ref{effectiveS}) and is given instead by a spherical shell of unit radius. However we could still reduce the spherical shell to one dimension. One can easily show that the spherical distribution (\ref{shell-p}) reduces to:
\begin{equation}\label{shell->1}
\rho_1^{p}(x)=\frac{\Gamma(\frac{p}{2})}{\pi^{1/2}\Gamma(\frac{p-1}{2})}\left(1-x^2\right)^{\frac{p-3}{2}}\ .
\end{equation}
Inspired by equation (\ref{shell->1}) we will assume that in general the reduced distribution is composed of terms of the form $(R^2-x^2)^{\alpha}$. If we define $\eta=R/x$, then for a very broad class of distributions the reduced distribution can be written as:
\begin{equation}\label{generic-rho1}
\rho(\eta)=\sum\limits_{n=1}^{\infty}c_n\, R^n\,\frac{\Gamma(\frac{n+3}{2})}{\pi^{1/2}\Gamma(\frac{n+2}{2})}\, (1-\eta^2)^{n/2}\ .
\end{equation}
The normalisation condition $\int\limits_{-R}^R dx\rho(x)=R\int\limits_{-1}^1 d\eta\rho(\eta)=1$ imposes the following constraint on the coefficients $c_n$ and $R$:
\begin{equation}\label{constr-1}
\sum\limits_{n=1}^{\infty}c_n\,R^{n+1}=1\ .
\end{equation}
It turns out that we can impose one more constraint on $c_n$ and $R$ without referring to the saddle point equation (see appendix B for the derivation for general potential). For the gaussian potential (\ref{Gaussian-p}) it reads:
\begin{equation}\label{gaussian-constraint}
\int d^px\rho_p(\vec x)\,\vec x^2=p\,\int\limits_{-R}^{R}dx\,\rho_1(x)\, x^2=1\ .
\end{equation}
Applying this to the distribution in (\ref{generic-rho1}) we obtain:
\begin{equation}\label{constr-2}
\sum\limits_{n=1}^{\infty}\frac{p}{n+3}c_n\,R^{n+3}=1\ .
\end{equation}
Clearly in general the two constraints in equations (\ref{constr-1}) and (\ref{constr-2}) are not sufficient to determine the coefficients $c_n$ and the radius $R$ in equation (\ref{generic-rho1}). However, they can determine these parameters for {\it pure states}, that is when only one of the coefficients $c_n$ is non-vanishing. If the non vanishing coefficient is $c_\alpha$ one easily obtains:
\begin{equation}\label{Ralpha}
R_\alpha^2=\frac{\alpha+3}{p}\ ,~~~c_\alpha=\left(\frac{p}{\alpha+3}\right)^{\frac{\alpha+1}{2}}\ .
\end{equation}
Let us consider such a pure state:
\begin{equation}\label{pure-state}
\tilde\rho_{\alpha}(x)=\frac{p^{\frac{\alpha+1}{2}}\Gamma(\frac{\alpha+3}{2})}{(\alpha+3)^{\frac{\alpha+1}{2}}\pi^{\frac12}\Gamma(\frac{\alpha+2}{2})}\, \left(\frac{\alpha+3}{p}-x^2\right)^{\alpha/2}\ .
\end{equation}
We will show that for a given $p$ the pure state with the lowest $\alpha\geq1$ has the lowest energy (Note also that in general we could take $\alpha$ to be continuous). To compare the energies of the different pure states we have to evaluate the reduced effective action $S_1$, however since the pure states are normalised to one and the potential term is fixed by the constraint (\ref{constr-2}) we need just to evaluate the term with the logarithmic kernel, thus we define:
\begin{equation}\label{Free-energy}
\mathcal{E}(\alpha)=-\frac{1}{2}\int\limits_{-R_\alpha}^{R_\alpha}dx\int\limits_{-R_\alpha}^{R_\alpha}dx'\,\tilde\rho_{\alpha}(x)\,\log(x-x')^2\,\tilde\rho_{\alpha}(x')\ .
\end{equation}
The easiest way to evaluate $\mathcal{E}_\alpha$ for integer $\alpha$ is to uplift the pure state (\ref{pure-state}) to $\alpha+3$ dimensions, where it is a spherical shell and use equation (\ref{SptoS1}) to evaluate $\mathcal{E}_\alpha$ (look at appendix C for a derivation). The result is:
\begin{equation}\label{Free-energy-final}
\mathcal{E}=\frac{3}{4}H_{\frac{\alpha+1}{2}}-\frac{1}{4}H_{\frac{\alpha}{2}}-\frac{1}{2}\,\log\frac{\alpha+3}{2\,p}\ .
\end{equation}
Let us calculate the derivative of $\mathcal{E}$ with respect to $\alpha$, we obtain:
\begin{equation}
\frac{\partial \mathcal{E}}{\partial\alpha}=-\frac{1}{2(\alpha+3)}-\frac{1}{8}\,\psi^{(1)}\left(\frac{\alpha+2}{2}\right)+\frac{3}{8}\,\psi^{(1)}\left(\frac{\alpha+3}{2}\right)\ ,
\end{equation}
where $\psi^{(1)}(x)$ is the polygamma function $\psi^{(m)}(x)\equiv\partial_x^m\log\Gamma(x)$. Note that $\frac{\partial \mathcal{E}}{\partial\alpha}$ is independent of $p$. One can also verify that: for $\alpha>1$ one has $\frac{\partial \mathcal{E}}{\partial\alpha}>0$, for $-2<\alpha<1$ one has $\frac{\partial\mathcal{E}}{\partial\alpha}<0$ and finally for $\alpha=1$ one has $\frac{\partial\mathcal{E}}{\partial\alpha}=0$. This clearly indicates that for $\alpha=1$, $\mathcal{E}$ has its minimum (look at figure \ref{fig:1}), as it should since for $\alpha=1$ the pure state is the Wigner semi-circle (\ref{Wigner}), which extremises the effective action $S_1$ and after uplift $S_p$. 
However, as we observed in the previous section for $p>4$ the joint eigenvalue is a shell of unit radius, which reduces to a pure state with $\alpha>1$. The reason is that the uplift of $\tilde\rho_{\alpha}$ from equation (\ref{pure-state}) is physical only up to $p=\alpha+3$ dimensions (when it is a shell). A further lift would produce either a negative shell (derivative of a delta function) or a non-integrable distribution. Therefore, for $p>4$ we cannot lift the Wigner semicircle and a pure state with $\alpha>1$ should be realised. Furthermore, since $\mathcal{E}$ is a monotonically increasing function of $\alpha$, for $\alpha>1$ we should always pick the lowest possible value of $\alpha$. This suggests that for $p=5$ we should pick $\alpha=2$, but this pure state can be lifted at most to $p=\alpha+3=5$ and therefore for $p=6$ we should pick the next one: $\alpha=3$. Following the same argument again, one concludes that in general for $p>4$ the pure state with $\alpha=p-3$ is realised, which is always a shell as equation (\ref{shell->1}) suggests. Furthermore, using equation (\ref{Ralpha}) for  the radius of the distribution we have that $R_{p-3}=(p-3+3)/p=1$. We arrive at the result that for $p\geq 4$ the radius is independent of the dimension and is equal to one, which is the same result that we obtained above using saddle point arguments. 

Now we have a better understanding why the spherical shell saddles
considered above equation (\ref{shell}) are not realised for $p<4$. It
is because the uplifts of the Wigner semi-circle (\ref{Wigner}) are
energetically preferred and whenever they are physical (correspond to
positive and integrable distribution) they are realised.

So far our analysis involved only pure states. In general we can have
a distribution which is a ``mixture'' of pure states (see equation
(\ref{generic-rho1})). However, the pure states have different
energies and it is plausible to assume that the pure state with the
lowest possible energy will have lower energy than any mixed state
since this will involve mixing with pure states of higher
energy. Generally this is not true for arbitrary potential. However,
for a gaussian potential the above considerations suggest that this is
the case. We also explicitly verified that pure states are
energetically more favoured than mixed states of two and three pure
states and believe that it is true for any mixed state.

\subsection{Monte Carlo simulation of the gaussian model}
In this subsection we perform Monte Carlo simulations of the gaussian
model with potential given in equation (\ref{Gaussian-p}). To this end
we implemented the algorithm of Metropolis into a C++ commuter
program. Over all we find excellent agreement with the distributions
derived in section \ref{gaussian-vardim}.

\begin{figure}[t]
  \centering
   \includegraphics[width=7.35cm]{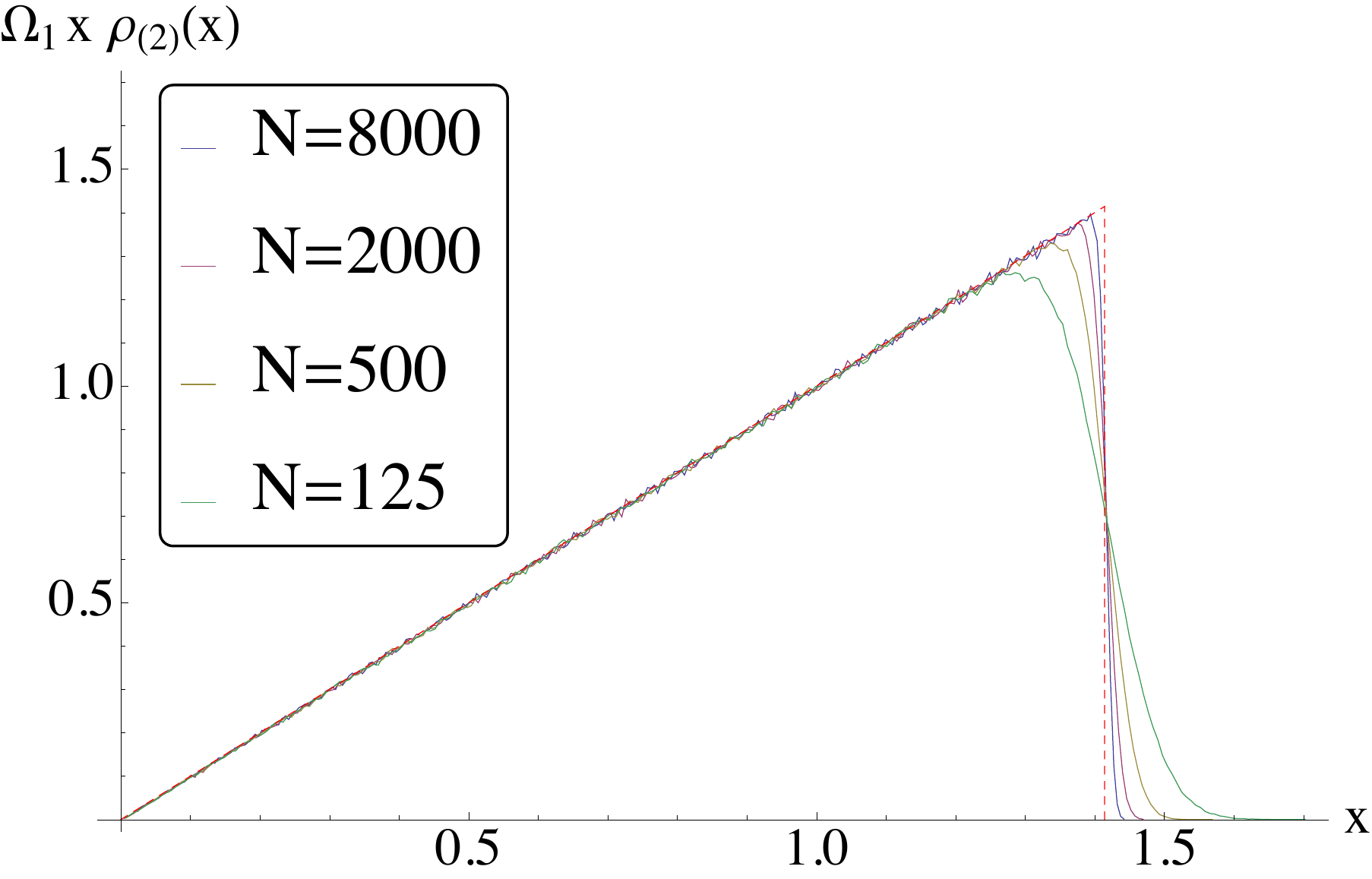} 
   \includegraphics[width=7.35cm]{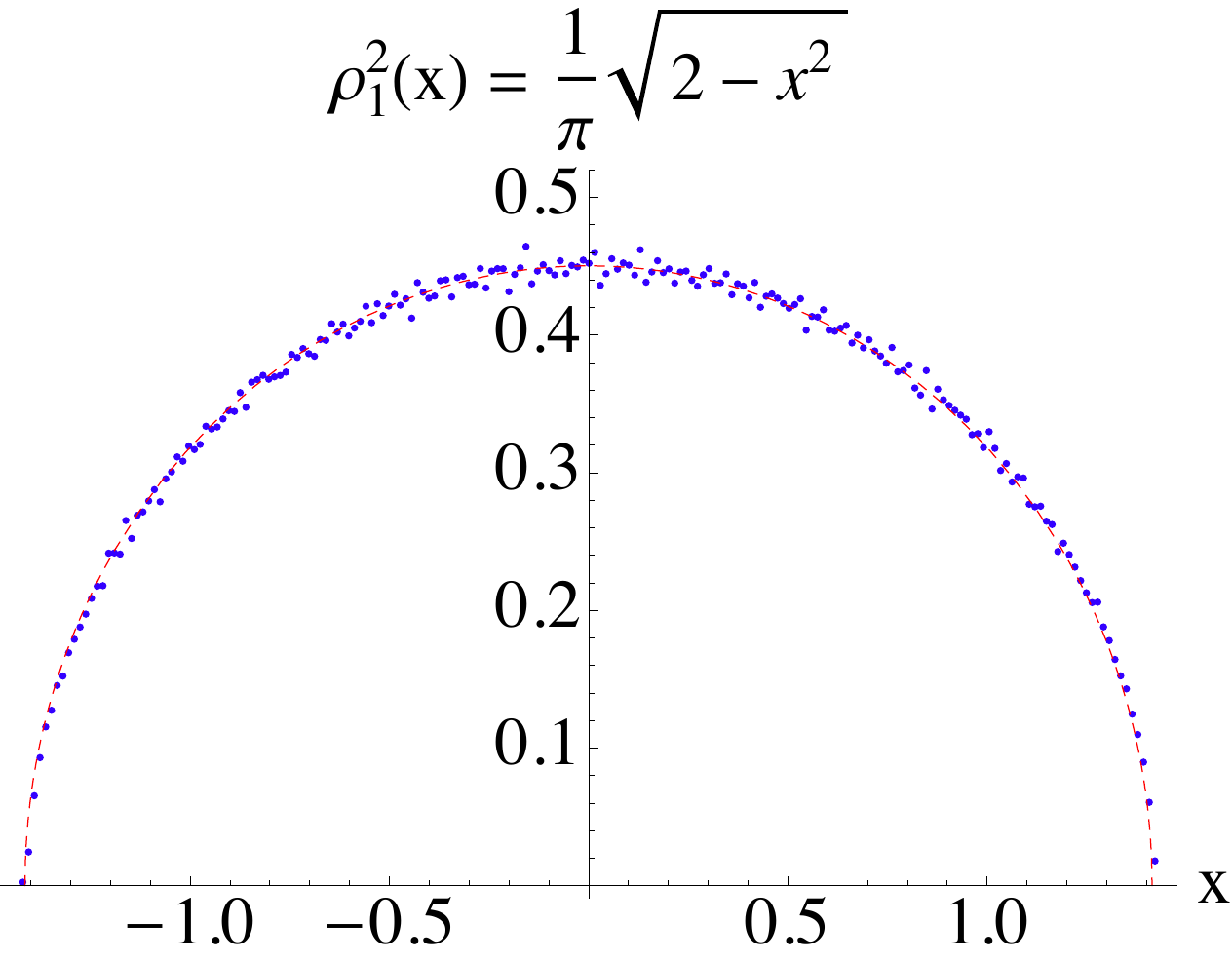} 
   \caption{Left panel:Plots of the radial distribution of two commuting matrices for $N=125,500,2000,800$. One can see that as $N$ increases the agreement with the theoretical result at $N\to\infty$ improves and at $N=8000$ it is already excellent. Right panel:A plot of the reduced distribution for $N=8000$. One can observe the excellent agreement with equation (\ref{Wigner}) for $p=2$.}
   \label{fig:2}
\end{figure}

For $p=1$ the model is just an ordinary one-matrix model with a Wigner
semi-circle distribution, therefore we will begin with the $p=2$
case. In this case the distribution is a uniform disk of radius
$\sqrt{2}$. Numerically it is easier to analyse the radial
distribution. Using equation (\ref{WG->2}) and that we are in two
dimensions we obtain:
\begin{equation}\label{radial-2}
\rho_{2}^{rad}(x)=2\pi\,x\rho_2(x)=x\,\Theta(2-x^2)\ .
\end{equation}
In the left panel of figure \ref{fig:2} we presented our numerical results for the radial distribution (\ref{radial-2}). The red dashed curve represents the $N\to\infty$ result (\ref{radial-2}). One can see that the agreement with the Monte Carlo simulations improves as the size of the matrices increases and at $N=8000$ it is already excellent.
In the right panel of figure \ref{fig:2} we present a plot of the reduced distribution (the distribution of one component of the eigenvalue). One can observe the excellent agreement of the numerical result for $N=8000$ with the Wigner semi-circle distribution from equation (\ref{Wigner}) for $p=2$.

\begin{figure}[t]
  \centering
   \includegraphics[width=7.49cm]{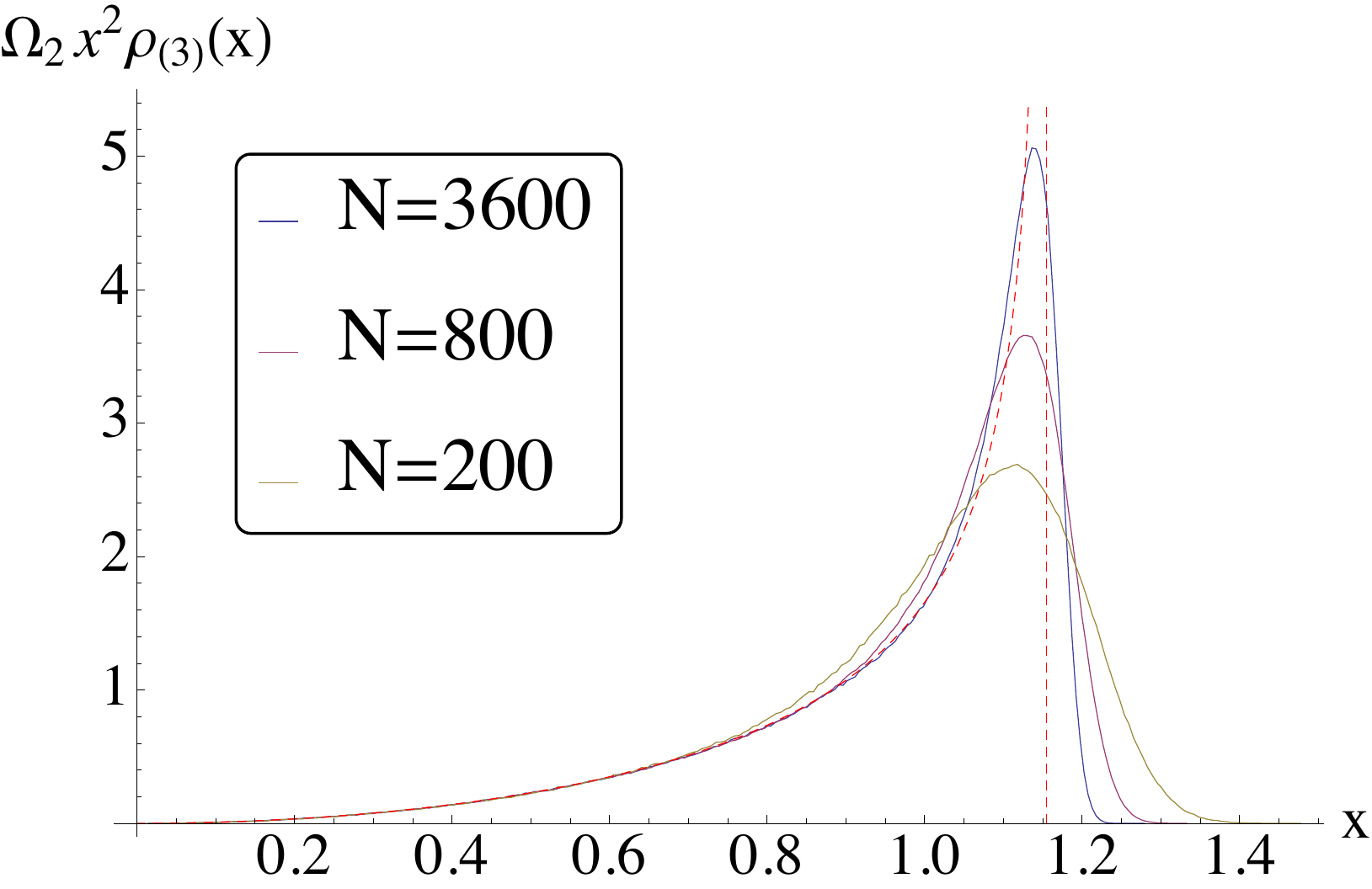} 
   \includegraphics[width=7.49cm]{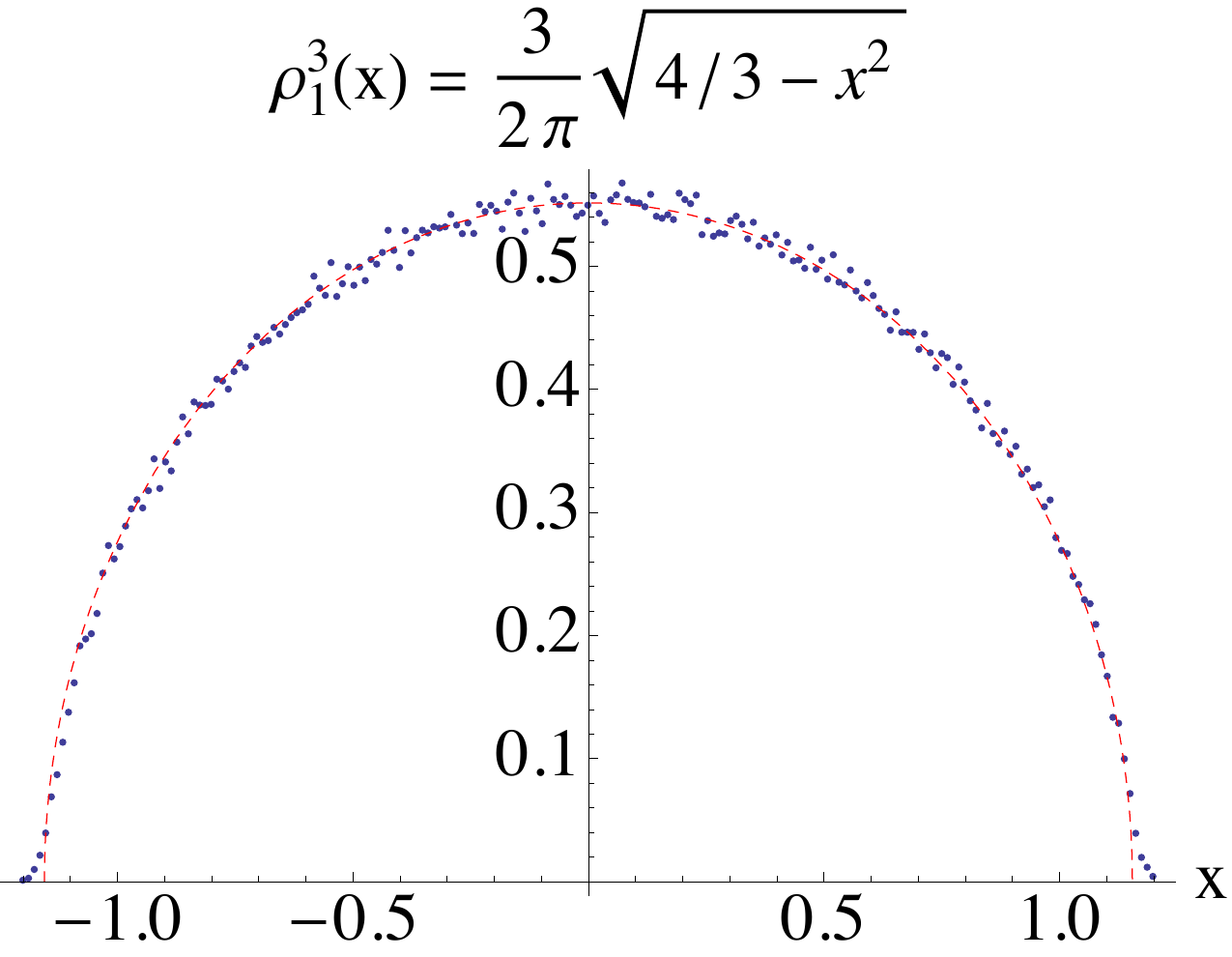} 
   \caption{Plots of the radial and reduced distributions for three commuting matrices. Left panel: One cans observe how the agreement with the theoretical result (\ref{radial-3}) agrees as the size of the matrices $N$ increases. Right panel: One can observe the excellent agreement of the reduced distribution  for $N=3600$ with the theoretical result (\ref{Wigner}).}
   \label{fig:3}
\end{figure}

Next we consider the $p=3$ case. Using equation (\ref{WG->3}) and that we are in three dimensions, for the radial distribution we obtain:
\begin{equation}\label{radial-3}
\rho^{rad}_3(x)=\frac{3}{\pi}\frac{x^2}{\sqrt{4/3-x^2}}\ .
\end{equation}
In figure \ref{fig:3} we have presented our numerical results for the radial distribution and for the reduced one. As one can see in the left panel of the figure the numerical results approach the theoretical curve (\ref{radial-3}) as the size of the matrices increases. In the left panel one can see the excellent agreement for the reduced distribution with the Wigner semi-circle (\ref{Wigner}) for $p=3$.

\begin{figure}[t]
  \centering
   \includegraphics[width=7.49cm]{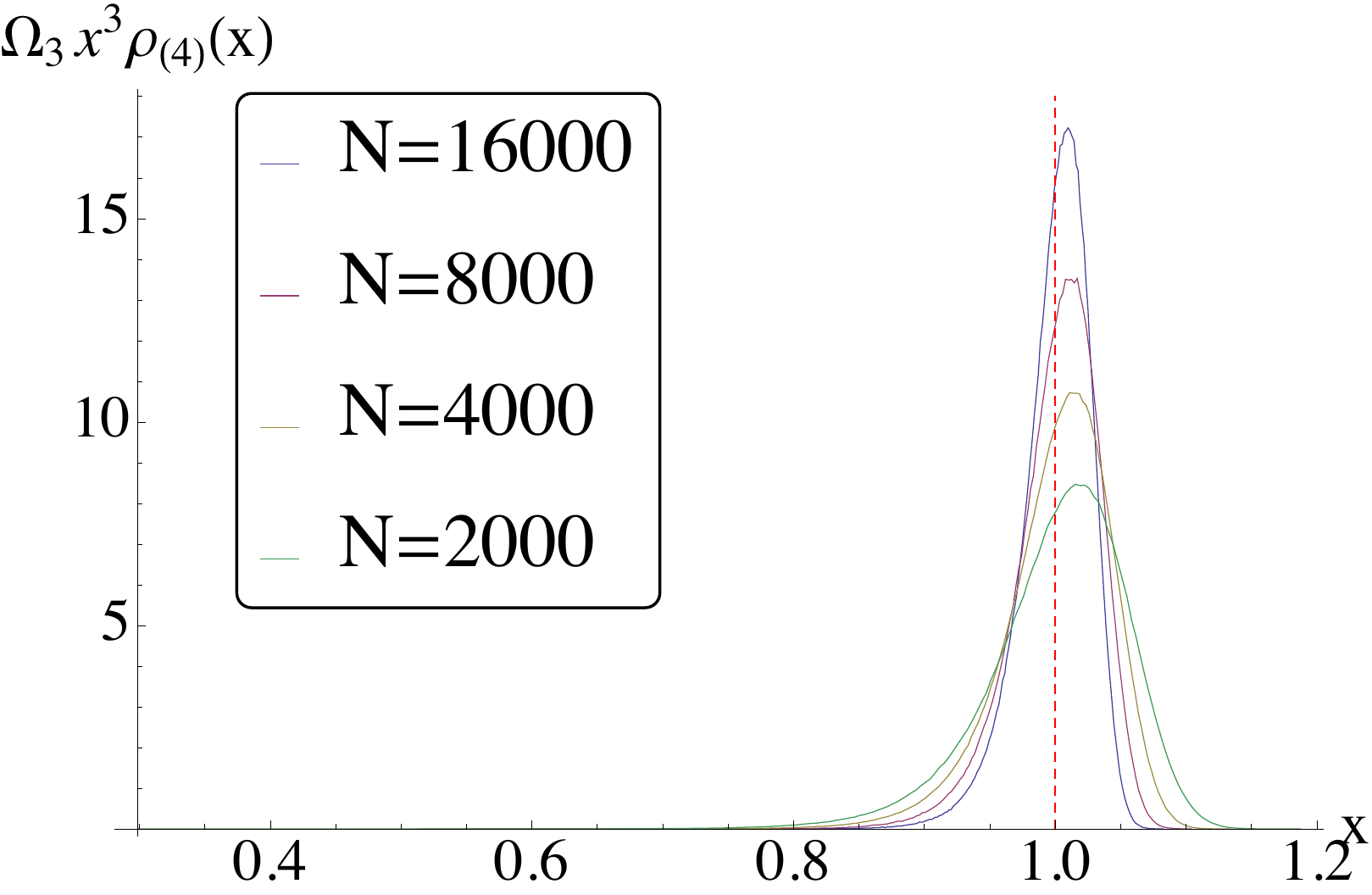} 
   \includegraphics[width=7.49cm]{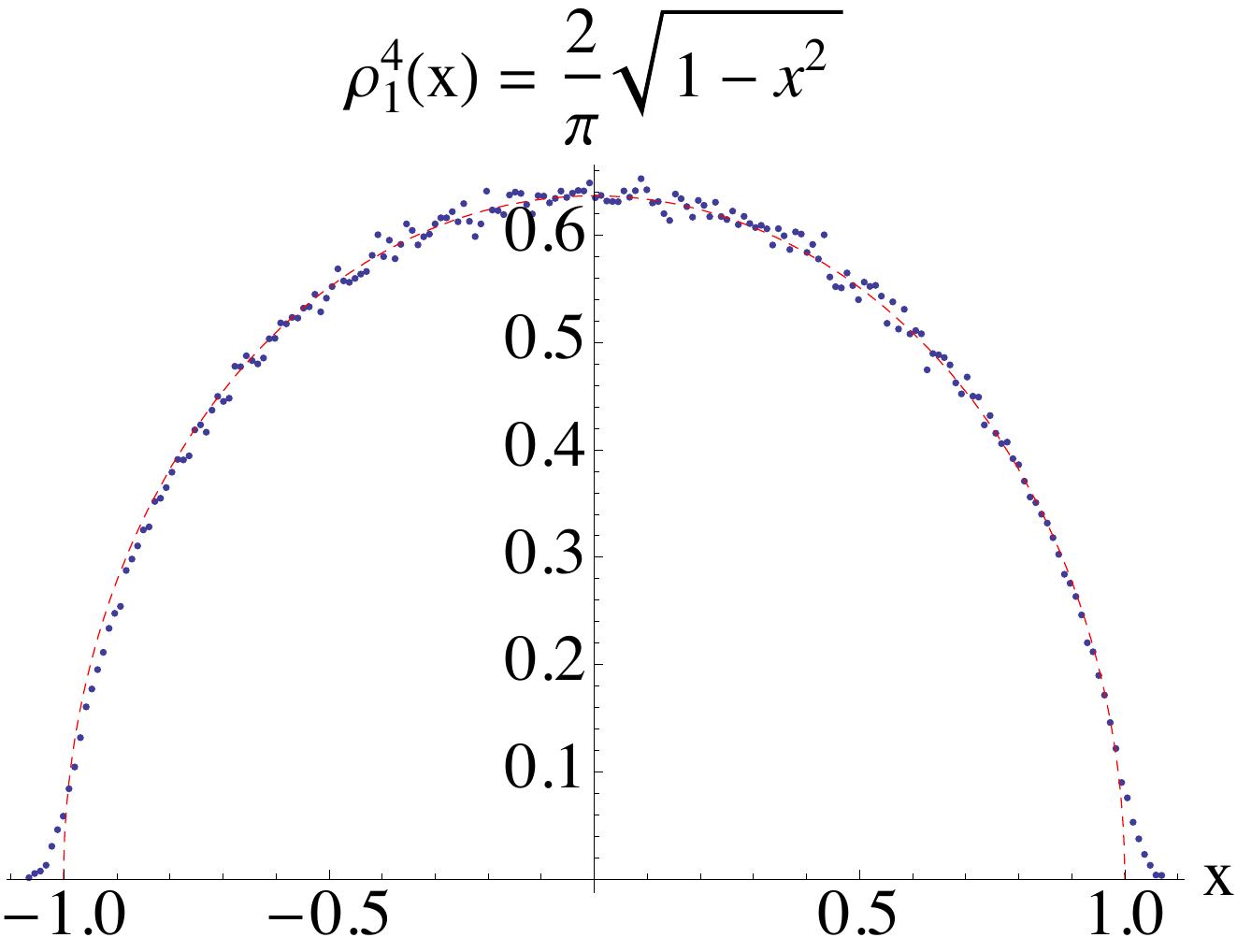} 
      \includegraphics[width=7.49cm]{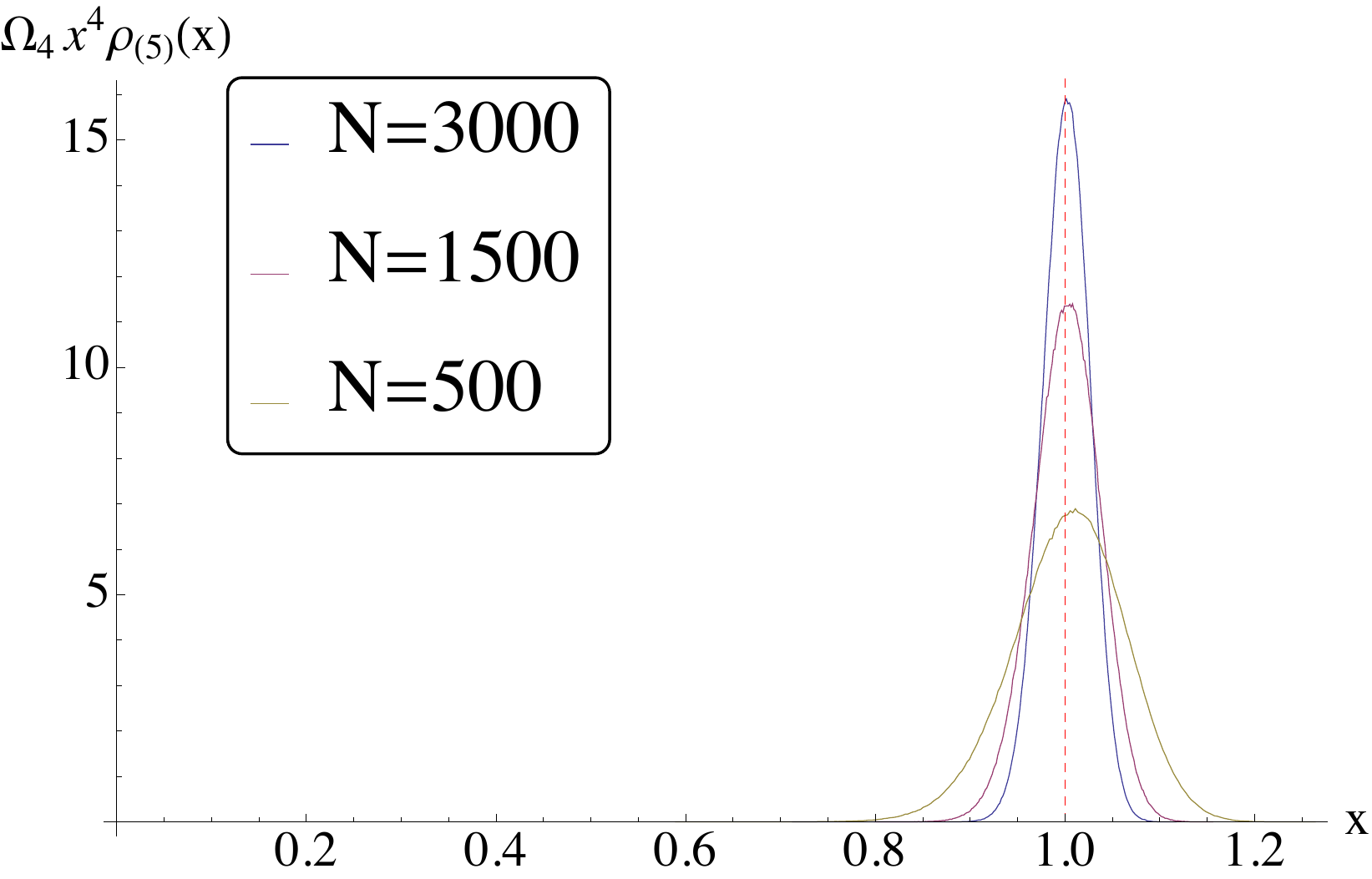} 
   \includegraphics[width=7.49cm]{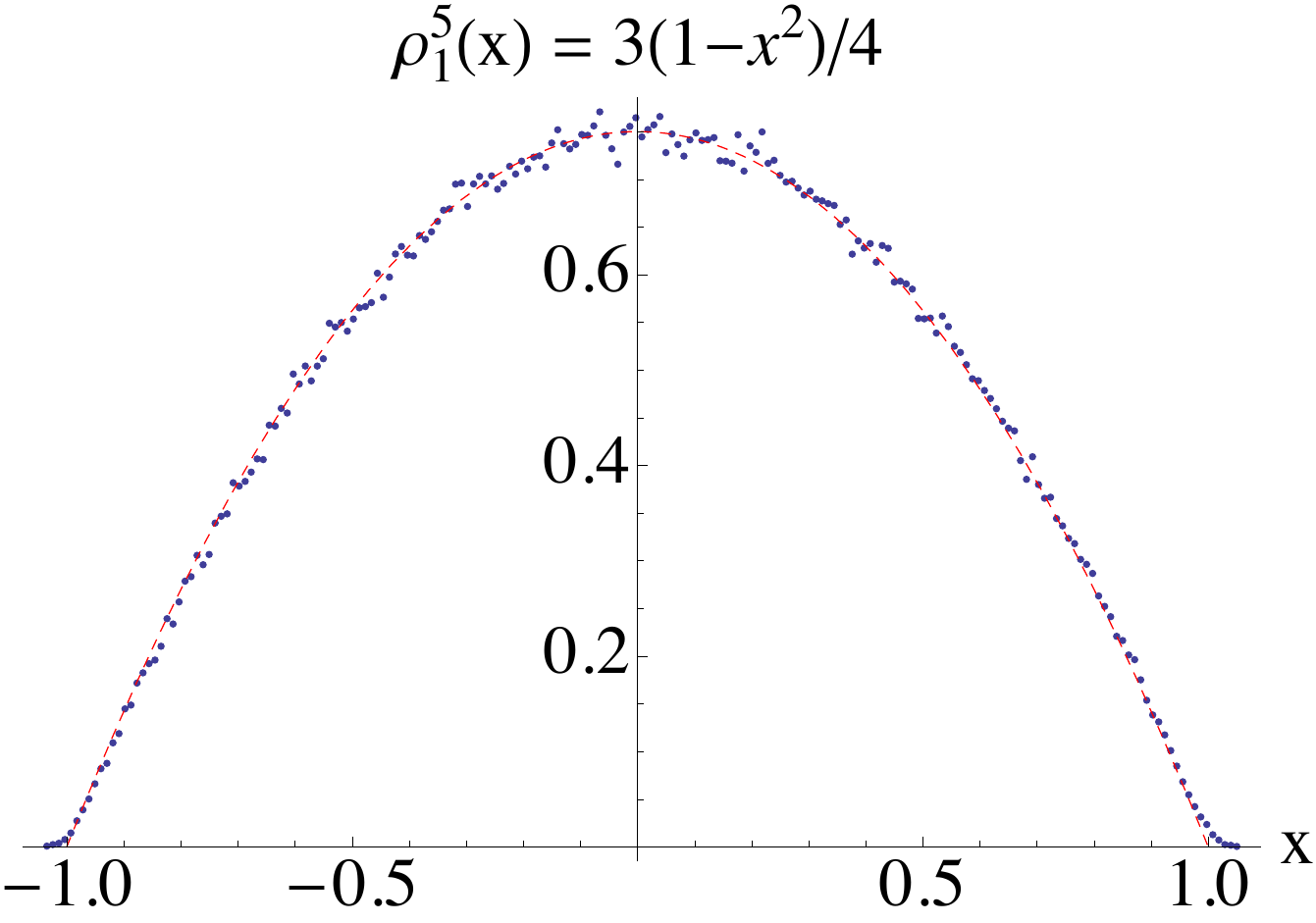} 
   \caption{Plots of the radial and reduced distributions for $p=4,5$. One can see that as $N$ is increased the radial distributions approach spherical shells of unit radius. One can also see an excellent agreement of the reduced distributions with equation (\ref{shell->1}).}
   \label{fig:4}
\end{figure}

\begin{figure}[t]
  \centering
   \includegraphics[width=7.49cm]{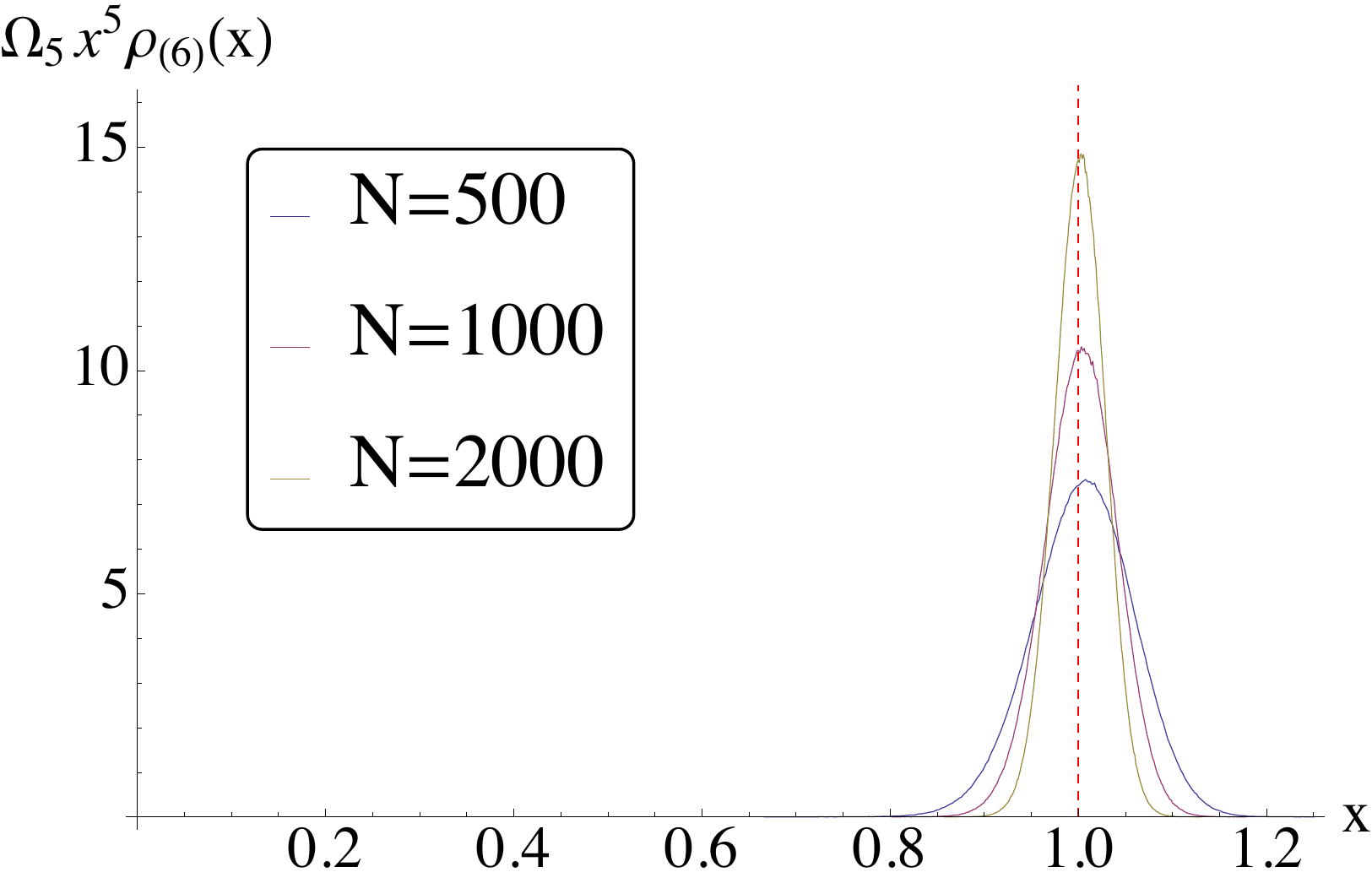}
   \includegraphics[width=7.49cm]{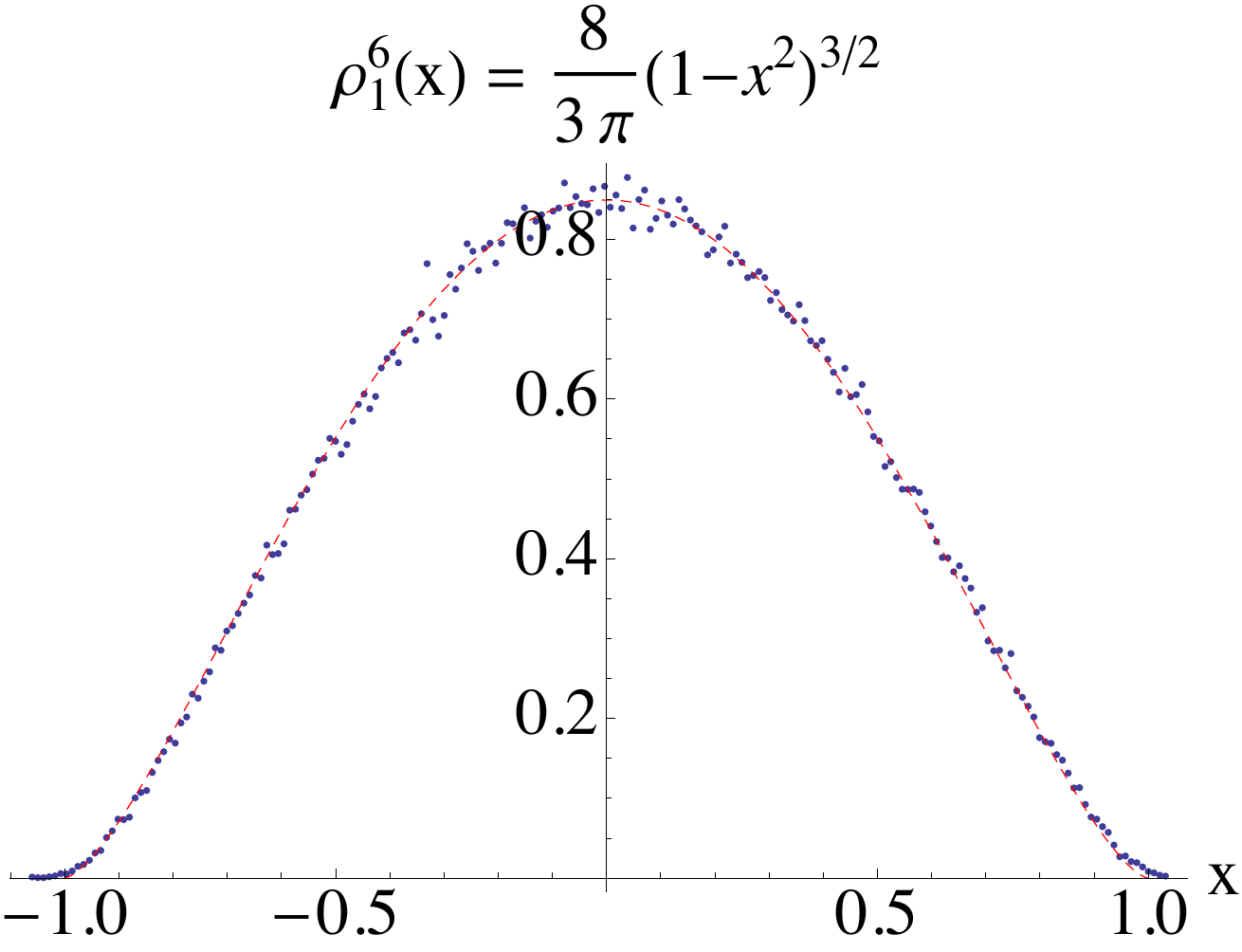} 
      \includegraphics[width=7.49cm]{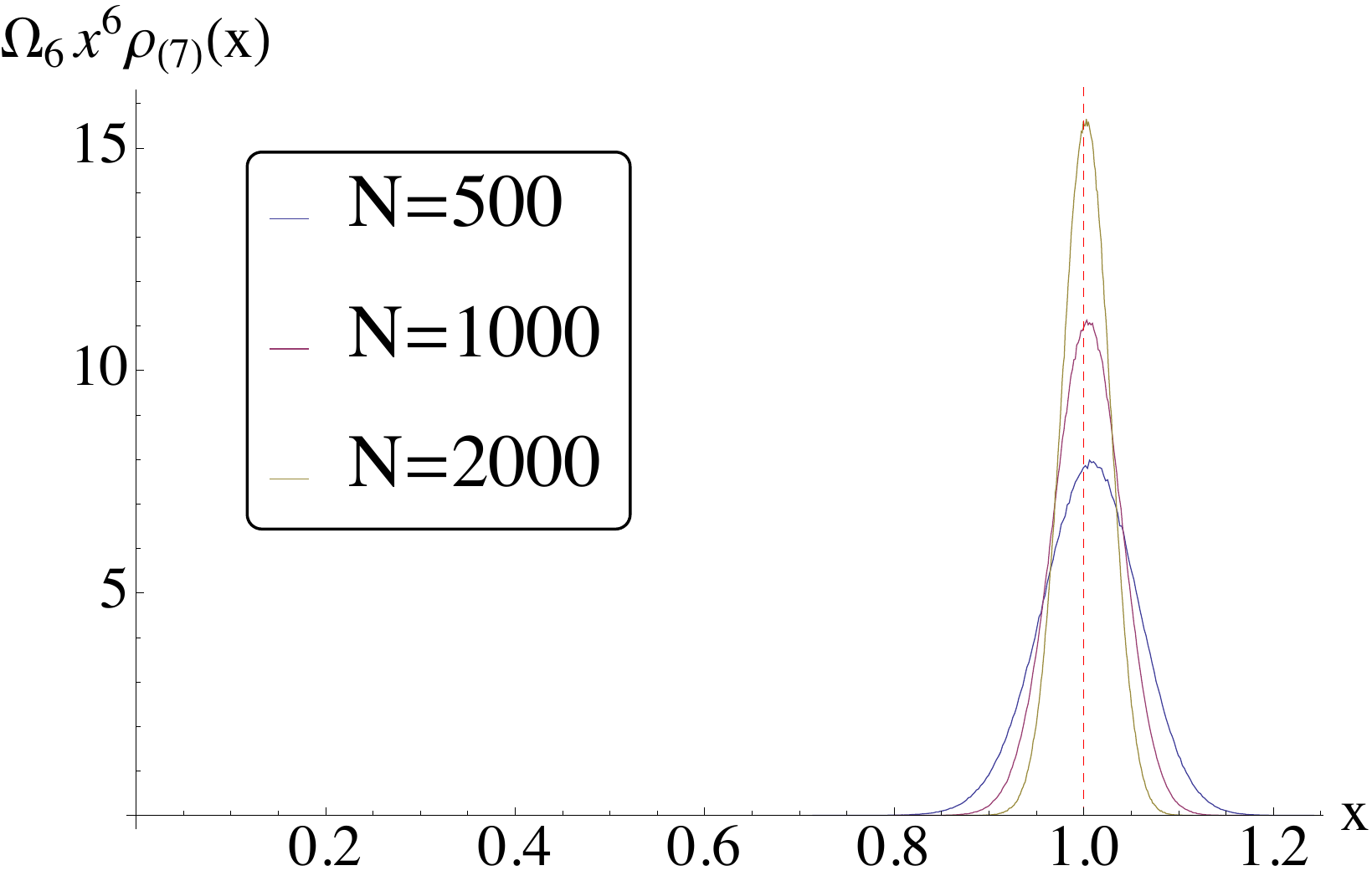} 
   \includegraphics[width=7.49cm]{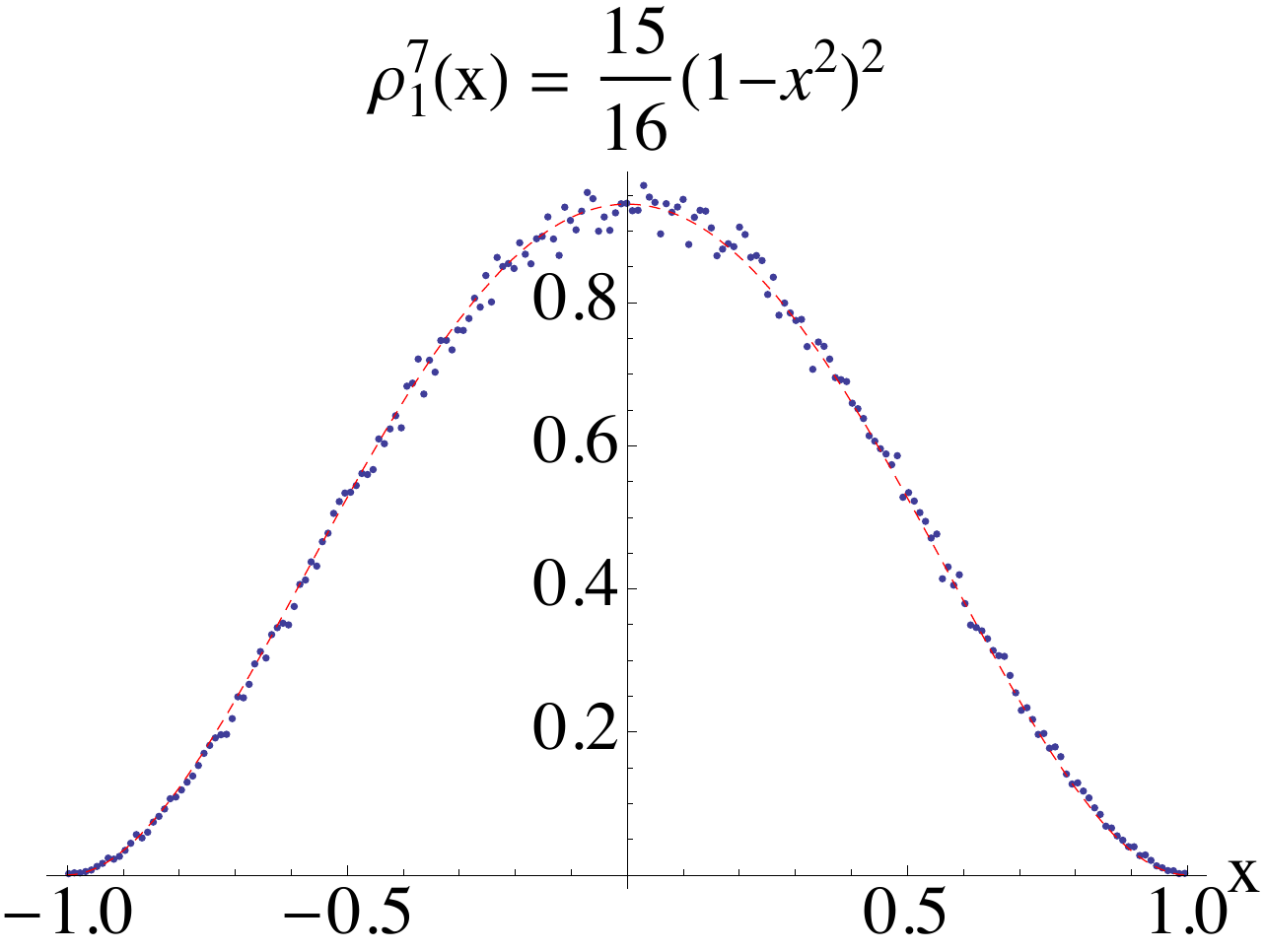} 
     \includegraphics[width=7.49cm]{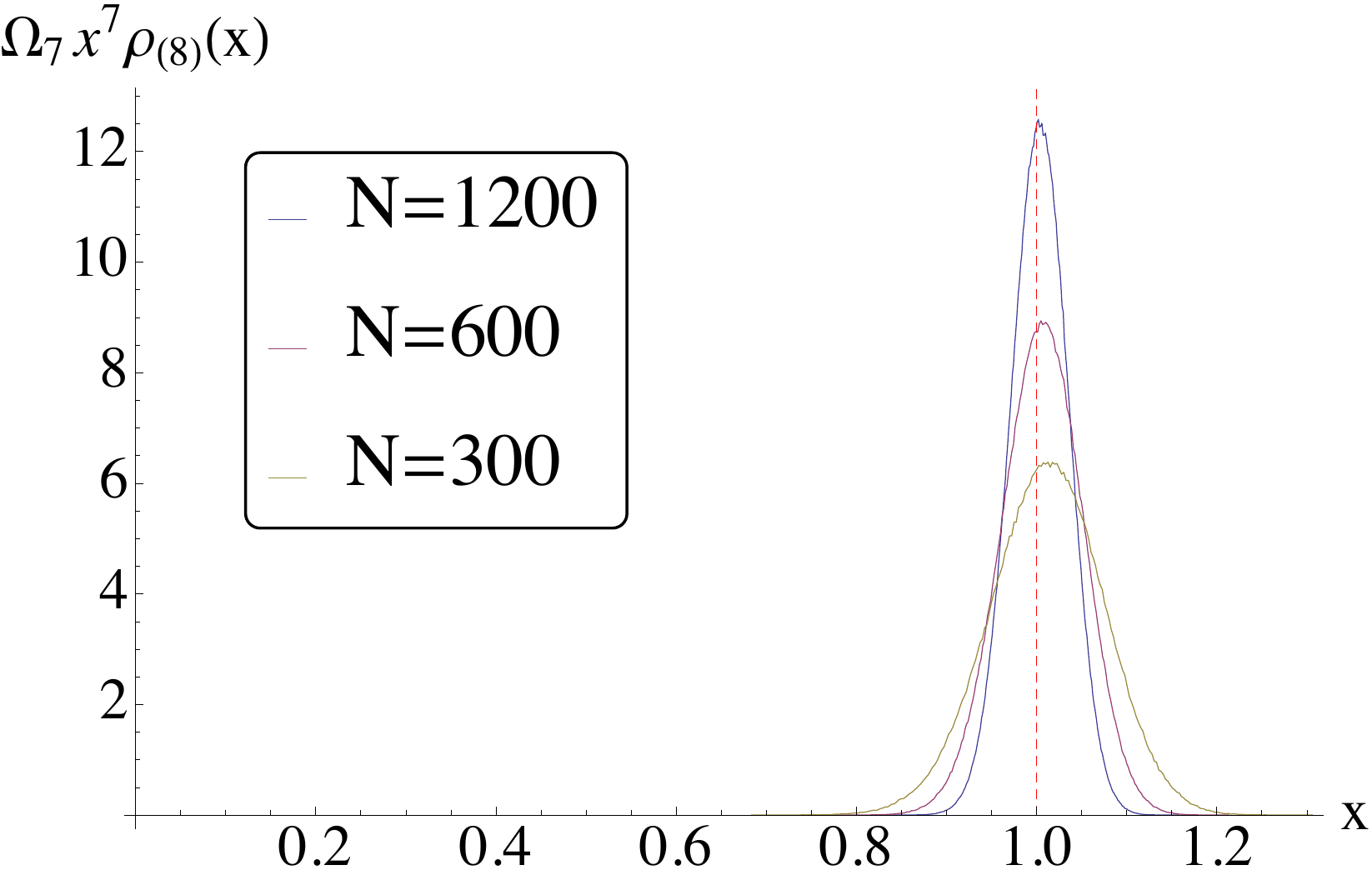} 
   \includegraphics[width=7.49cm]{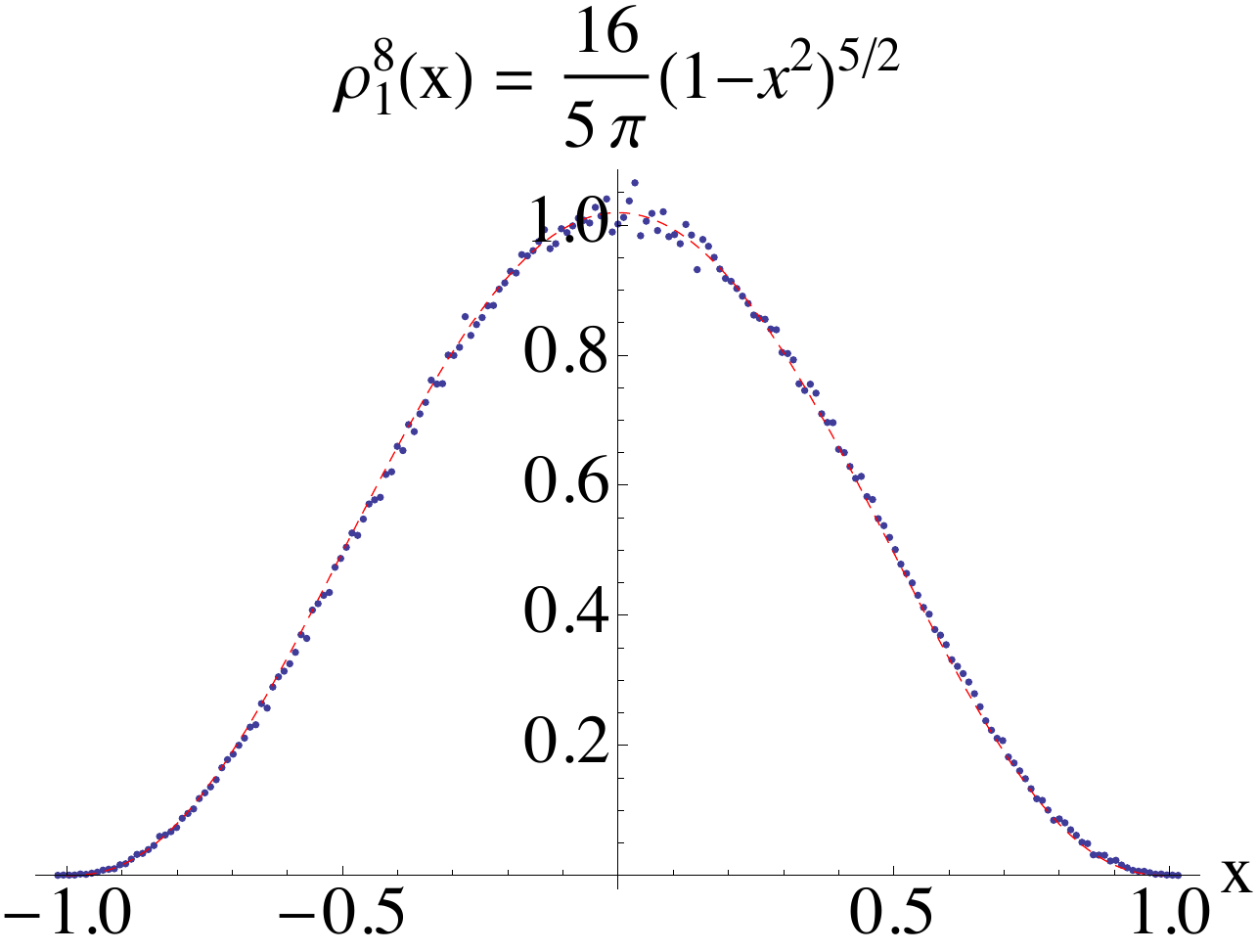} 
\caption{Plots of the radial and reduced distributions for $p=6,7,8$. One can see that as $N$ is increased the radial distributions approach spherical shells of unit radius. One can also see an excellent agreement of the reduced distributions with equation (\ref{shell->1}).}
   \label{fig:5}
\end{figure}

Our next focus is the case $p\geq4$. In section \ref{gaussian-vardim} we showed that for $p\geq4$ the joint eigenvalue distribution is a spherical shell of unit radius. We also learned that the reduced one-dimensional distribution is given by equation (\ref{shell->1}), which for $p=4$ agrees with a Wigner semi-circle, but for $p>4$ differs significantly. In figure \ref{fig:4} and figure \ref{fig:5} we have presented our numerical results for $p=4,5$ and $p=6,7,8$. The left panels represent the radial distributions. One can see that as the size of the matrices is increased the radial distributions approach spherical shells of unit radii. In the right panels we have presented the reduced distributions. One can see the excellent agreement with equation (\ref{shell->1}) for $p=4,5,6,7,8$.

These results support the analysis of the previous chapters and that of ref. \cite{Berenstein:2005aa}. We experimented with higher values of $p>8$ and found the same behaviour confirming that there are only two classes of solutions the Wigner semi-circle family
for $p\leq4$ and the spherical shell distributions for $p\geq4$.

\section{Non-Gaussian potentials}\label{NonGaussianPotentials}

In this section we consider non-gaussian potentials. We will focus on potentials of the form:
\begin{equation} \label{quartic}
V(\vec x) = a |\vec x|^2+b |\vec x|^4\ ,
\end{equation}
containing a quartic term. Note that in order for the model to be stable
we have to impose the restriction $b\geq 0$, where the value $b=0$ is
allowed only if $a$ is positive\footnote{Note however, the case of $b<0$ with $a>0$ is also of possible interest for the one matrix model where the transition
at the critical value where the eigenvalues spill out of the well at the origin corresponds to two dimensional quantum gravity \cite{DiFrancesco:2004qj}.}.

\subsection{Quartic potential in one dimension.}
It is instructive to review the properties of a one dimensional matrix
model with potential of the form (\ref{quartic}). 
The one dimensional random matrix version of this model has been
extensively studied in the literature
\cite{Brezin:1977sv,Feinberg:1997if} and it has been shown that as the
parameters of the potential are varied, the model undergoes a phase
transition. This phase transition is reflected in a change of the
topology of the eigenvalue distribution. Let us describe the solution
to the one matrix model in some details. We will then discuss the
generalisation to our setting of $p$-commuting matrices.

The integral equation determining the
eigenvalue distribution is given by:
\begin{equation}\label{int-eq-nonint}
a\,x+2b\,x^3=\int\limits_{-R}^R dx'\,\frac{\rho_1(x')}{x-x'}\ .
\end{equation}
The potential (\ref{quartic}) is even, which implies that the eigenvalue distribution should also be even. This allows us to rewrite the integral equation (\ref{int-eq-nonint}) as:
\begin{equation}
a+2b\,x^2=2\int\limits_{0}^R dx'\,\frac{\rho_1(x')}{x^2-x'^2}\ ,
\end{equation}
which can be brought to a Cauchy form by the reparametrisation: 
\begin{equation}
z=a+2b x^2,~~~y(z)=\rho_1(x(z))/x(z) \ .
\end{equation}
We obtain:
\begin{equation}\label{reducedCy}
z =\int\limits_{c_1}^{c_2}dz'\frac{y(z')}{z-z'}\ ,
\end{equation}
where $c_1$ and $c_2$ are given by:
\begin{equation}
c_1=a+2b\,r^2; ~~~c_2=a+2b\,R^2;\ , 
\end{equation}
here $r=0$ for connected distribution and $R>r>0$ for disconnected distributions. Let us first consider the case of a connected distribution, in this case the boundary of the eigenvalue distribution is at $x =\pm R$ and we seek a solution to equation (\ref{reducedCy}), which is bounded at $c_2$ and unbounded at $c_1$. The unique such solution is given by:
\begin{equation}
y(z)=\frac{1}{\pi}\frac{\sqrt{c_2-z}}{\sqrt{z-c_1}}\left(z+\frac{c_2-c_1}{2}\right)
\end{equation}
and for the eigenvalue distribution we obtain:
\begin{equation}\label{connected-dist}
\rho_1(x)=\frac{a+b\,(R^2+2x^2)}{\pi}\,\sqrt{R^2-x^2}
\end{equation}
The radius can be determined by normalising the distribution to one, we obtain:
\begin{equation}
R^2=\frac{\sqrt{a^2+12b}-a}{3b}\ .
\end{equation}
Note that the distribution (\ref{connected-dist}) is well defined only for a certain range of the parameter $a$. Indeed, it is easy to show that the minimum of the distribution is achieved at $x=0$ and then requiring that the distribution is positive at its minimum results in the restriction:
\begin{equation}
a >-2\sqrt{b}\ .
\end{equation}
At $a=-2\sqrt{b}$ we have a ``critical'' distribution which vanishes at $x=0$:
\begin{equation}
\rho_1^{cr}(x)=\frac{2b\,x^2}{\pi}\sqrt{\frac{2}{{b}^{1/2}}-x^2}\ ,
\end{equation}
a further reduction of $a$ results in a phase transition to a disconnected distribution. To find the form of the distribution we have to search for solutions of equation (\ref{reducedCy}) that are bounded at both ends. In fact we can look for solutions symmetric with respect to $z=0$. Substituting $c_1=-z_0$ and $c_2=z_0$, which implies $z_0=b(R^2-r^2)$ and $a=-b(R^2+r^2)$ , for the unique such solution we obtain:
\begin{equation}
y(z)=\frac{1}{\pi}\,\sqrt{z_0^2-z^2}\ .
\end{equation}
Going back to variables $x$ and $\rho_1$ for the eigenvalue
distribution we obtain:
\begin{equation}\label{disconnected-dist}
\rho_1(x)=\frac{2b|x|}{\pi}\sqrt{(R^2-x^2)(x^2-r^2)}\ .
\end{equation}
requiring that $\rho_1$ is normalised to one and using the relation $a=-b(R^2+r^2)$ for the outer and inner radii we obtain:
\begin{equation}
R^2=\frac{2\sqrt{b}-a}{2b};~~~r^2=-\frac{2\sqrt{b}+a}{2b}\ .
\end{equation}
One can see that at the critical distribution, when $a=-2\sqrt{b}$,
one has $r=0$. This justifies the name ``critical" since it belongs to
both classes: the connected and the disconnected distributions which
are more commonly referred to as the ``one-cut'' and ``two-cut''
solutions respectively. One can also see that for $a <-2\sqrt{b}$,
which is the regime when the ``one-cut'' solution
(\ref{connected-dist}) is inconsistent, both radii of the ``two-cut''
solution are well defined and the system is described by the
``two-cut'' solution (\ref{disconnected-dist}).  The system in fact
goes through a 3-rd order phase transition at $a=-2\sqrt{b}$. To show this we have to analyse the behaviour of the specific heat of the model across the phase transition. The easiest way to calculate the heat capacity is to calculate the derivative of the internal energy with respect to the ``temperature''. To this end we calculate the expectation value of the potential (\ref{quartic}) for the eigenvalue distributions (\ref{connected-dist}) and (\ref{disconnected-dist}) with rescaled couplings $a\to a/T,\,b\to b/T$. The next step is to calculate the derivative with respect to $T$ and then take $T=1$. We obtain:
\begin{eqnarray}\label{CvD1}
C_v^1&=&\frac{a^4+54b^2-a(a^2-6b)\sqrt{a^2+12b}}{216b^2}\ ;~~~\text{for}~~ a \geq -2\sqrt{b}\nonumber\\
C_v^2&=&\frac{1}{4}\ ;~~~\text{for}~~ a \leq -2\sqrt{b}\ ,
\end{eqnarray} 
where $C_v^1$ and $C_v^2$ are the specific heats of the `one-cut' and `two-cut' solutions, respectively. One can easily see that at $a=-2\sqrt{b}$ we have $C_v^1=C_v^2=1/4$, while $\partial_a C_v^1\neq\partial_a C_v^2$ at this point. This confirms that the phase transition is of a third order.

\begin{figure}[t]
  \centering
   \includegraphics[width=4.9cm]{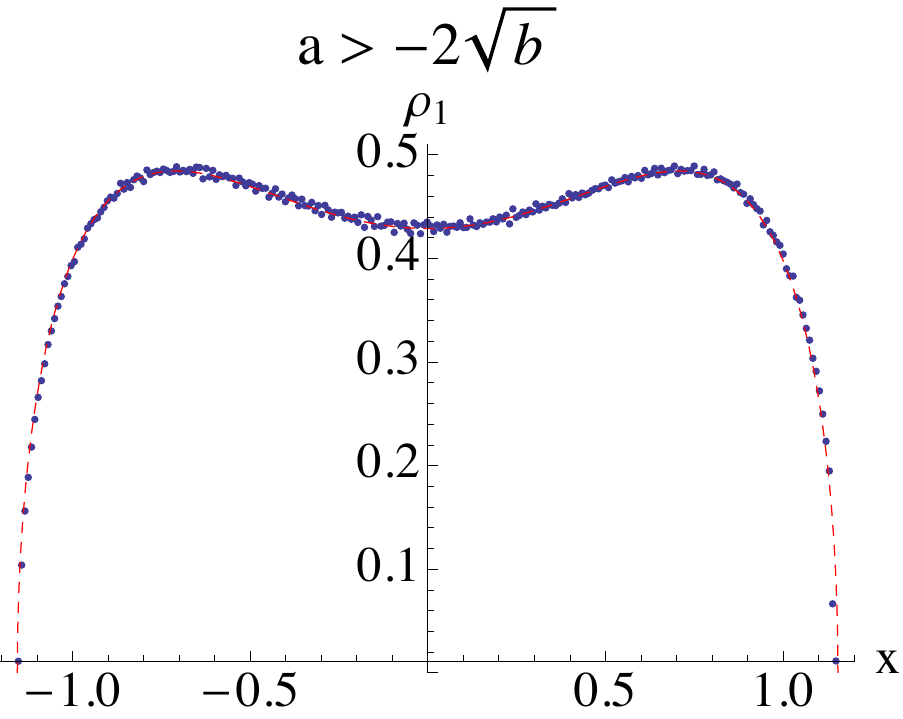} 
   \includegraphics[width=4.9cm]{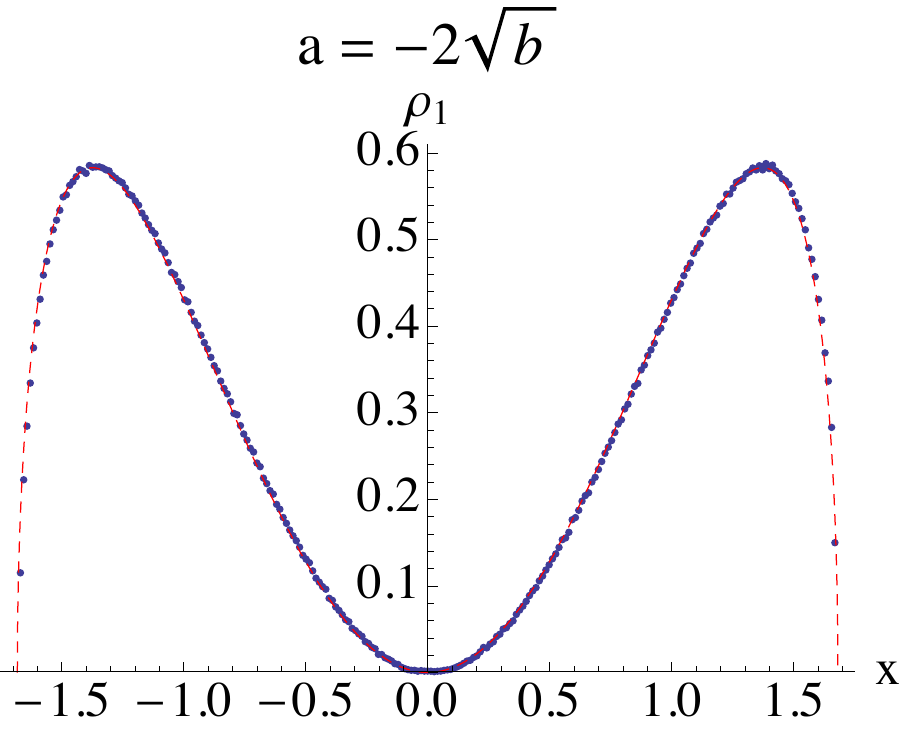} 
      \includegraphics[width=4.9cm]{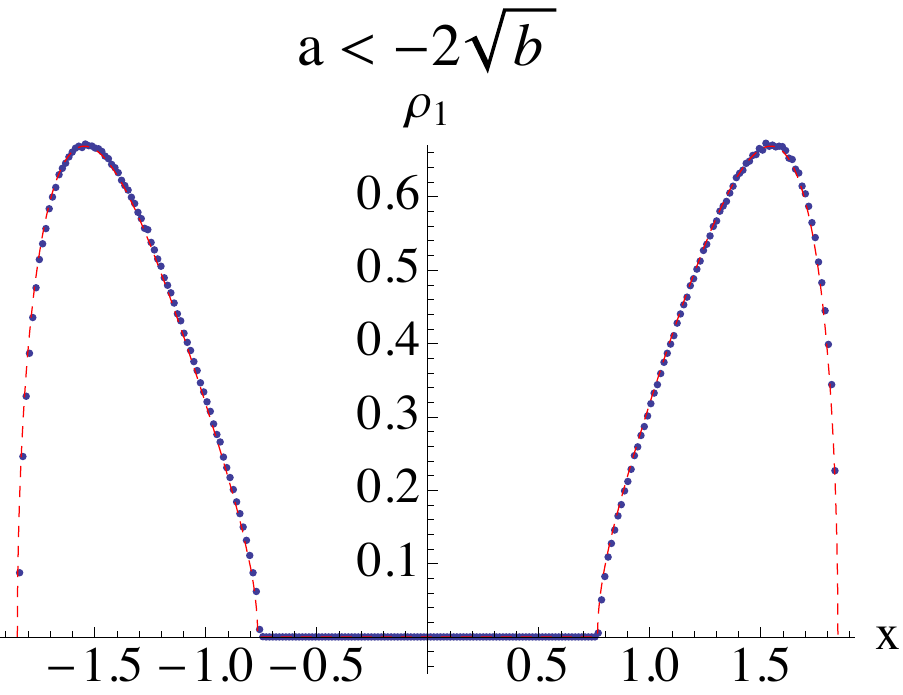} 
\caption{Comparison between the exact result for large $N$ and Monte Carlo simulations for $N=800$ and $b=1/2$. The blue dotted curve represents the numerical results and the red dashed curves correspond to the theoretical predictions. The first plot from left to right is for $a=1/2>-2\sqrt{b}$. The second plot represents the critical distribution with $a=-\sqrt{2}=-2\sqrt{b}$. Finally, the third plot is for $a=-2 <-2\sqrt{b}$. In all cases one can observe an excellent agreement of the numerical simulations with the theoretical predictions. }
   \label{fig:6}
\end{figure}

\begin{figure}[t]
  \centering
    \includegraphics[width=12cm]{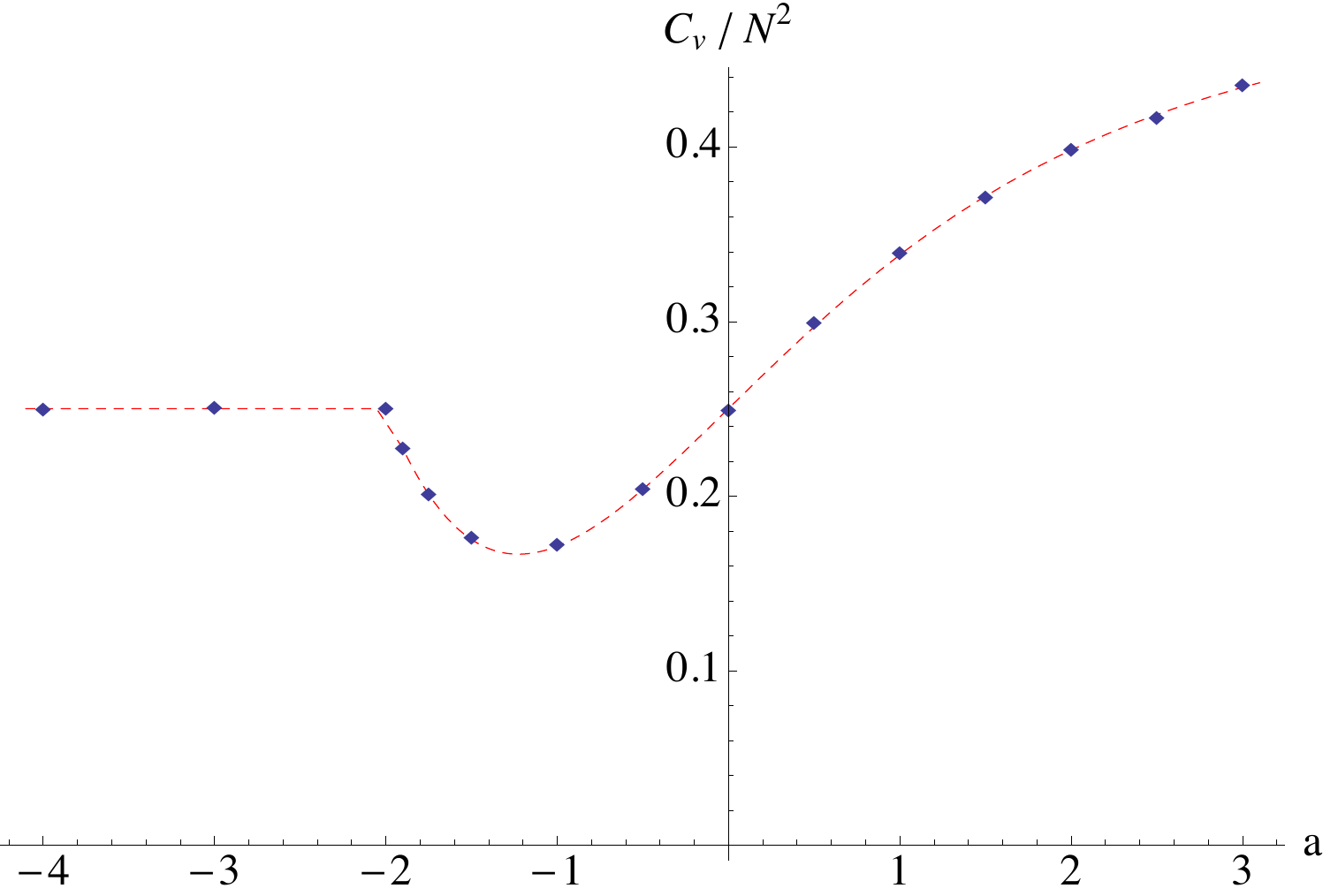} 
\caption{Plot of the specific heat of the model $C_v$ as a function of the parameter $a$ for fixed $b=1$. One can see the cusp at the critical value $a=-2\sqrt{b}$. The red dashed curve corresponds to the analytic expressions (\ref{CvD1}), while the blue diamonds represents results of numerical simulations for $N=100$, one can see the excellent agreement between the two.}
   \label{fig:6a}
\end{figure}

In figure \ref{fig:6} we have compared the large $N$ analytic expressions (\ref{connected-dist}) and (\ref{disconnected-dist}) to Monte Carlo
simulations of the model for $N=800$ and for definiteness we have set
$b=1/2$. The figure shows the excellent agreement with the theoretical
large $N$ results. Furthermore, one can see that at $a=-2\sqrt{b}$ the
critical embedding is realised, which confirms the phase transition
is continuous. 
In figure \ref{fig:6a} we have compared the analytic expressions for the specific heat (\ref{CvD1}) to Monte Carlo simulations for $N=100$ and $N=400$.

\subsection{Commuting matrix model with quartic potential in two dimensions}

There are many possible extensions of the one matrix model to
rotationally invariant two matrix models. The most obvious extension
would be to consider (\ref{quartic}) where $\vec{x}$ are two random
matrices, which do not commute. To our knowledge this model has not
been solved. An alternative approach is to build a non-hermitian
matrix from $\Phi=X+i Y$ and consider a non-Hermitian model with
Hermitian Hamiltonian and quartic potential built from
$\Phi^\dag\Phi$. Such a system was solved by \cite{Feinberg:1997if} (see also ref.~\cite{Feinberg:2001vj})
using their method of Hermitization. Because the matrices don't
commute and $\Phi^\dag\Phi=X^2+Y^2+i[X,Y]$ this model is significantly
different from those we consider but should reduce to our model if the
contribution from the commutator is forced to zero. To our knowledge
the commuting matrix models described below are new.

Here we perform an analogous
investigation with emphasis on the relation between the two-matrix
model and the reduced one matrix model. Our starting point is the
integral equation:
\begin{equation}
\mu_2+a |\vec x|^2+b|\vec x|^4 =\int d^2x'\,\rho_2(\vec x)\,\log(\vec x-\vec x')^2\ .
\end{equation}
Applying the Laplacian on both sides of the equation and using 
the two dimensional identity $\nabla^2\log(\vec x-\vec x')^2=4\pi\,\delta^{(2)}(\vec x-\vec x')$, one arrives at:
\begin{equation}\label{quartic-2D}
\rho_2(x)=\frac{4b\,x^2+a}{\pi} ~~~\text{for}~\vec x\in {\cal D},
\end{equation}
where ${\cal D}$ is the domain of the distribution. Rotational
invariance requires that the domain is either a disk or an annulus or
a more exotic configuration of numerous concentric disks. The
intuition that we gained from the one dimensional model suggest that
for a quartic potential only the disk and the annulus are
realised. Indeed, stability of the model requires that $b\geq 0$,
where $b=0$ is allowed only for positive $a$\footnote{Again it may be
  of interest to study the case of positive $a$ and negative but small
  $b$ up to the transition where the eigenvalues spill out of the
  local well at the origin.}. For $a>0$ the
distribution (\ref{quartic-2D}) is positive and well defined for all
$\vec x$ inside a disk of radius $R$. Normalising the distribution to one, we obtain:
\begin{equation}
R^2=\frac{\sqrt{a^2+8b}-a}{4b}\ .
\end{equation}
The eigenvalue distribution of the disk phase is then:
\begin{equation}\label{disk-dist}
\rho_2(x)=\frac{4b\,x^2+a}{\pi}\,\Theta(R^2-x^2)\ . 
\end{equation}

If $a=0$ we have critical distribution, which goes to zero at
$x=0$. For $a<0$ the expression in equation (\ref{quartic-2D}) is
negative at $x=0$ and vanishes for some $x>0$. It is therefore
unphysical in this region and we need to modify our expression for the
distribution. However, the functional form of the distribution
(\ref{quartic-2D}) is independent on the shape of the domain ${\cal D}
$, this is a special property of the logarithmic kernel in two
dimensions. Because of this property we are free to modify only the
range of the distribution. A natural choice is to keep the same outer
radius $R$ and choose the inner radius $r$ in such a way that the
integral $\int\limits_{|\vec x|<r} d^2x\,\rho_2(x)$ vanishes. This
results in:
\begin{equation}
\int\limits_{|\vec x|<r}d^2 x\,\rho_2(x)=a\,r^2+2b\,r^4=0~ \therefore~ r^2=-\frac{a}{2b}\ 
\end{equation}  
and we can write the eigenvalue distribution of the annulus phase as:
\begin{equation}\label{annulus-dist}
\rho_2(x)=\frac{4b\,x^2+a}{\pi}\,\Theta(R^2-x^2)\Theta(x^2-r^2)\ . 
\end{equation}
Let us now calculate the reduced distribution:
\begin{equation}
\rho_2^{(1)}(x)=\int\limits_{-\sqrt{R^2-x^2}}^{\sqrt{R^2-x^2}} dy\,\rho_2(x,y)=2\int\limits_x^Rd\xi\,\xi\,\frac{\rho_2(\xi)}{\sqrt{\xi^2-x^2}}\ .
\end{equation}
For $a>0$ we reduce the disk distribution (\ref{disk-dist}) to obtain:
\begin{equation}\label{connected-reduced}
\rho_2^{(1)}(x)=\frac{2a+\frac{8b}{3}\,(R^2+2x^2)}{\pi}\sqrt{R^2-x^2}
\end{equation}
which as expected looks like equation (\ref{connected-dist}) for the
connected distribution in one dimension. In fact, if we reduce the
potential according equation (\ref{V2toV1}) we obtain:
\begin{equation}\label{reducedV2->1}
V_{2\to1}(x)=2a\,x^2+\frac{8b}{3}\,x^4\ .
\end{equation}
It is easy to convince oneself that to derive the connected one
dimensional distribution for the reduced potential
(\ref{reducedV2->1}), one has to take $a\to2a$ and $b\to8b/3$ in
equation (\ref{connected-dist}). In doing so one arrives at equation
(\ref{connected-reduced}), confirming that indeed the disk phase of
the commuting two-matrix model maps to the connected phase of the
one-matrix model, which is what we expect.

Let us now reduce the annulus phase. Naively we might expect this
phase to map to the disconnected phase of the one-matrix
model. However, this is not the case. For the reduced distribution of
(\ref{annulus-dist}) we obtain:
\begin{equation}
\rho_2^{(1)}(x)=\frac{2a+\frac{8b}{3}\,(R^2+2x^2)}{\pi}\sqrt{R^2-x^2}\,-\,\Theta(r^2-x^2)\,\frac{2a+\frac{8b}{3}\,(r^2+2x^2)}{\pi}\sqrt{r^2-x^2}\ ,
\end{equation}
which is profoundly different from the disconnected distribution
(\ref{disconnected-dist}). This is an important observation. In all
previous examples the ``shadow" of the higher dimensional model
(namely the reduced distribution) corresponded to the physical
distribution of the lower dimensional problem (with the reduced
potential). Now we see that this does not hold uniformly. In
particular for phases with non-trivial topology, the shadow of the
higher dimensional problem does not reduce to the physics of the lower
dimensional one. One should not be surprised by this result. Indeed,
although the annulus phase has a non-trivial topology, it still
corresponds to a connected distribution, this is clearly not the case
for the disconnected phase of the one-dimensional model which has two
disconnected components and is thus quite different.

Physically, this can be understood, because the quartic potential in
one dimension is a double well and thus drives the theory into two
disconnected phases associated to the different vacua, i.e. the moduli
space of vacua is two points. The rotationally invariant quartic
potential in two dimensions corresponds to a Mexican hat and the
associated moduli space of vacua is the circle.  So all vacua are
connected and hence one expects the distribution to remain
connected. With this revised intuition we can correct our na\"ive
expectation to anticipate that the topology of the space of
eigenvalues undergoes a transition from a disc to an annulus, in
accord with the observation above.

These differences between the one-dimensional and two-dimensional
models are not manifest when the theories are in the trivial vacuum
(at the origin) and both distributions are topologically an interval
and a disk, respectively. This is the reason we can map the dynamics
of the disk phase of the two-matrix model to the dynamics of the
connected phase of the one-matrix model.

One may wonder what happens in the interval $a\in
[-\sqrt{\frac{8b}{3}},0]$, which still corresponds the one-cut
distribution of the one-dimensional model, and whether there is
anything special happening at $a=-\sqrt{\frac{8b}{3}}$ i.e. to the
parameter value of the one-dimensional transition. It turns out that 
in the two dimensional model there is no further non-analyticity at this
value. The two dimensional transition is shifted to $a=0$ and it is 
at this value that a hole appears in the eigenvalue distribution. 
What is special about $a=-\sqrt{\frac{8b}{3}}$ is that 
the inner radius occurs at the maximum
of the reduced distribution, but we find no further non-analyticity 
at this parameter value.

\begin{figure}[h]
  \centering
   \includegraphics[width=12cm]{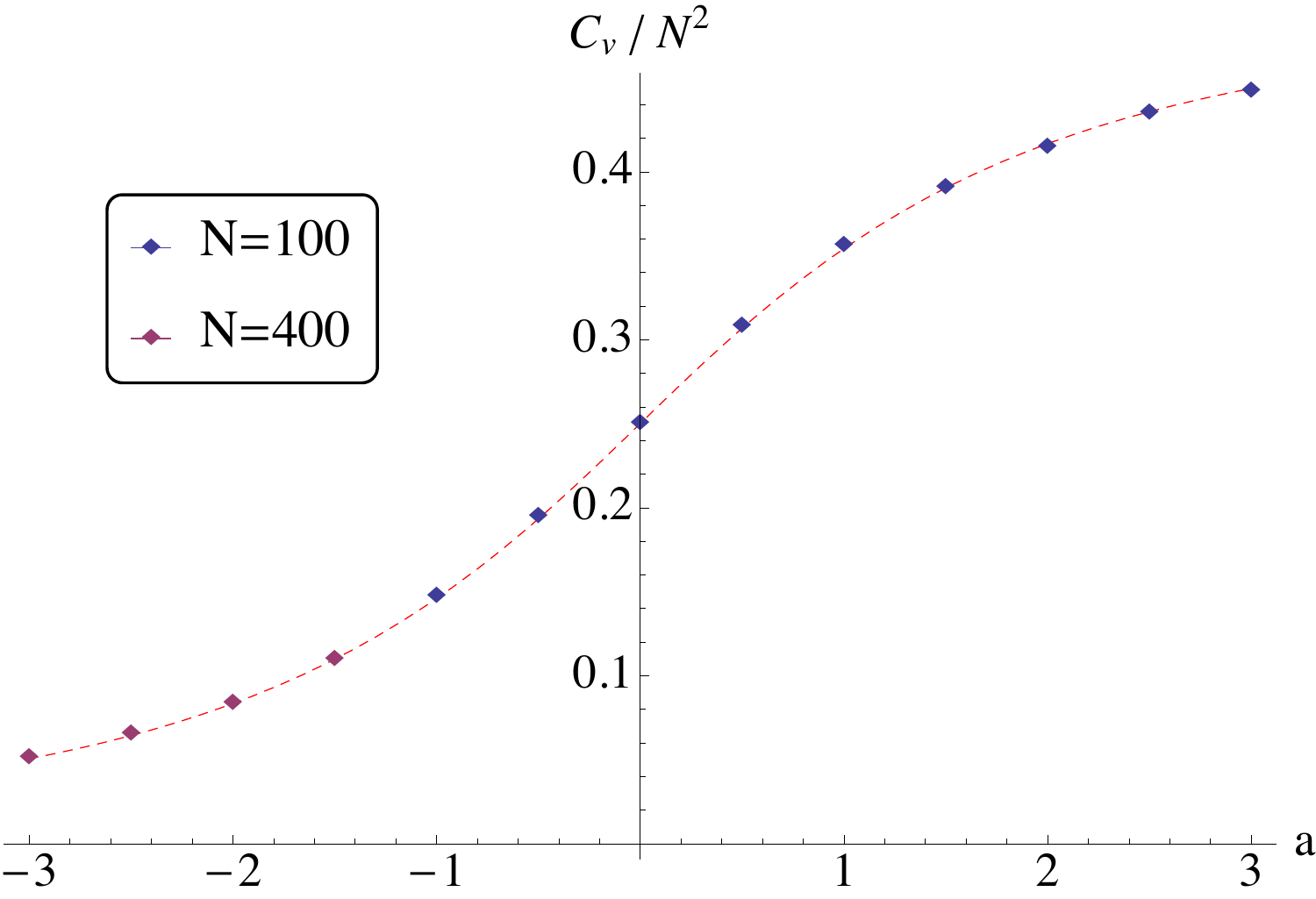} 
\caption{ Plot of the
large $N$ result for the heat capacity (\ref{CvD2}) compared with numerical simulations for $N=100$ (blue
diamonds) and $N=400$ (red diamonds). }
   \label{fig:8a}
\end{figure}

To emphasise the different physics described by the one- and
two-matrix models let us calculate the specific heat and explore its
behaviour across the disk-annulus phase transition. Following the same
path as in the analysis of the one matrix model we arrive at the
following result for the heat capacity:
\begin{eqnarray}\label{CvD2}
C_v^1&=&\frac{1}{4}+\frac{a^4}{96b^2}-\frac{(a^3-4ab)\sqrt{a^2+8b}}{96b^2}\ ;~~~\text{for}~~ a \geq 0\nonumber\\
C_v^2&=&\frac{1}{4}-\frac{a^4}{96b^2}-\frac{(a^3-4ab)\sqrt{a^2+8b}}{96b^2}\  ;~~~\text{for}~~ a \leq -0\ ,
\end{eqnarray} 
One can see that $C_v^1-C_v^2={a^4}/({48b^2})$. This shows that at
$a=0$ the heat capacity and its first three derivatives are
discontinuous at $a=0$ and it is the fourth derivative of the heat
capacity which has a finite jump. Since the heat capacity is already a
second derivative of the free energy this suggests that the phase
transition is of sixth order. The heat capacity has another intriguing property, it is
exactly $1/4$ at $a=0$, just like in the one-matrix model case. This
is due to the constraint (\ref{constr-contin}). Furthermore, it is odd
with respect to the point $(0,1/4)$ (see figure \ref{fig:8a}).
\begin{figure}[h]
  \centering
   \includegraphics[width=4.9cm]{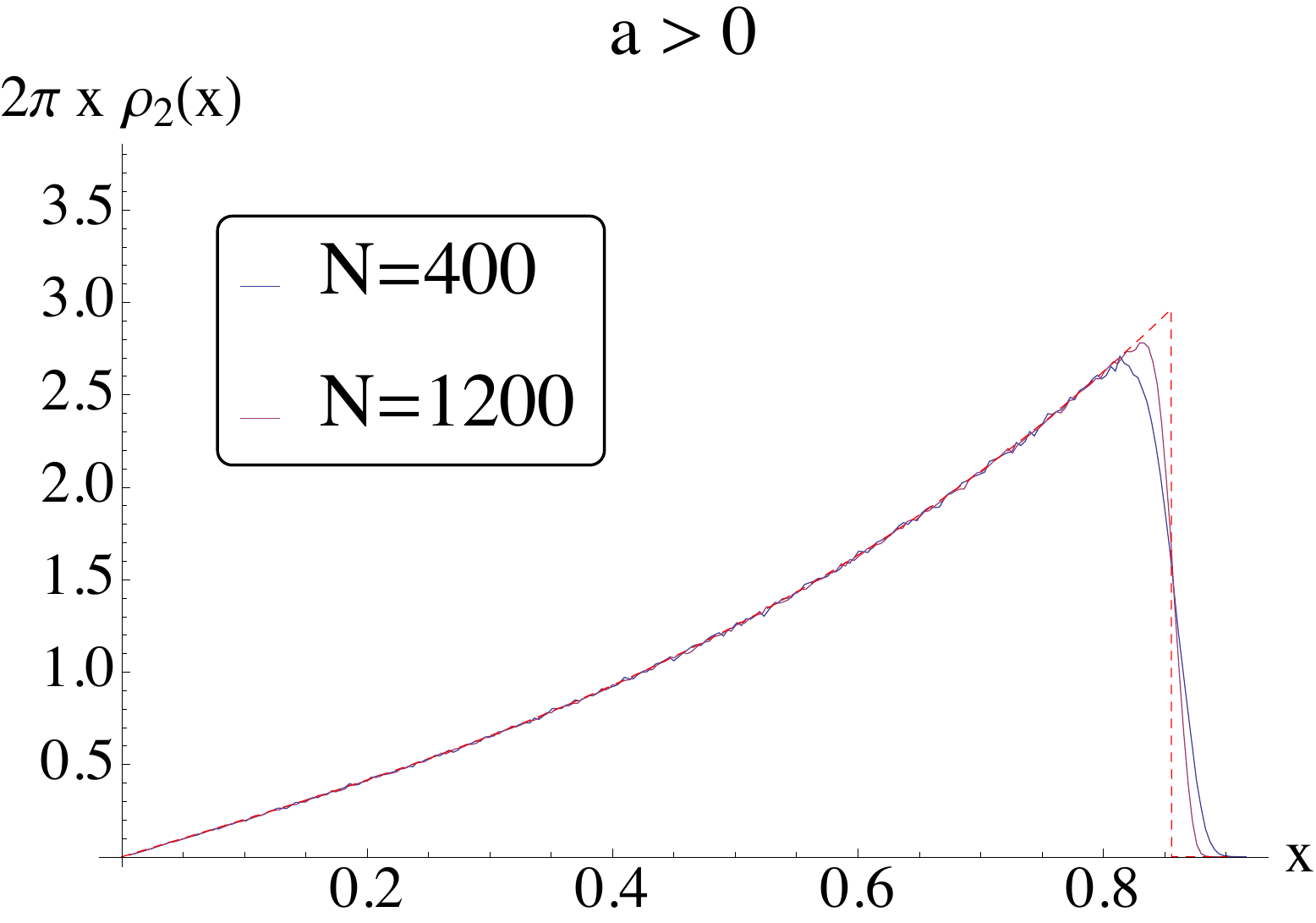} 
   \includegraphics[width=4.9cm]{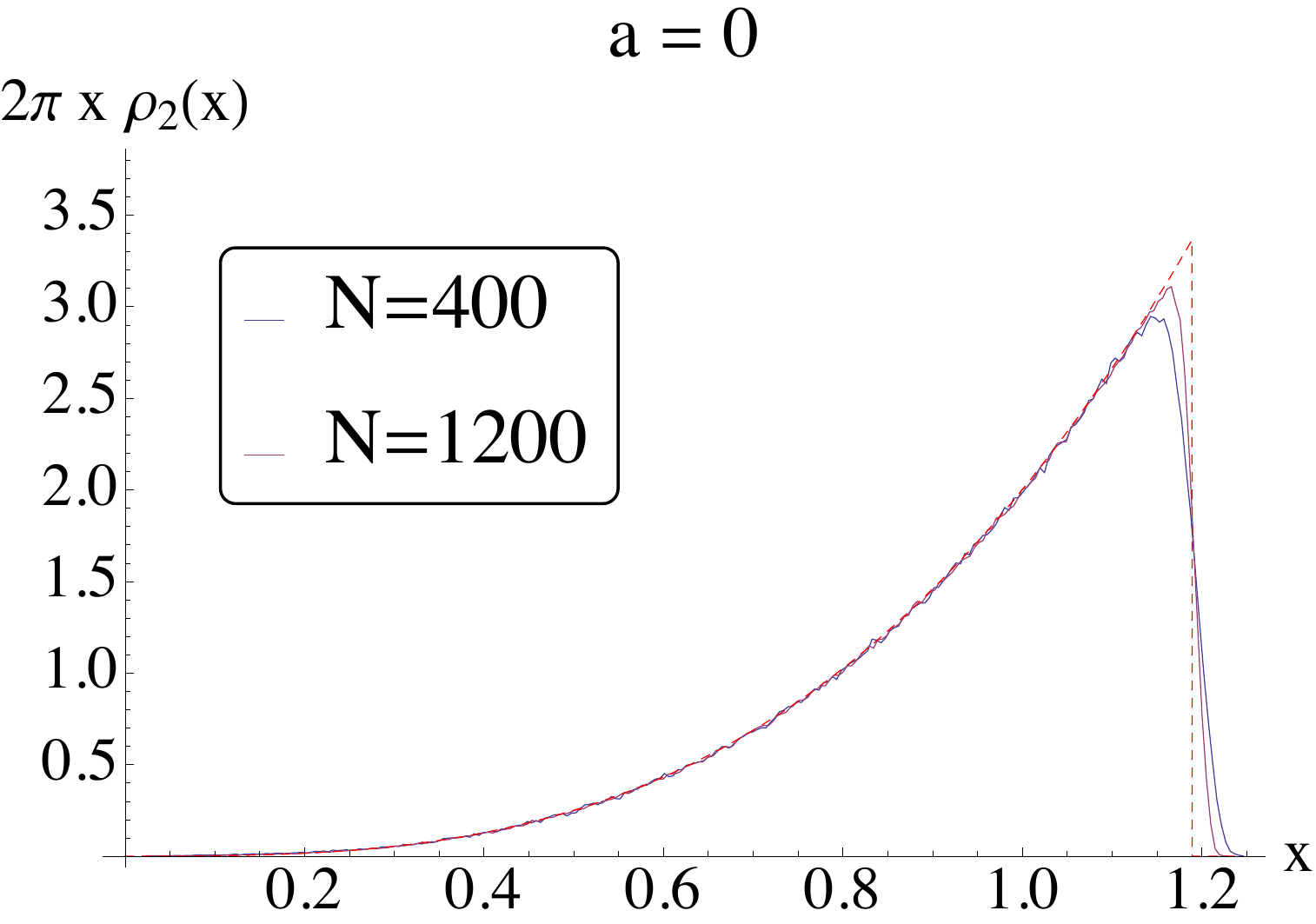} 
   \includegraphics[width=4.9cm]{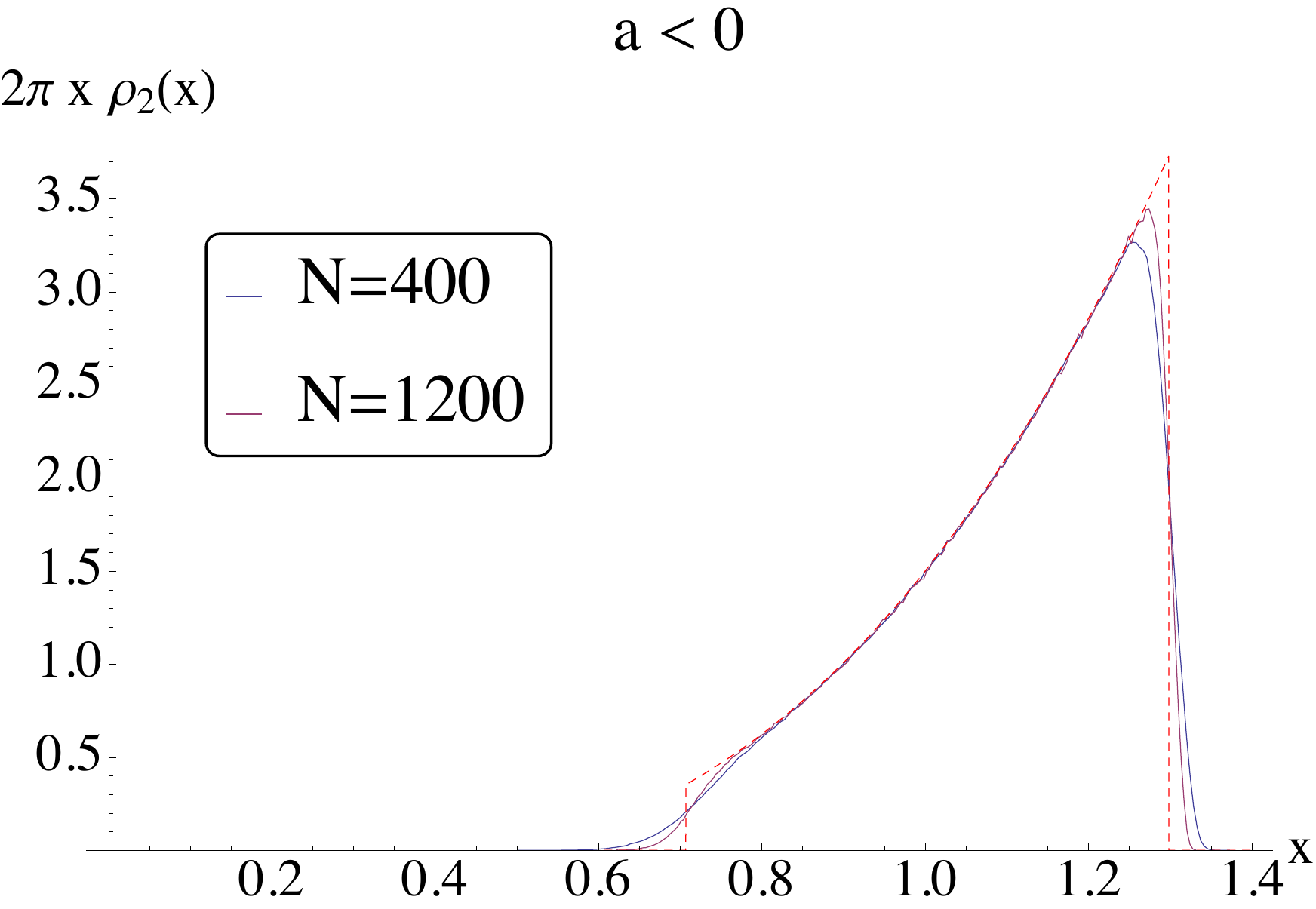} 
   \includegraphics[width=4.9cm]{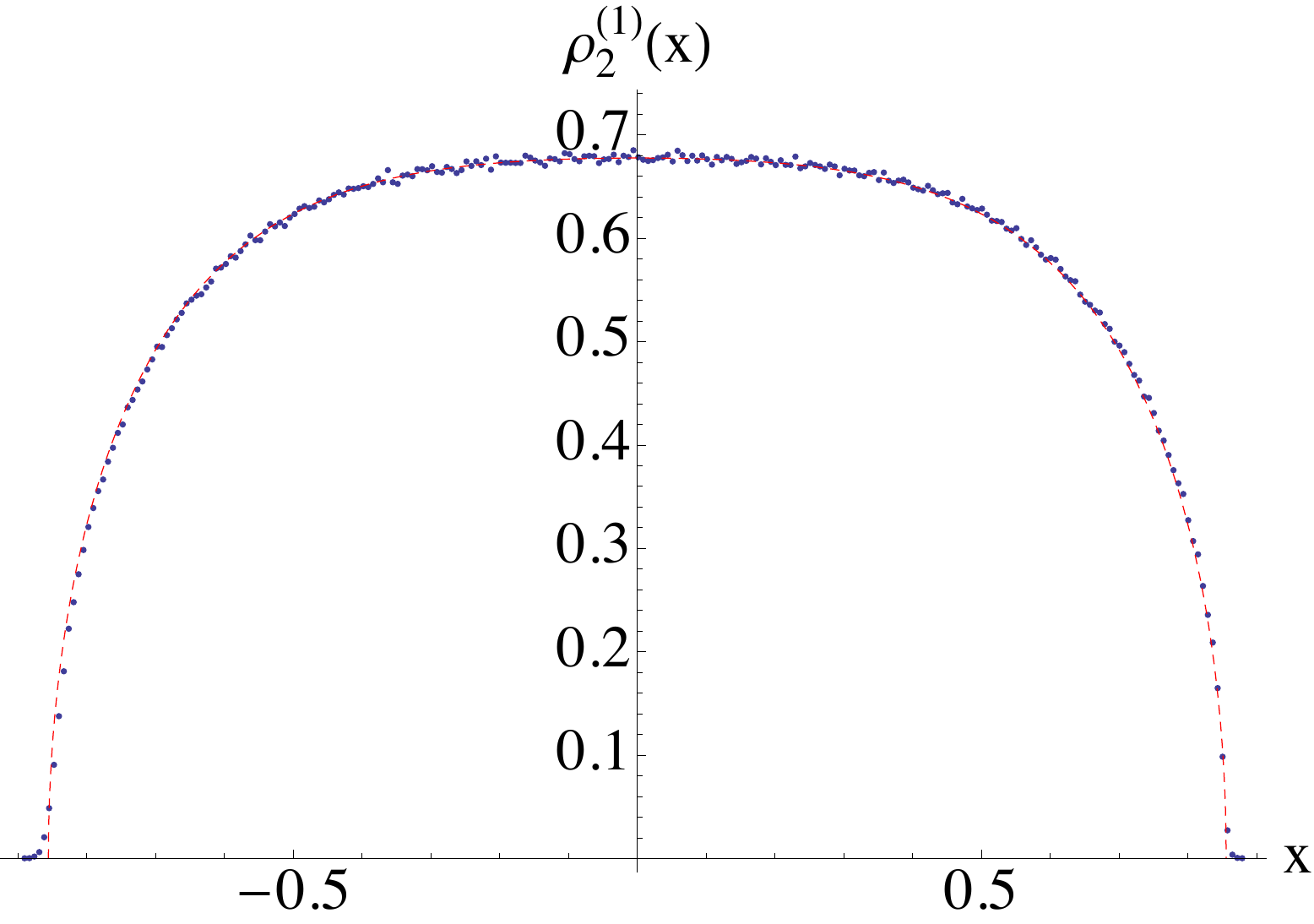} 
   \includegraphics[width=4.9cm]{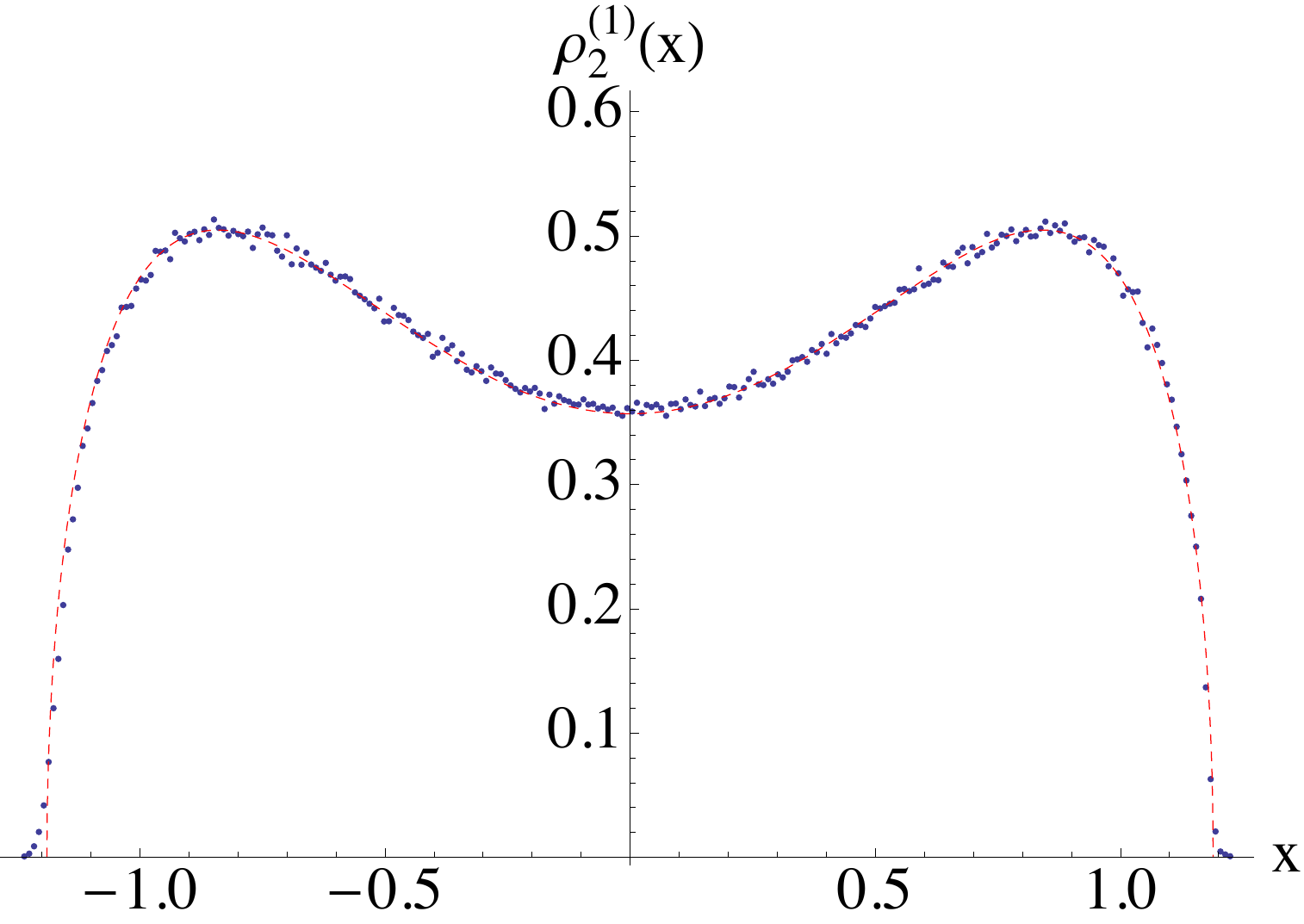} 
   \includegraphics[width=4.9cm]{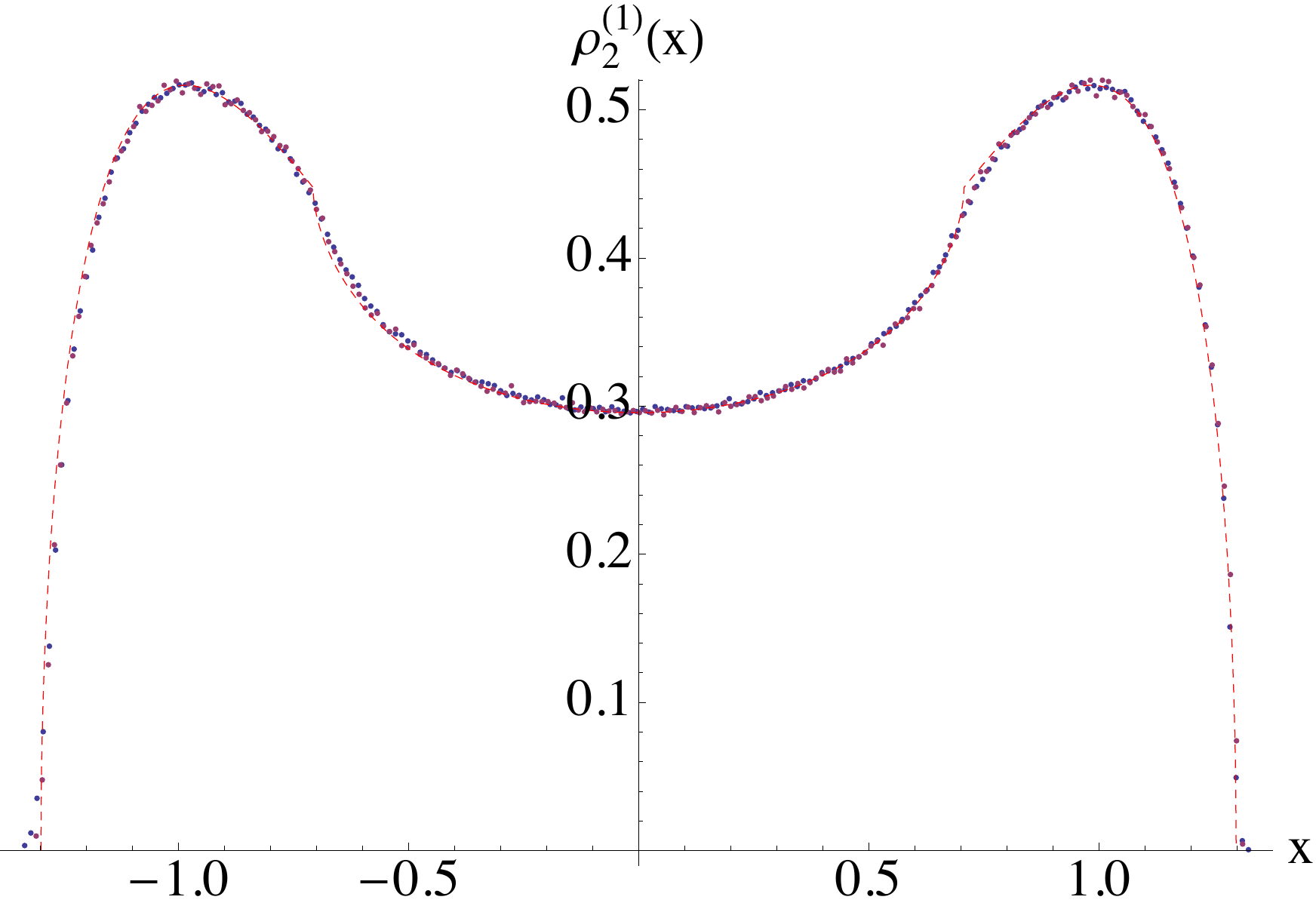} 
\caption{Comparison of numerical simulation with theoretical
  predictions. In all plots $b=1/2$. The first pair of plots from left
  to right represent the 2D and the reduced eigenvalue distributions
  in the disk phase for $a=1$. The second pair corresponds to the
  critical case $a=0$. Finally, the last pair represents the 2D and
  the reduced distributions in the annulus phase for $a=-1/4$. }
   \label{fig:7}
\end{figure}
\begin{figure}[h]
  \centering
   \includegraphics[width=4.9cm]{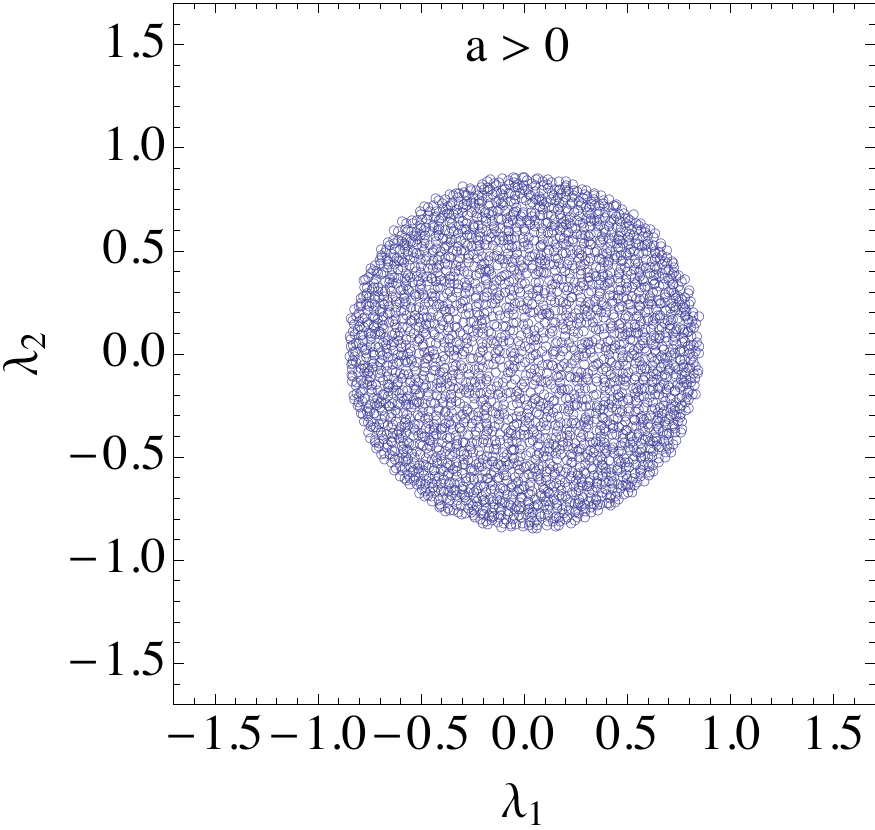}
     \includegraphics[width=4.9cm]{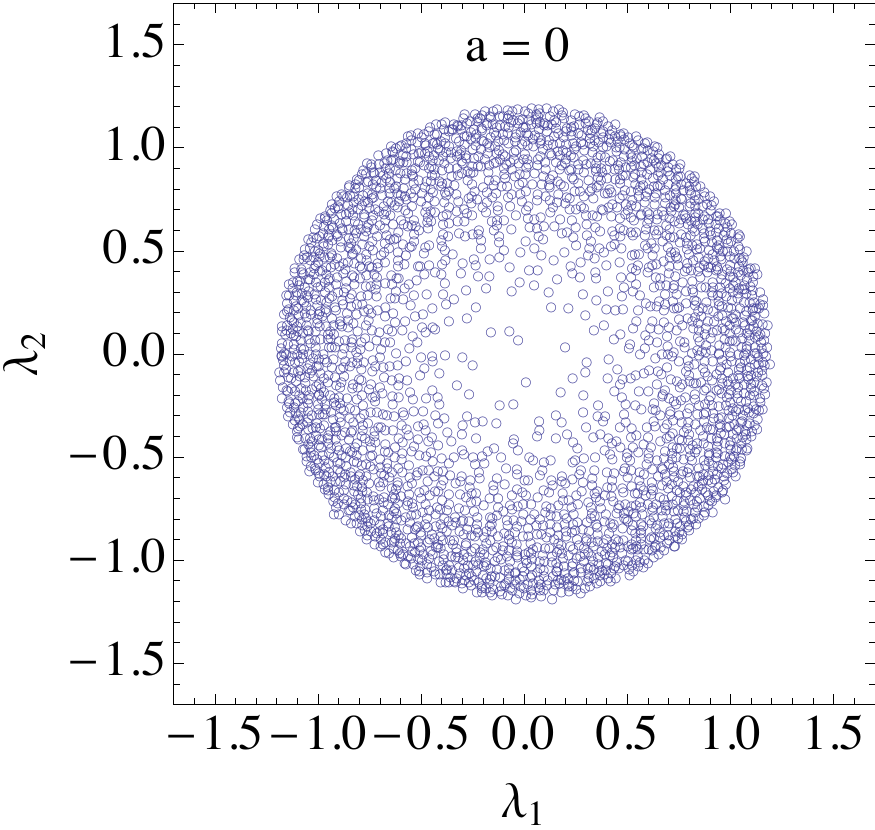} 
     \includegraphics[width=4.9cm]{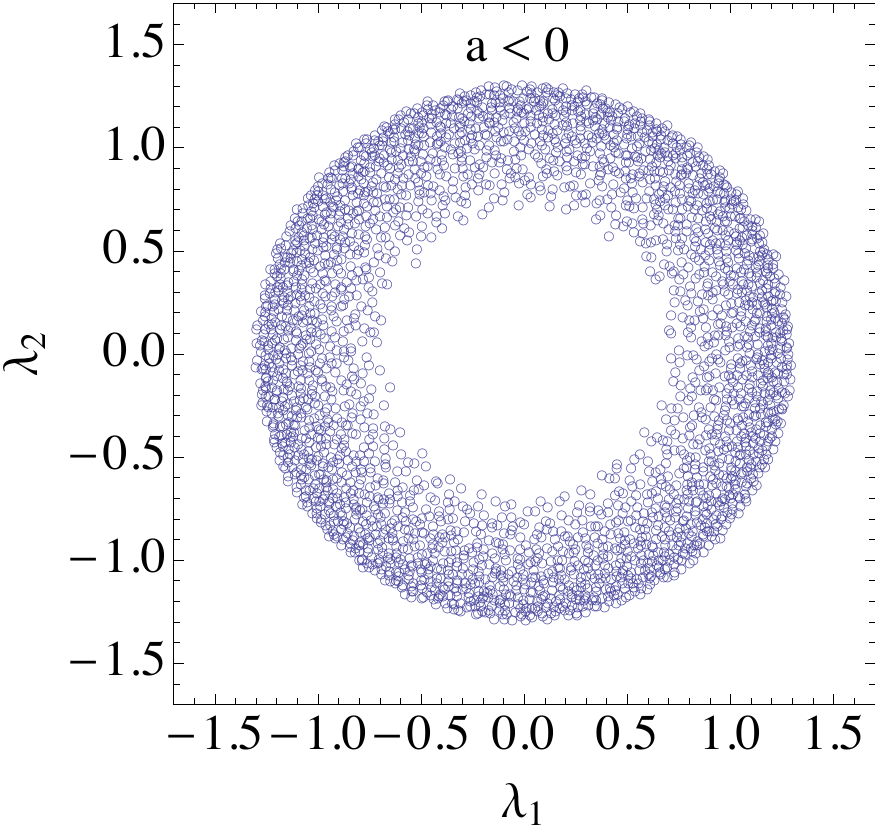}
     \caption{Plots of the spread of eigenvalues for the disk and annulus phases for $N=3000$. The first plot form left to right represents the disk phase for $a=1$. The middle plot represents a critical disk for $a=0$ and the last plot represents the annulus phase for $a=-1/4$. }
   \label{fig:8}
  \end{figure}

Let us conclude this subsection by comparing our results to the
results of Monte Carlo simulations. In figure \ref{fig:7} we have
presented plots of the 2D and the reduced eigenvalue distributions for
the disk and annulus phases as well as for the critical
distribution. One can observe an excellent agreement between the large
$N$ theoretical predictions and the numerical results. Figure
\ref{fig:8} represents the spread of the eigenvalues for the disk and
annulus phases. The middle plots represents a critical disk for
$a=0$. Finally, in figure \ref{fig:8a} we have compared the plot of the
large $N$ result for the heat capacity (\ref{CvD2}) with the results
for the heat capacity from numerical simulations for $N=100$ (blue
diamonds) and $N=400$ (red diamonds) and one can see the excellent
agreement with the theoretical large $N$ results.

\subsection{Quartic potential in three dimensions}\label{QuarticPotential}

In this subsection we investigate the properties of a three-matrix commuting model with quartic potential. Our starting point is the integral equation:
\begin{equation}
\mu_3+a |\vec x|^2+b|\vec x|^4 =\int d^3x'\,\rho_3(\vec x)\,\log(\vec x-\vec x')^2\ .
\end{equation}
Applying the operator $|\vec x|\,\nabla_x^2$ on both sides of the equation and using that $\rho_3$ is spherically symmetric to perform the angular integrals we obtain:
\begin{equation}\label{3DlogK}
\frac{3a}{\pi}x+\frac{10b}{\pi}x^3=\int\limits_r^R dx'\,x'\,\rho_3(x')\,\log\left(\frac{x+x'}{x-x'}\right)^2 ,
\end{equation}
where $r=0$ for a ``one-cut'' solution with the topology of a ball and
$r>0$ for a ``two-cut'' solution with the topology of an annulus. The
easiest way to solve equation (\ref{3DlogK}) is to reduce it to an
integral equation with a Cauchy kernel. To this end we differentiate
with respect $x$ and change variables to $z=3a/\pi+30b\,x^2/\pi$
and $y(z)=2x(z)\,\rho_3(x(z))$ we obtain:
\begin{equation}\label{3DCauchy}
z =\int\limits_{c_1}^{c_2}dz'\frac{y(z')}{z'-z}\ ,
\end{equation}
where $c_1=3a/\pi+30b\,r^2/\pi$ and $c_2=3a/\pi+30b\,R^2/\pi$. Our intuition from the gaussian case suggest that we look for a solution to (\ref{3DCauchy}) which is unbounded at $z=c_2$ (corresponding to $x=R$) and bounded at $z=c_1$ ($x=r$). The unique such solution is given by:
\begin{equation}
y(z)=\frac{1}{\pi}\frac{\sqrt{z-c_1}}{\sqrt{c_2-z}}\left(z-\frac{c_2-c_1}{2}\right)\ .
\end{equation}
Going back to $x$ and $\rho_3$ we obtain:
\begin{equation}\label{two-cut-3D}
\rho_3(x)=\frac{3}{2\pi^2 x}\frac{\sqrt{x^2-r^2}}{\sqrt{R^2-x^2}}\left(a+5b\,(2x^2+r^2-R^2)\right)\ .
\end{equation}
The corresponding reduced one dimensional distribution is given by:
\begin{equation}\label{reduceddistD3-D1}
\rho_3^{(1)}(x)=2\pi\int\limits_x^R dx' x' \rho_3(x')\,\Theta(x'^2-r^2)\ ,
\end{equation}
where $\Theta(x)$ is the step function. The explicit form of the distribution for $r\neq 0$ can be obtained in terms of elliptic integrals, we will use this solution to compare to numerical simulations.

Equation (\ref{two-cut-3D}) is our candidate for the ``two-cut'' solution. To get the ``one-cut'' solution we simply take the limit $r\to0$ in (\ref{two-cut-3D}) obtaining:
\begin{equation}\label{one-cut-3D}
\rho_3(x)=\frac{3}{2\pi^2}\frac{a+5b\,(2x^2-R^2)}{\sqrt{R^2-x^2}}\ .
\end{equation}
Using equation (\ref{reduceddistD3-D1}) with $r=0$ for the corresponding reduced one dimensional distribution we obtain:
\begin{equation}
\rho_3^{(1)}=\frac{1}{\pi}\left(3a+5b(R^2+2x^2)\right)\,\sqrt{R^2-x^2}\ .
\end{equation}
To obtain the radius of the one-cut solution we normalise it to one. For the radius we find:
\begin{equation}
R^2=\frac{\sqrt{9a^2+60 b}-3a}{15b}\ .
\end{equation}

One can check that with this radius the one-cut distribution also satisfies the constraint (\ref{constr-contin}). 

Obtaining the inner and outer radii of the two-cut solution is more
subtle. The normalisation condition for $\rho_3$ can be used to find
the outer radius as a function of $a, b$ and the inter radius $r$. We
obtain:
\begin{equation}\label{R-two-cut-3D}
R^2=\frac{\sqrt{9(a+10b\,r^2)^2+60b}-3a-15b\,r^2}{15b}\ .
\end{equation}
To specify completely $R$ and $r$ we need to use the constraint (\ref{constr-contin}), for the two-cut distributions it is given by:
\begin{gather}
\frac{3}{16}(R^2-r^2)\left(4a^2(r^2+3R^2)+4ab\left(11(r^2+R^2)^2+4R^4\right)+\nonumber \right.\\
\left.+5b^2\left(9(R^2+r^2)^3-14r^4R^2+6R^6\right)\right)=1\label{R-r-3Dim}
\end{gather}
Solving equations (\ref{R-two-cut-3D}) and (\ref{R-r-3Dim}) for $R,r$ results in complex algebraic expressions, which we do not write explicitly, but we will keep in mind that in principle $R$ and $r$ are known as functions of $a$ and $b$.

Note that the one-cut distribution (\ref{one-cut-3D}) achieves its minimum at $x=0$ and hence is well defined when $\rho_3(0)\geq0$, which implies:
\begin{equation}
a \geq\frac{\sqrt{20b}}{3}\ .
\end{equation}
At $a=\sqrt{20b}/3$ we have a critical solution and for $a
<\sqrt{20b}/3$ we expect a phase transition from a ball phase (the
one-cut solution) to an annulus phase (the two-cut solution). Let us
analyse the heat capacity of the model. To calculate the heat capacity
of the model we need the internal energy of the system as a function of
both $a$ and $b$. The internal energy is given by the expectation
value of the potential with respect to the eigenvalue distribution
$\rho_3$. Next we multiply the internal energy by $T$ rescale $a\to
a/T,b\to b/T$ and find the derivative with respect to $T$ setting
$T\to1$ afterwords. Note that this procedure requires knowing the
derivatives of $R$ and $r$ with respect to $a$ and $b$. While we
didn't provide an explicit solution for the radii, the derivatives can
be easily obtained indirectly by differentiating equations
(\ref{R-two-cut-3D}) and (\ref{R-r-3Dim}). Our final expression for the
heat capacity is:
\begin{eqnarray}\label{CvD3}
C_v^1&=&\frac{1}{4}+\frac{9a^4+(10 a b-3 a^3)\sqrt{9 a^2+60 b}}{600b^2}\ ;~~~\text{for}~~~a \geq\frac{\sqrt{20b}}{3}\ , \\
C_v^2&=&\frac{1}{4}+\frac{12R^4a^3-3a(R^2+r^2)^2(a^2+5b^2((R^2+r^2)^2-4R^4))}{16(a+b(3R^2+r^2))}+\\
&&+\frac{6a^2b(R^4-r^4)(5r^2+7R^2)}{16(a+b(3R^2+r^2))};~~~\text{for}~~~a\leq \frac{\sqrt{20b}}{3}\ .\nonumber
\end{eqnarray}
Next using equations (\ref{R-two-cut-3D}) and (\ref{R-r-3Dim}) we obtain the following expansion for $C_v^1$ and $C_v^2$ near  $a={\sqrt{20b}}\,/\,{3}\,$:
\begin{eqnarray}
C_v^1&=&\frac{43}{108}+\frac{\sqrt{5}}{36\sqrt{b}}\left(a-\frac{\sqrt{20b}}{3}\right)-\frac{7}{320b}\left(a-\frac{\sqrt{20b}}{3}\right)^2+\dots\ ,\\
C_v^2&=&\frac{43}{108}+\frac{\sqrt{5}}{36\sqrt{b}}\left(a-\frac{\sqrt{20b}}{3}\right)-\frac{1}{40b}\left(a-\frac{\sqrt{20b}}{3}\right)^2+\dots\ .
\end{eqnarray}
Therefore we conclude that the heat capacity and its first derivative are continuous at $a=\sqrt{20b}/3$, while the second derivative has a finite jump. Therefore, the phase transition is of a fourth order.  

\begin{figure}[h]
  \centering
   \includegraphics[width=12cm]{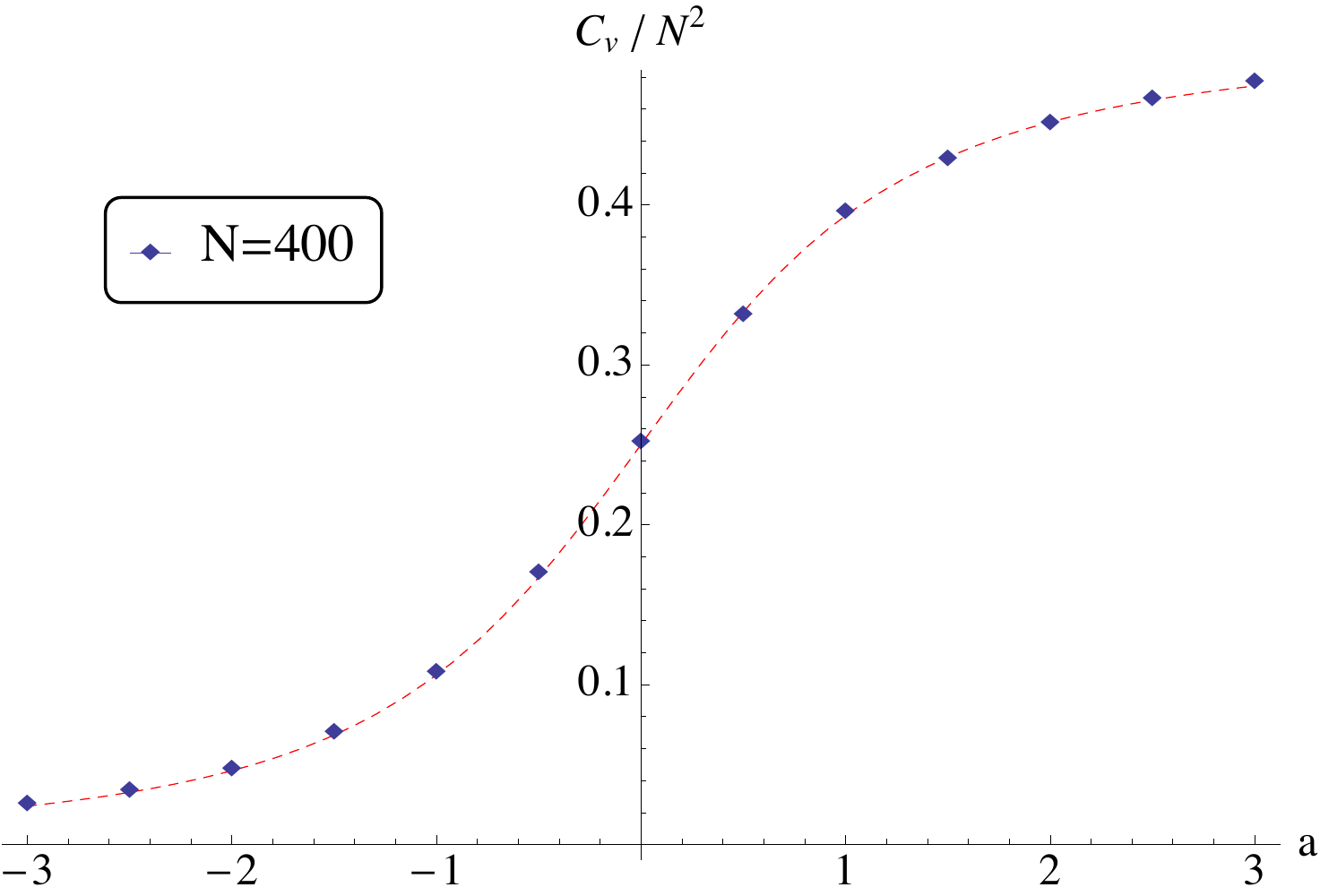} 
\caption{The heat capacity for the three dimensional model. The diamonds 
represent Monte Carlo simulations and the red-dashed line the analytic 
expressions (\ref{CvD3}). The critical value occurs are $a=\frac{\sqrt{20}}{3}\simeq 1.49$.  }
   \label{fig:11}
\end{figure}

In figure \ref{fig:11} we present a plot of the heat capacity as a
function of $a$ for $b=1$. An interesting property that stands out is
that $C_v=1/4$ at $a=0$ (just like in the one- and two- matrix
models), which is a consequence of the constraint
(\ref{constr-contin}). Furthermore, the heat capacity appears odd with
respect to the point $(0,1/4)$. In fact by expanding $C_v^2$ near
$a=0$ one can show that it is indeed odd with respect to the point
$(0,1/4)$. Remarkably this symmetry persist as an approximate symmetry
even across the phase transition for $a>\sqrt{20b}/3$. There is also
a striking similarity with the heat capacity of the two-matrix model
(look at figure \ref{fig:8a}). The diamonds in the figure \ref{fig:11}
represent the results of Monte Carlo numerical simulations. One can
see the good agreement between numerical results and the large $N$
predictions (\ref{CvD3}).

\begin{figure}[h]
  \centering
    \includegraphics[width=4.9cm]{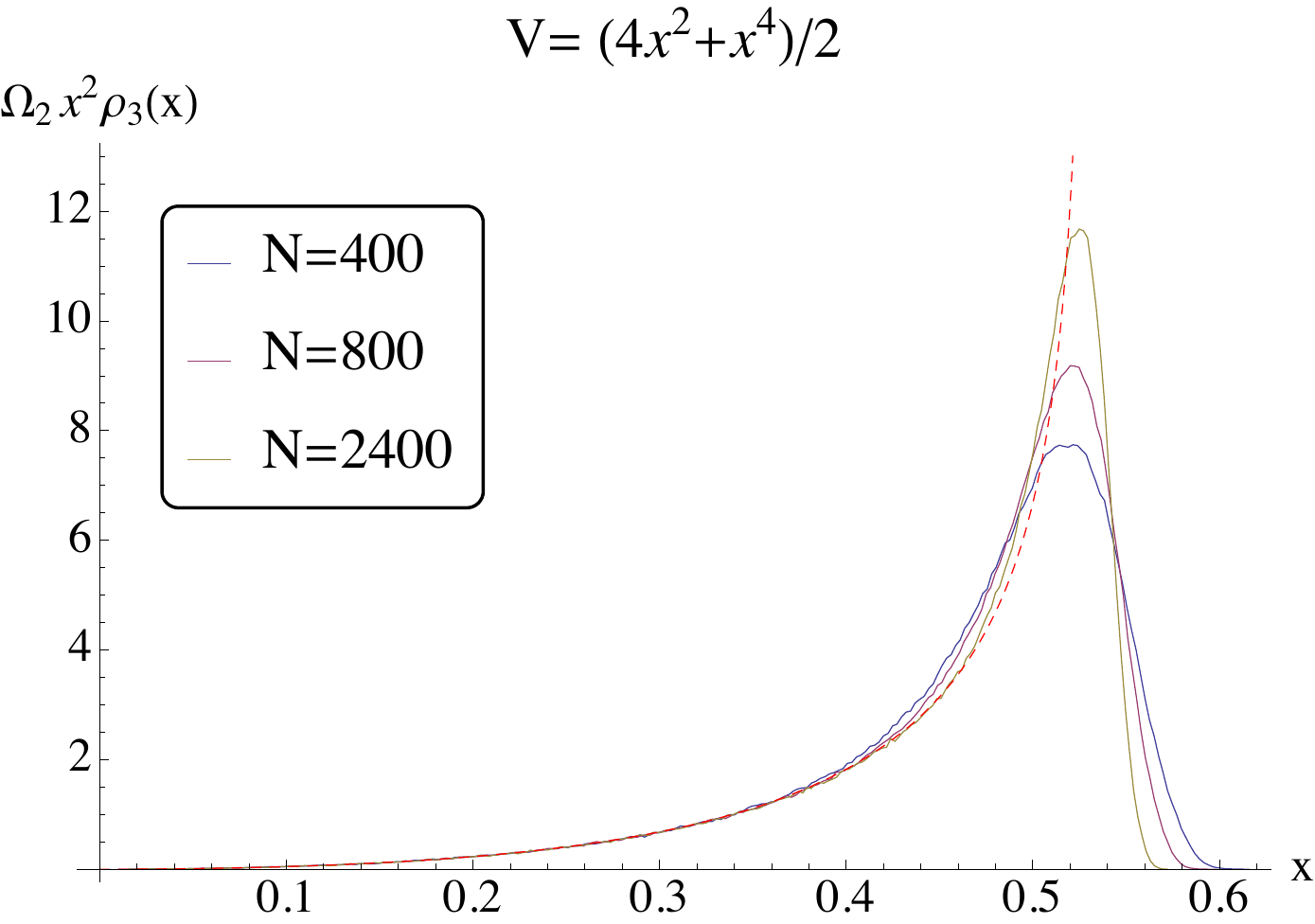}
    \includegraphics[width=4.9cm]{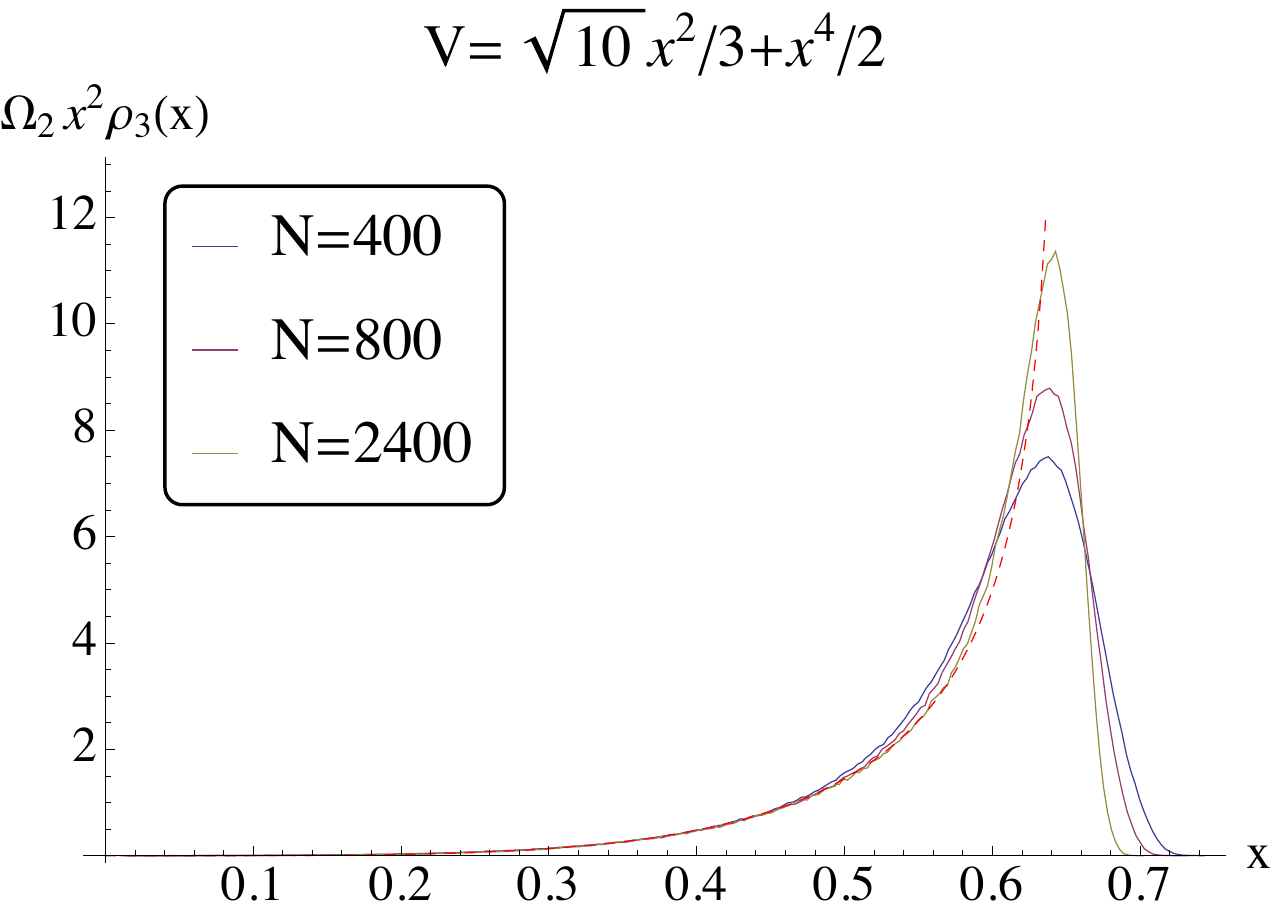} 
    \includegraphics[width=4.9cm]{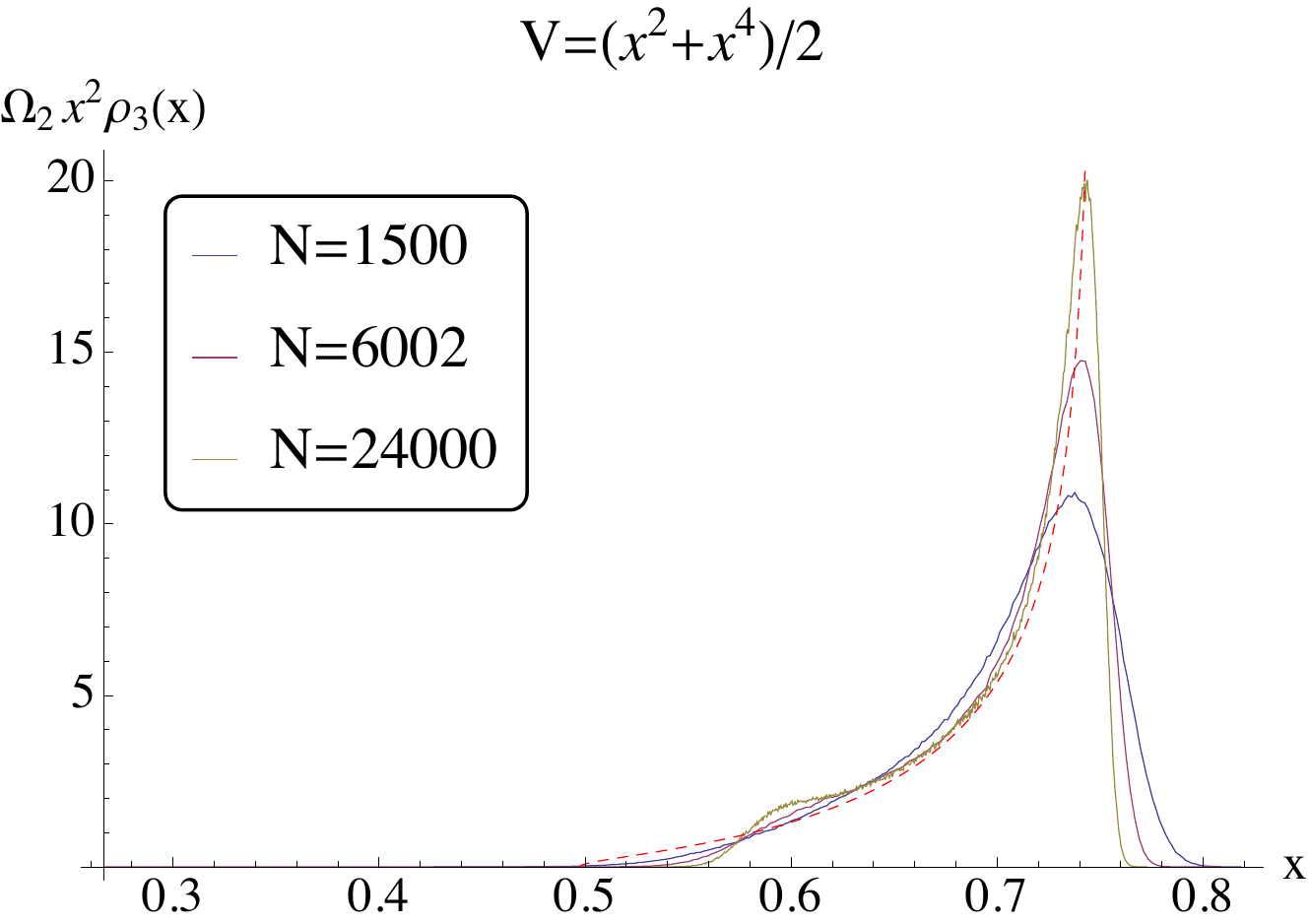}
    
    \includegraphics[width=4.9cm]{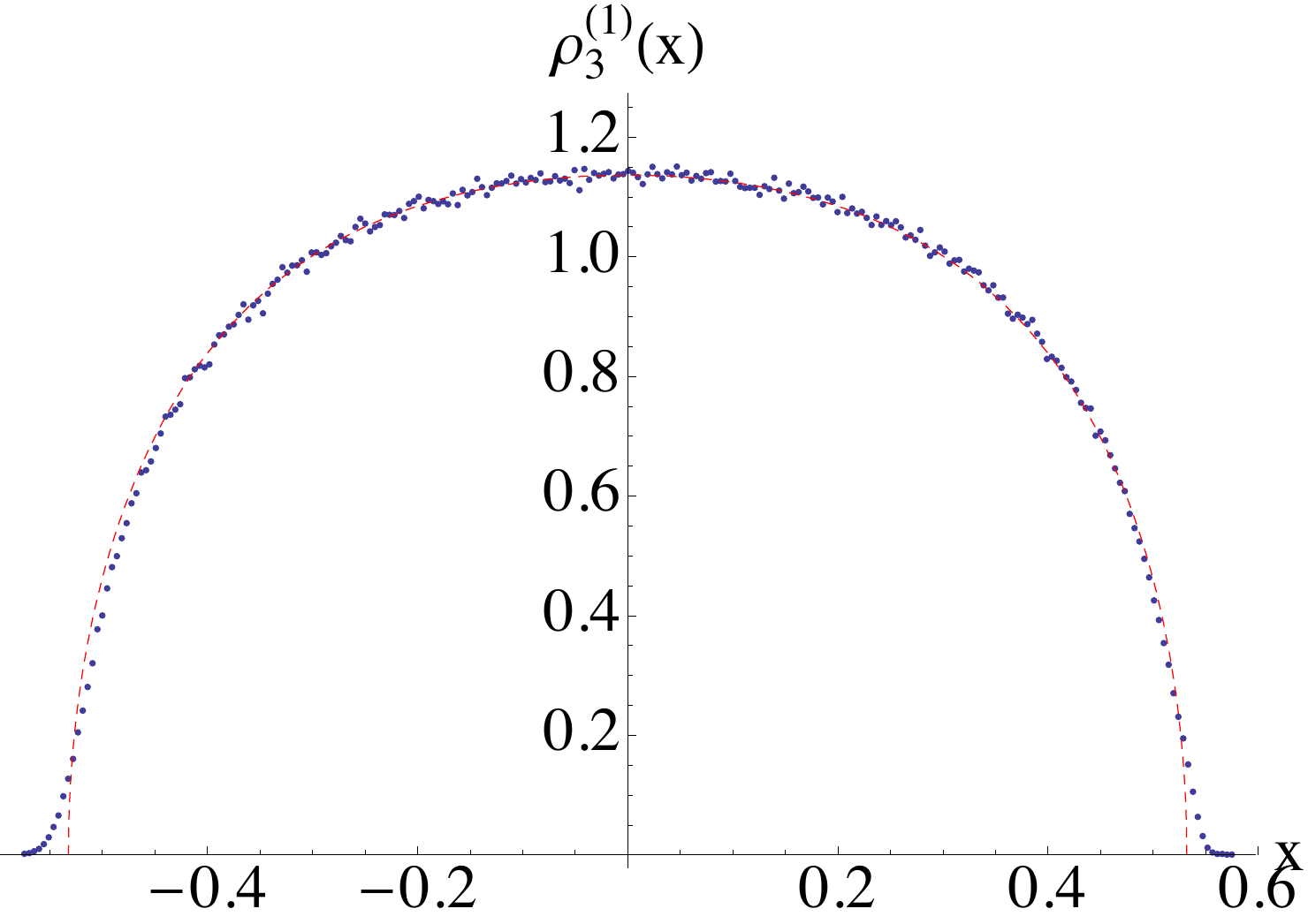} 
    \includegraphics[width=4.9cm]{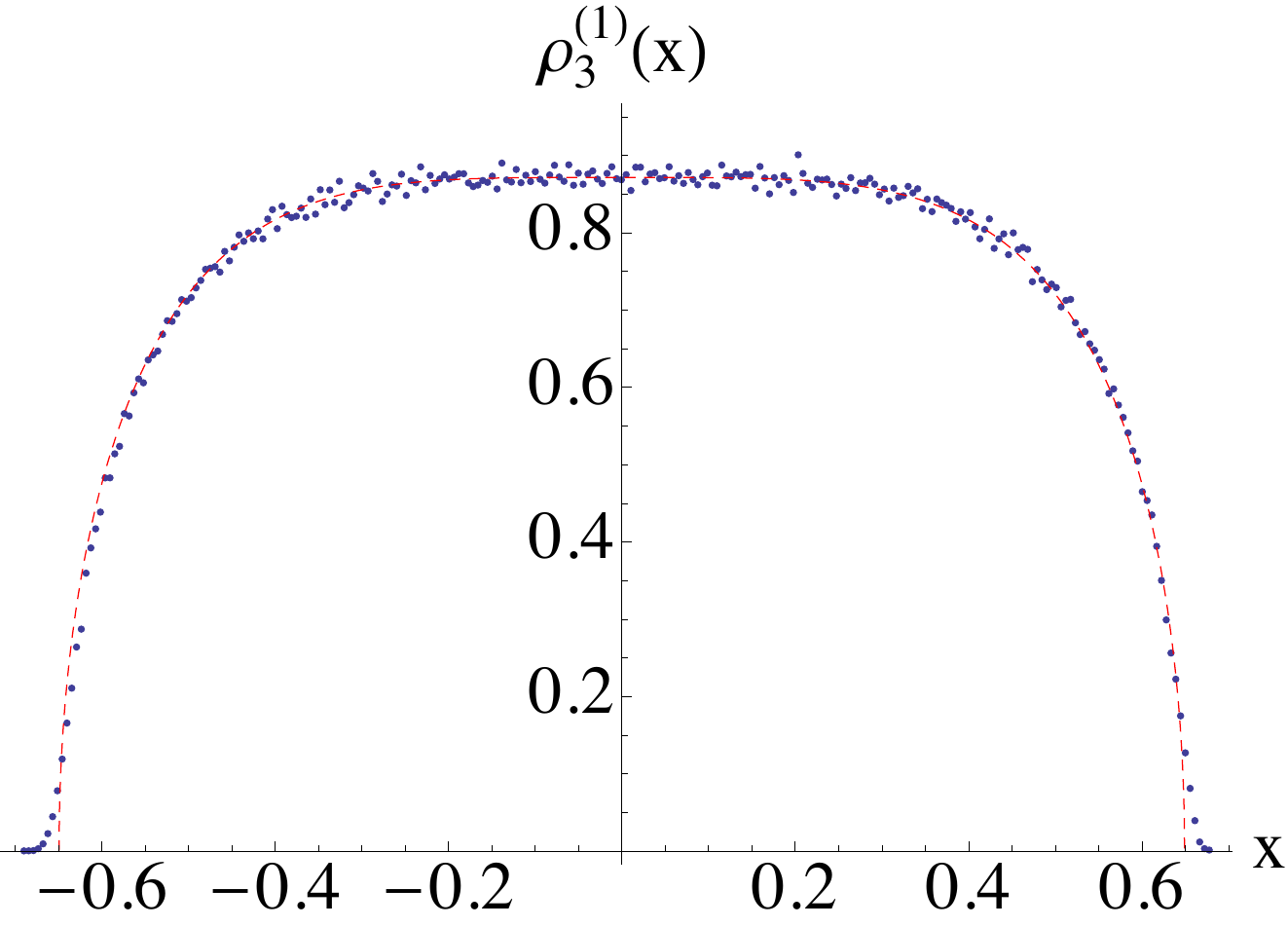} 
    \includegraphics[width=4.9cm]{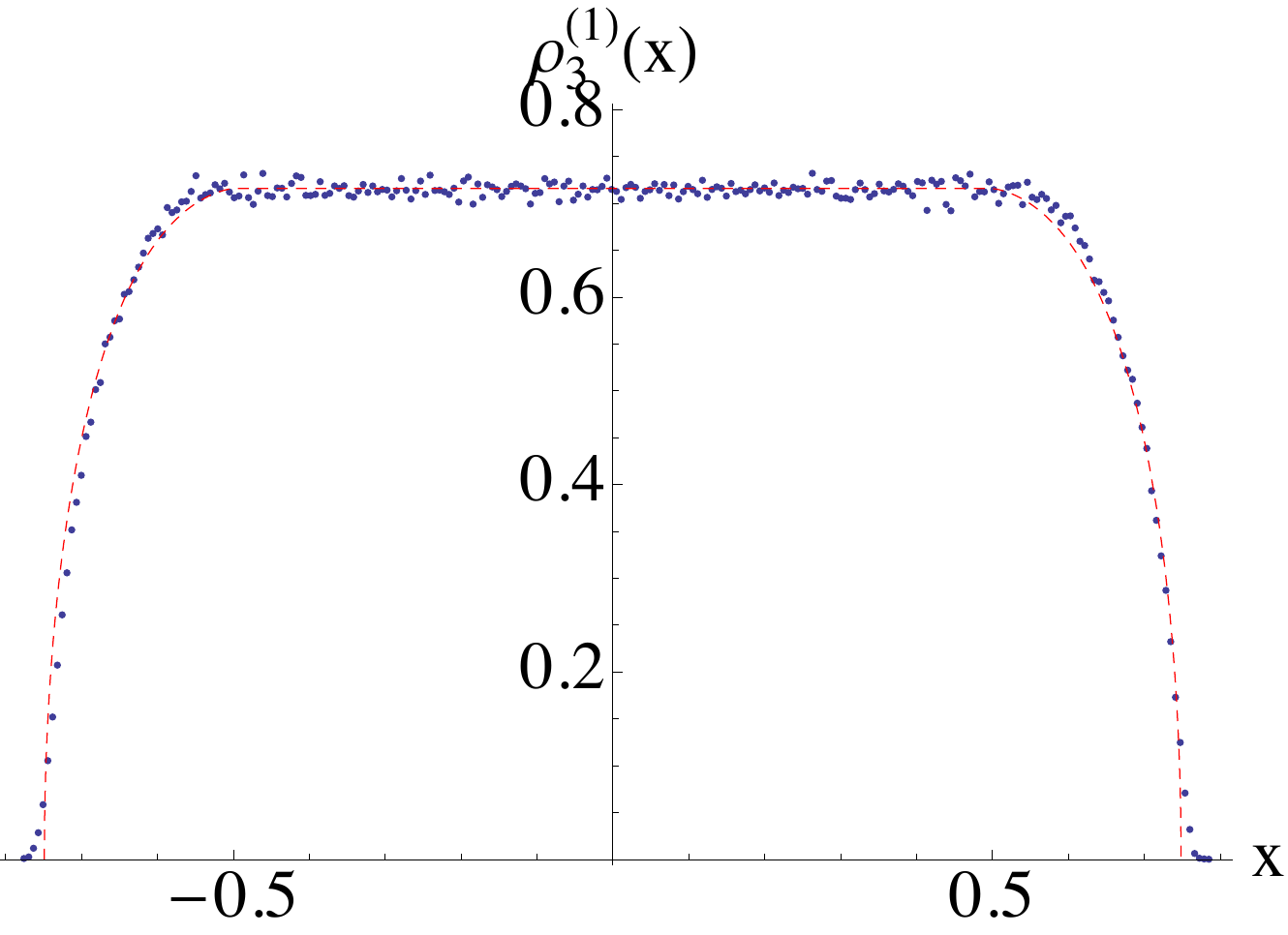} 
\caption{Comparison of numerical simulation with theoretical predictions. In all plots $b=1/2$. The first pair of plots from left to right represent the 3D and the reduced eigenvalue distributions in the ball phase for $a=2$. The second pair corresponds to the critical case $a=\sqrt{10}/3$. Finally, the last pair represents the 3D and the reduced distributions in the annulus phase for $a=1/2$. One can observe a very good agreement, except for the behaviour of the 3D distribution near the inner radius. }
   \label{fig:12}
\end{figure}

Let us also compare our results for the eigenvalue distribution with
numerical simulations. In figure \ref{fig:12} we present plots of the
one-eigenvalue distributions for the ball phase, the annulus phase and
for the critical distribution (at $a=\sqrt{20b}/3$). While one can see
very good agreement in the ball phase (for the one-cut solution), one
can see that for the annulus phase (the two-cut solution) the
agreement is good only away from the inner radius of the
distribution. Near the inner radius numerical simulations imply a
sharp fall off, and a probable jump, in the distribution (similar to the one
in the two-matrix model), while the analytic expression (\ref{two-cut-3D}) falls
gradually. This discrepancy is enhanced as $N$ is increased. At
present we don't have a theoretical way of describing such a sharp
fall, since the bounded solutions of the Cauchy kernel integral
equation (\ref{3DCauchy}) necessarily vanish at the boundary. A
possible way would be to attack numerically the integral equation
(\ref{3DlogK}), but such studies are beyond the scope of this
paper. Furthermore, the very good agreement of the heat capacity of
the annulus phase obtained using the two-cut solution implies that it is
very close to the real saddle point.

\subsection{Quartic potential in four dimensions}

In four dimensions out starting point is the integral equation:
\begin{equation}\label{inteq-4d}
\mu_4+a |\vec x|^2+b|\vec x|^4 =\int d^4x'\,\rho_4(\vec x)\,\log(\vec x-\vec x')^2\ .
\end{equation}
Using the fact $(\nabla^2)^2\log(\vec x-\vec x')^2\propto \delta^{(4)}(\vec x- \vec x')$ in four dimensions as well as the result from section \ref{gaussian-vardim} that the solution to (\ref{inteq-4d}) for $b=0$ is a spherical shell, one arrives at the following result for the solution for general $a$ and $b$:
\begin{eqnarray}\label{mixed-saddle-4d}
\rho_4(x)&=&-\frac{12b}{\pi^2}\Theta(R^2-\vec x^2)+\frac{2\sqrt{-6b}\sqrt{a^2+6b}}{(a-\sqrt{a^2+6b})^{1/2}\pi^2}\delta(R^2-x^2)\ ,\\
\text{where}~~R^2&=&\frac{a-\sqrt{a^2+6b}}{-6b}\ . \nonumber
\end{eqnarray}
As one can see the eigenvalue distribution is a mixture of an uniform
distribution with density proportional to $-b$ and a spherical shell
distribution. One can also see that the distribution is physical only
for $b<0$ and $a^2>|6b|$.
However, for $b<0$ the potential is unstable. Therefore this solution
can be realised, for large $N$, only as a metastable phase trapped
near the local minimum of the potential at $x=0$. The absence of
tunnelling stabilises this phase in the large $N$ limit. Since
increasing $a$ broadens the well of the potential, while lowering the
radius of the distribution, for sufficiently large $a$ the eigenvalues
spill out of this local well. This transition occurs at the upper
bound at $a^2=-6b$ which represents the critical value and corresponds
to the quantum gravity transition of the one dimensional model
\cite{DiFrancesco:2004qj}. We will not investigate  
this transition further in this paper.

For $b>0$ the model is stable for any value of $a$, but the solution
(\ref{mixed-saddle-4d}) is unphysical, therefore we expect that it is the
shell saddle (\ref{shell}) that is realised.
 
We conclude that if $a$ is sufficiently large one should encounter a
phase transition at $b=0$ from the spherical shell phase to a mixed
phase comprising of a spherical shell distribution and an uniform
distribution inside the shell.
 
\begin{figure}[h]
  \centering
    \includegraphics[width=7cm]{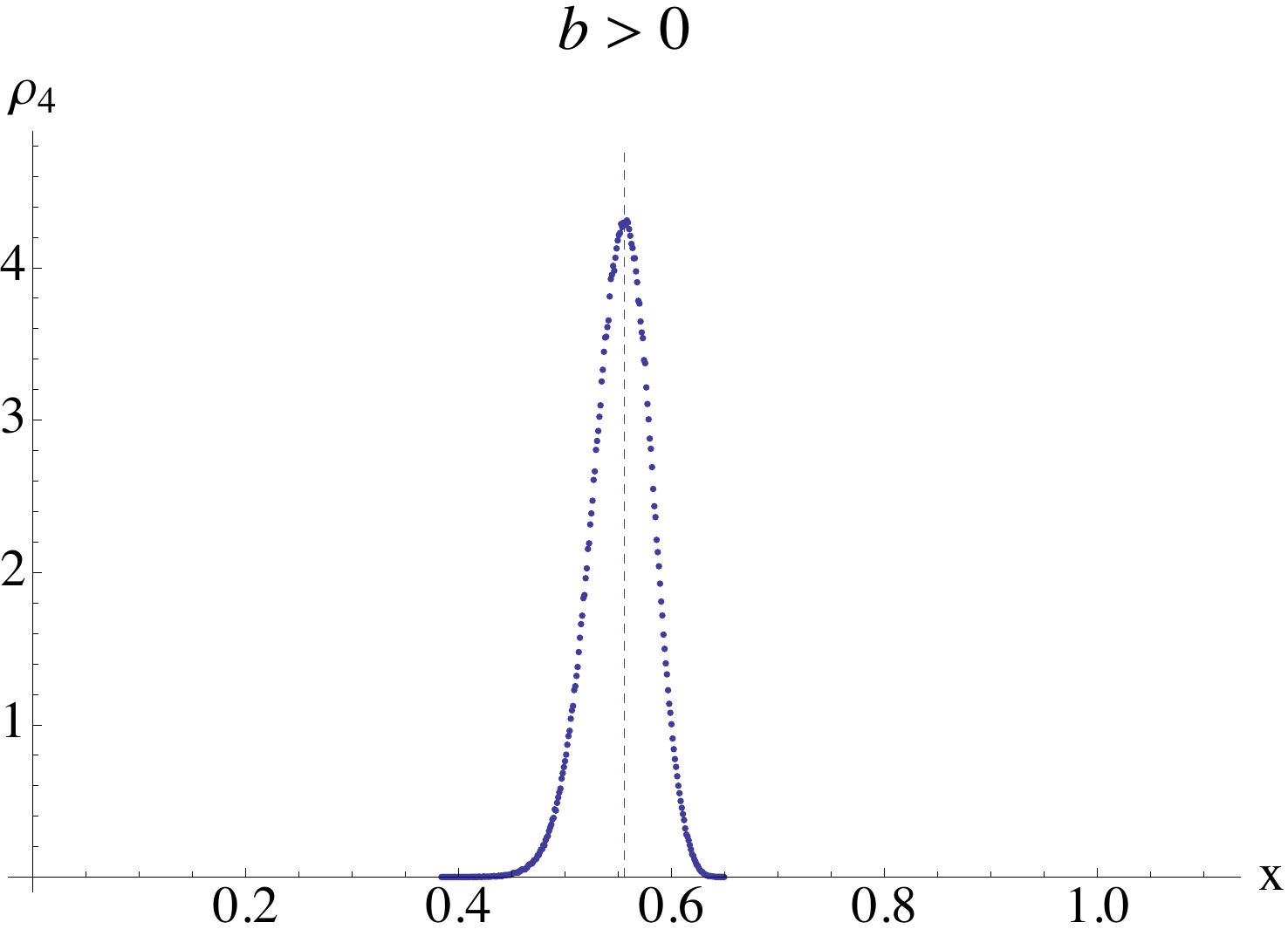}
    \includegraphics[width=7cm]{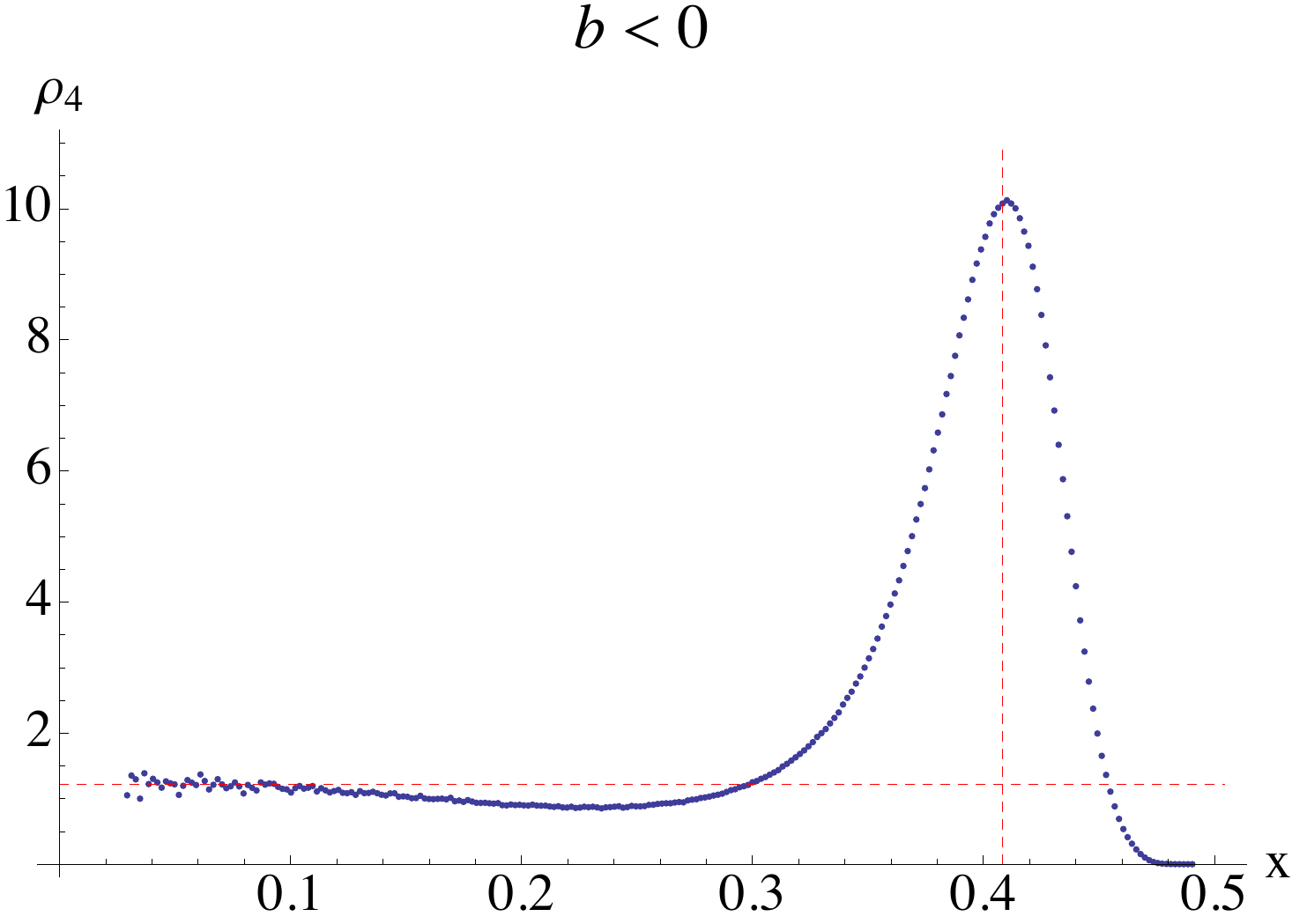} 
\caption{The first plot from left to right represents the spherical
  shell phase for $b=1, a=1$ and $N=400$. The vertical dashed line
  represents the radius of the shell determined by equation
  (\ref{shell}). The second plot represents the mixed phase for $b=-1,
  a=3/2$. The vertical dashed line represents the radius of the shell,
  while the horizontal dashed line represents the density of the
  uniform distribution both determined by equation
  (\ref{mixed-saddle-4d}). One can observe a very good agreement of
  the numerical results with the large $N$ predictions }
   \label{fig:12a}
\end{figure}

In figure \ref{fig:12a} we have presented our results of Monte Carlo
simulations. The first plot from left to right represents the
spherical shell phase for $b=1, a=1$ and $N=400$. The vertical dashed
line represents the radius of the shell determined by equation
(\ref{shell}). The second plot represents the mixed phase for $b=-1,
a=3/2$. The vertical dashed line represents the radius of the shell,
while the horizontal dashed line represents the density of the uniform
distribution both determined by equation (\ref{mixed-saddle-4d}). One
can observe a very good agreement of the numerical results with the
large $N$ predictions.
 
It would be interesting to explore deeper the onset of instability in
the mixed phase as $a$ approaches $\sqrt{|6b|}$. It would be also
interesting to study the heat capacity of the system and determine the
order of the phase transition. We leave these interesting studies for
future work.

\section{Conclusions}\label{Conclusions}

We have performed a systematic study of commutative $SO(p)$ invariant
matrix models with quadratic and quartic potentials. We found that the
physics of these systems depends crucially on the number of matrices
with a critical r\^ole played by $p=4$. For $p\leq4$ and a quartic
potential the system undergoes a phase transition, while for $p > 4$
the system is always in the low temperature phase.

In terms of the joint eigenvalue distribution of the matrices, for
$p=2$ the transition is from a disc distribution to an annular one at
the critical value $a_c=0$.  This is precisely where one would expect
the transition in the absence of fluctuations. The physics here is
straightforward: for $a>0$ the potential $V(\vec{x})$ has a unique
ground state and the resulting eigenvalue distribution is a disc. The
precise distribution is given by (\ref{disk-dist}),
i.e. $\rho_2(x)=\frac{4b\,x^2+a}{\pi}\,\Theta(R^2-x^2)$ with
$R^2=\frac{\sqrt{a^2+8b}-a}{4b}$ and becomes the uniform distribution
for $b=0$ with $R^2=\frac{1}{a}$. For $a<0$ the moduli space of ground
states of $V(\vec{x})$ is the circle of radius
$r=\sqrt{\frac{-a}{2b}}$.  The eigenvalue distribution then spreads
into an annulus around this circle.  The surprise is that eigenvalue
repulsion is sufficiently strong that the annular phase emerges even
at $a=0$ corresponding to the pure quartic potential. Furthermore 
in contrast to the one-dimensional model the transition 
in which the eigenvalue distribution changes from a disk to a shell is
in fact {\it sixth order}.

For $p=3$ the physics is very similar to that for $p=2$: One has an
eigenvalue ball for large positive $a$. For negative $a$ the moduli
space of vacua of the potential is now a sphere and the eigenvalues
spread about this sphere to give a spherical shell distribution.  The
transition between the two occurs at the surprisingly positive
critical value $a_c=\sqrt{20b}/3$, so that even a small quartic
potential is not sufficient to guarantee some eigenvalues near the
origin. Also the transition in this case turns out to be {\it fourth
  order}.

Surprisingly for $p=4$ there is no longer a standard ball to annulus
phase transition.  For positive $b$, when the quartic potential is
stable, the spherical shell phase (given by (\ref{shell-p})) is the
only possible phase, since the effective action's saddle is
unphysical.  However, for negative\footnote{We only consider $b<0$ for
  $p=4$ due to critical r\^ole played by $p=4$. It would of course be
  interesting to consider $b<0$ for all $p$.}  $b$ and sufficiently
large $a$ (for $a^2>|6b|$) there is a mixed metastable phase
comprising of a spherical shell with an uniform distribution inside
the shell.  As a result for $a^2<|6b|$, there is a phase transition
at $b=0$ from the spherical shell phase to the mixed phase. Since the
mixed phase contains an uniform ball, this transition can also be
viewed as a topology changing phase transition for the eigenvalue
distribution.

For $p>4$ and quartic potential there are no transitions and one is always
in the ``broken''-symmetry phase. In fact for $p> 4$ the joint eigenvalue
distribution is the infinitely thin spherical shell given by (\ref{shell-p})
i.e. $\rho_p(\vec
x)=\frac{2}{\Omega_{p-1}}\,\delta(1-\vec x^2)$ where $\Omega_{p-1}$ is
the volume of the unit $p-1$-sphere.

There are several generalisations of this work that can be undertaken.
One is to consider supersymmetric systems, this should be quite
straightforward.  A second is to consider the matrix quantum mechanics
of commuting matrix models. Further generalisations are to consider
non-rotationally invariant systems and more general potentials. 
We hope to return to these topics in the near future.
\\ \\
{\bf Acknowledgements:} The work of VF was partially supported my an INSPIRE IRCSET-Marie Curie International Mobility Fellowship.

\newpage
\appendix

\section{Reducing from two dimensions}\label{AppendixA}
In this section we derive equation (\ref{S2toS1}) relating the effective action in two and one dimensions. Let us write the effective action in two dimensions:
\begin{eqnarray}\label{S2}
S_{2}[\rho_2]&=&\int d^2 x\,\rho_{2}(\vec x)V_2(|\vec x|)-\frac{1}{2}\int\int d^2 x \, d^2 x'\rho_{2}(\vec x)\,\rho_{2}(\vec x')\log(\vec x -\vec x')^2\nonumber\\
&+&\mu_2\left(\int d^2 x\,\rho_{2}(\vec x)-1\right)
\end{eqnarray}
We start with the first term in equation (\ref{S2}). Using equation (\ref{p-1top}) and integrating by parts we obtain:
\begin{eqnarray}\label{potential2to1}
&&\int d^2 x\,\rho_{2}(\vec x)V_2(|\vec x|)=2\int\limits_0^R d x\, V_2'(x)\,\int\limits_x^R dr\frac{r\rho_1(r)}{\sqrt{r^2-x^2}}-2\int\limits_0^R d r\,\rho_1(r)\,V_2(0)=\nonumber\\
&&=2\int\limits_0^R dr\,\rho_1(r)\left(\int\limits_0^r dx\frac{r\,V_2'(x)}{\sqrt{r^2-x^2}}-V_2(0)   \right)=\int\limits_{-R}^{R}dx\,\rho_1(x)\,V_1(x)\ ,
\end{eqnarray}
where we defined the reduced potential $V_1$:
\begin{equation}\label{V2toV1}
V_1(x)=\int\limits_0^x dx'\frac{x\,V_2'(x')}{\sqrt{x^2-x'^2}}-V_2(0) 
\end{equation}
Using again equation (\ref{p-1top}) it is easy to show that:
\begin{equation}\label{last}
\int d^2\, x\rho_{2}(\vec x)=\int\limits_{-R}^{R} d x\,\rho_{1}(x)\ ,
\end{equation}
which takes care for the last term in equation (\ref{S2}). Finally, we focus on the term containing the logarithmic kernel. Defining:
\begin{equation}
F(x)=\int\limits_R^x dr\frac{r\,\rho_1(r)}{\sqrt{r^2-x^2}}\ ,
\end{equation}
and using equation (\ref{p-1top}) and the kernel (\ref{Kp}) for $p=2$ we can write:
\begin{align}\label{middle}
&\int\int d^2 x  d^2 x'\rho_{2}(\vec x)\rho_{2}(\vec x')\log(\vec x -\vec x')^2=4\int\limits_0^R\int\limits_0^R dxdx'F'(x)F'(x')\log\frac{(x^2+x'^2+|x^2-x'^2|)}{2}=\nonumber\\
&=8\int\limits_0^R dx\log x^2F'(x)\int\limits_0^x dx'F'(x')=4\int\limits_0^R\frac{d}{dx}(F(x)-F(0))^2\log x^2=\nonumber\\
&=-8\int\limits_0^R\frac{dx}{x}(F(x)-F(0))^2+4F(0)^2\log R^2=4F(0)^2\log R^2-\nonumber\\
&-8\int\limits_0^R\int\limits_0^Rdr dr'\rho_1(r)\rho_1(r')\int\limits_0^R\frac{dx}{x}\left(1-\frac{\Theta(r-x)\,r}{\sqrt{r^2-x^2}}\right)\left(1-\frac{\Theta(r'-x)\,r'}{\sqrt{r'^2-x^2}}\right)=\nonumber\\
&=4\int\limits_0^R\int\limits_0^Rdr dr'\rho_1(r)\rho_1(r')\left(\log|r^2-r'^2|+2\log2\right)=\nonumber\\
&=\int\limits_{-R}^R\int\limits_{-R}^Rdx dx'\rho_1(x)\rho_1(x')\log(x-x')^2+2\log2\left(\int\limits_{-R}^Rdx\rho_1(x)\right)^2
\end{align}
Combining equations (\ref{potential2to1}), (\ref{last}) and (\ref{middle}) we arrive at equation (\ref{S2toS1}), which we duplicate bellow:
\begin{equation}
\label{S2toS1appendix}
S_2[\rho_p]=S_1[\rho_1]-\log2\,\left(\int d x\, \rho_{1}(x)\right)^2\ ,
\end{equation}
were $S_1$ is given by:
\begin{eqnarray}\label{S1appendix}
S_{1}[\rho_2]&=&\int d x\,\rho_{1}(x)V_1(x)-\frac{1}{2}\int\int d x \, d x'\rho_{1}( x)\,\rho_{1}( x')\log(x-x')^2\nonumber\\
&+&\mu_1\left(\int d x\,\rho_{1}( x)-1\right)\ ,
\end{eqnarray}
with $\mu_1=\mu_2$

\section{General constraint}
In this section we derive a general constraint for the model (\ref{partition2}), which in the special case of gaussian potential and in the $N\to\infty$ limit reduces to equation (\ref{gaussian-constraint}). Our starting point is the mathematical identity:
\begin{equation}\label{identity-1}
\int \prod_{i} d^p\lambda_i \,\sum\limits_{k}\sum\limits_{\mu=1}^p\frac{\partial}{\partial\lambda_k^{\mu}}\left(\lambda_k^{\mu}\, e^{-N^2\,S_{\rm eff}[\vec\lambda]}\right)=0\ ,
\end{equation}
where $S_{\rm eff}$ is the action (\ref{eff-action}). The identity can by proven by integrating by parts and using that the integrant vanishes at $|\vec\lambda_i|\to\infty$. Performing the differentiation in (\ref{identity-1}) we obtain:
\begin{equation}
\int \prod_{i} d^p\lambda_i \,\left[p\,N-N^2\,\sum\limits_k\vec\lambda_k \cdot\frac{\partial S_{\rm eff}[\vec\lambda]}{\partial \vec\lambda_k}   \right]e^{-N^2\,S_{\rm eff}[\vec\lambda]}=0,
\end{equation}
which after dividing by the partition function (\ref{partition2}) can be written as:
\begin{equation}\label{identity-2}
\left\langle \sum\limits_k\vec\lambda_k \cdot\frac{\partial S_{\rm eff}[\vec\lambda]}{\partial \vec\lambda_k}  \right\rangle =\frac{p}{N}\ .
\end{equation}
Using equation (\ref{eff-action}) it is easy to check that:
\begin{eqnarray}
\sum\limits_k\vec\lambda_k \cdot\frac{\partial S_{\rm eff}[\vec\lambda]}{\partial \vec\lambda_k} &=&\frac{1}{N}\sum\limits_i\,|\vec\lambda_i|V_p'(|\vec\lambda_i|)-\frac{1}{N^2}\sum\limits_{k\neq i}\left(\frac{\vec\lambda_k\cdot(\vec\lambda_k-\vec\lambda_i)}{(\vec\lambda_k-\vec\lambda_i)^2}-\frac{\vec\lambda_i\cdot(\vec\lambda_k-\vec\lambda_i)}{(\vec\lambda_k-\vec\lambda_i)^2}\right)=\nonumber\\
&&=\frac{1}{N}\sum\limits_i\,|\vec\lambda_i|V_p'(|\vec\lambda_i|)-\left(1-\frac{1}{N}\right)\ .
\end{eqnarray}
Substituting in equation (\ref{identity-2}) we obtain the constraint:
\begin{equation}\label{constraint-anyN}
\langle \frac{1}{N}\sum\limits_i\,|\vec\lambda_i|V_p'(|\vec\lambda_i|)\rangle =1+\frac{p-1}{N}\ ,
\end{equation}
which holds for any $N$. In the continuous $N\to\infty$ limit equation (\ref{constraint-anyN}) reduces to:
\begin{equation}\label{constr-contin}
\int d^p x\,\rho_p(\vec x)\, |\vec x|\,V'(|\vec x|) =1\ .
\end{equation}
For the gaussian potential (\ref{Gaussian-p}) we obtain equation (\ref{gaussian-constraint}), which we duplicate bellow:
\begin{equation}
\int d^p x\,\rho_p(\vec x)\,\vec x^2 =1\ .
\end{equation}

\section{Analytic expression for the free energy}

In this section we obtain an analytic expression for the non-constant part of the free energy $\mathcal{E}$ defined in equation (\ref{Free-energy}). The idea is to uplift the calculation to $\alpha+3$ dimensions, where the pure state $\tilde\rho_\alpha$ lifts to a spherical shell distribution. It is also convenient to rescale the distribution $\tilde\rho_\alpha$ to the range $(-1,1)$, To this end we define:
\begin{equation}
\hat\rho_\alpha(\eta)=R_\alpha\,\tilde\rho_\alpha (\eta R_\alpha)=\frac{\Gamma(\frac{\alpha+3}{2})}{\pi^{1/2}\,\Gamma(\frac{\alpha+2}{2})}(1-\eta^2)^{\alpha/2}\ ,
\end{equation}
where $R_\alpha=(\alpha+2)/p$ as given in equation (\ref{Ralpha}). Next we write $\mathcal{E}$ in terms of $\hat\rho_\alpha$:
\begin{equation}\label{E-eta}
\mathcal{E}=-\frac{1}{2}\log\,R_\alpha^2-\frac{1}{2}\int\limits_{-1}^1d\eta\int\limits_{-1}^1d\eta'\,\hat\rho_\alpha(\eta)\,\log(\eta-\eta')^2\,\hat\rho_\alpha(\eta')
\end{equation}
and uplift the calculation of the second term in (\ref{E-eta}) to $\alpha+3$ dimensions. The distribution $\hat\rho_\alpha$ lifts to:
\begin{equation}\label{eta-shell}
\rho_{\alpha+3}^{\rm shell}(\vec\eta)=\frac{2}{\Omega_{\alpha+2}}\,\delta(1-\vec\eta^{\,2})\ ,
\end{equation}
where $\Omega_{\alpha+2}$ is the volume of the unit $\alpha+2$ dimensional sphere. The crucial step is to use equation (\ref{reduced2}) and the same considerations that lead to equation (\ref{SptoS1}) to write:
\begin{eqnarray}\label{C4}
&&\int\limits_{-1}^1d\eta\int\limits_{-1}^1d\eta'\,\hat\rho_\alpha(\eta)\,\log(\eta-\eta')^2\,\hat\rho_\alpha(\eta')=\int d^{\alpha+3}\eta\int d^{\alpha+3}\eta'\rho_{\alpha+3}^{\rm shell}(\vec\eta)\,\log(\vec\eta-\vec\eta\,')^2\rho_{\alpha+3}^{\rm shell}(\vec\eta\,')\nonumber \\
&&-\left(2\log 2+H_{\frac{\alpha+1}{2}}\right)\left(\int\limits_{-1}^1d\eta\,\hat\rho_{\alpha}(\eta)\right)^2\ ,
\end{eqnarray}
where $H_n$ is the harmonic number. Let us deal first with the first term on the right-hand side of equation (\ref{C4}). Using equation (\ref{eta-shell}) we obtain:
\begin{equation}\label{C5}
\int d^{\alpha+3}\eta\int d^{\alpha+3}\eta'\rho_{\alpha+3}^{\rm shell}(\vec\eta)\,\log(\vec\eta-\vec\eta\,')^2\rho_{\alpha+3}^{\rm shell}(\vec\eta\,')=\frac{K_{\alpha+3}(1,1)}{\Omega_{\alpha+2}^2}=\frac{H_{\frac{\alpha}{2}}-H_{\frac{\alpha+1}{2}}}{2}+\log2\ ,
\end{equation}
where we have used equation (\ref{Kp}). Now substituting equation (\ref{C5}) into equation (\ref{C4}) and using that $\hat\rho_\alpha$ is normalised to one, we obtain:
\begin{equation}\label{C6}
\int\limits_{-1}^1d\eta\int\limits_{-1}^1d\eta'\,\hat\rho_\alpha(\eta)\,\log(\eta-\eta')^2\,\hat\rho_\alpha(\eta')=\frac{1}{2}H_{\frac{\alpha}{2}}-\frac{3}{2}H_{\frac{\alpha+1}{2}}-\log2\ .
\end{equation}
Finally, substituting equation (\ref{C6}) into equation (\ref{E-eta}) and using equation (\ref{Ralpha}) we arrive at equation (\ref{Free-energy-final}), which we duplicate bellow:
\begin{equation}
\mathcal{E}=\frac{3}{4}H_{\frac{\alpha+1}{2}}-\frac{1}{4}H_{\frac{\alpha}{2}}-\frac{1}{2}\log\frac{\alpha+3}{2\,p}\ .
\end{equation}


\begin{thebibliography}{99}

\bibitem{Ishibashi:1996xs}
  N.~Ishibashi, H.~Kawai, Y.~Kitazawa and A.~Tsuchiya,
  ``A Large N reduced model as superstring,''
  Nucl.\ Phys.\ B {\bf 498} (1997) 467
  [hep-th/9612115].
  
\bibitem{Kim:2011cr}
  S.~-W.~Kim, J.~Nishimura and A.~Tsuchiya,
  ``Expanding (3+1)-dimensional universe from a Lorentzian matrix model for superstring theory in (9+1)-dimensions,''
  Phys.\ Rev.\ Lett.\  {\bf 108} (2012) 011601
  [arXiv:1108.1540 [hep-th]].
  
\bibitem{Banks:1996vh}
  T.~Banks, W.~Fischler, S.~H.~Shenker and L.~Susskind,
   ``M theory as a matrix model: A Conjecture'',
   {\it Phys. Rev.} {\bf D55} (1997) 5112
        [hep-th/9610043].
          
\bibitem{Townsend:1995kk}
  P.~K.~Townsend,
  ``The eleven-dimensional supermembrane revisited'',
  {\it Phys. Lett.} {\bf B350} (1995) 184
  [hep-th/9501068].

\bibitem{Dijkgraaf:1997vv}
  R.~Dijkgraaf, E.~P.~Verlinde and H.~L.~Verlinde,
  ``Matrix string theory,''
  Nucl.\ Phys.\ B {\bf 500} (1997) 43
  [hep-th/9703030].

\bibitem{Banks:1996my}
  T.~Banks and N.~Seiberg,
  Nucl.\ Phys.\ B {\bf 497} (1997) 41
  [hep-th/9702187].
         
\bibitem{Berenstein:2003gb}
  D.~E.~Berenstein, J.~M.~Maldacena and H.~S.~Nastase,
  ``Strings in flat space and pp waves from N=4 Super Yang Mills,''
AIP Conf.\ Proc.\  {\bf 646} (2003) 3.

\bibitem{Aharony:2008gk}
  O.~Aharony, O.~Bergman and D.~L.~Jafferis,
  ``Fractional M2-branes,''
  JHEP {\bf 0811} (2008) 043
  [arXiv:0807.4924 [hep-th]].
  
\bibitem{Kovacs:2013una}
  S.~Kovacs, Y.~Sato and H.~Shimada,
  ``Membranes from monopole operators in ABJM theory: large angular momentum and M-theoretic $AdS_4/CFT_3$,''
  arXiv:1310.0016 [hep-th].
    
 \bibitem{Hoppe:PhDThesis1982} Hoppe, J.~R.\ 1982, 
Ph.D.~Thesis.
 
\bibitem{de Wit:1988ig}
  B.~de Wit, J.~Hoppe and H.~Nicolai,
  ``On the Quantum Mechanics of Supermembranes,''
  Nucl.\ Phys.\ B {\bf 305} (1988) 545. \\

  B.~de Wit, U.~Marquard and H.~Nicolai,
   ``Area Preserving Diffeomorphisms And Supermembrane 
   Lorentz Invariance'',
    {\it Commun. Math. Phys.}  {\bf 128} (1990) 39.

\bibitem{Connes:1997cr}
  A.~Connes, M.~R.~Douglas and A.~S.~Schwarz,
  ``Noncommutative geometry and matrix theory: Compactification on tori,''
  JHEP {\bf 9802} (1998) 003
  [hep-th/9711162].

\bibitem{Kazakov:1998ji}
  V.~A.~Kazakov, I.~K.~Kostov and N.~A.~Nekrasov,
  ``D-particles, matrix integrals and KP hierarchy,''
  Nucl.\ Phys.\  B {\bf 557}, 413 (1999)
  [arXiv:hep-th/9810035].

\bibitem{DelgadilloBlando:2007vx}
  R.~Delgadillo-Blando, D.~O'Connor and B.~Ydri,
  ``Geometry in Transition: A Model of Emergent Geometry,''
 Phys.\ Rev.\ Lett.\  {\bf 100} (2008) 201601
 [arXiv:0712.3011 [hep-th]].

\bibitem{DelgadilloBlando:2012xg}
  R.~Delgadillo-Blando and D.~O'Connor,
  ``Matrix geometries and Matrix Models,''
  JHEP {\bf 1211} (2012) 057
  [arXiv:1203.6901 [hep-th]].

\bibitem{Steinacker:2012ra}
  H.~Steinacker,
  ``Gravity and compactified branes in matrix models,''
 JHEP {\bf 1207} (2012) 156 
[arXiv:1202.6306 [hep-th]].

\bibitem{Blaschke:2010ye}
  D.~N.~Blaschke and H.~Steinacker,
  ``Schwarzschild Geometry Emerging from Matrix Models,''
 Class.\ Quant.\ Grav.\  {\bf 27} (2010) 185020
 [arXiv:1005.0499 [hep-th]].



\bibitem{Krauth:1998xh}
  W.~Krauth, H.~Nicolai and M.~Staudacher,
  ``Monte Carlo approach to M theory,''
  Phys.\ Lett.\ B {\bf 431} (1998) 31
  [hep-th/9803117].

\bibitem{Krauth:1998yu}
  W.~Krauth and M.~Staudacher,
  ``Finite Yang-Mills integrals,''
  Phys.\ Lett.\ B {\bf 435} (1998) 350
  [hep-th/9804199].

\bibitem{Hotta:1998en}
  T.~Hotta, J.~Nishimura and A.~Tsuchiya,
  ``Dynamical aspects of large N reduced models,''
  Nucl.\ Phys.\ B {\bf 545} (1999) 543
  [hep-th/9811220].

\bibitem{Ambjorn:2000bf}
  J.~Ambjorn, K.~N.~Anagnostopoulos, W.~Bietenholz, T.~Hotta and J.~Nishimura,
  ``Large N dynamics of dimensionally reduced 4-D SU(N) superYang-Mills theory,''
  JHEP {\bf 0007} (2000) 013
  [hep-th/0003208].

\bibitem{Ambjorn:2000dx}
  J.~Ambjorn, K.~N.~Anagnostopoulos, W.~Bietenholz, T.~Hotta and J.~Nishimura,
  ``Monte Carlo studies of the IIB matrix model at large N,''
  JHEP {\bf 0007} (2000) 011
  [hep-th/0005147].

\bibitem{Azuma:2004zq}
  T.~Azuma, S.~Bal, K.~Nagao and J.~Nishimura,
  ``Nonperturbative studies of fuzzy spheres in a matrix model with the Chern-Simons term,''
  JHEP {\bf 0405} (2004) 005
  [hep-th/0401038].

\bibitem{Filev:2013pza} 
  V.~G.~Filev and D.~O'Connor,
  ``Multi-matrix models at general coupling,''
 J.\ Phys.\ A {\bf 46}, 475403 (2013)
  [arXiv:1304.7723 [hep-th]].

\bibitem{O'Connor:2012vr}
  D.~O'Connor and V.~G.~Filev,
  ``Near commuting multi-matrix models,''
  JHEP {\bf 1304} (2013) 144
  [arXiv:1212.4818 [hep-th]].

\bibitem{Berenstein:2005aa} 
  D.~Berenstein,
  ``Large N BPS states and emergent quantum gravity,''
  JHEP {\bf 0601}, 125 (2006)
  [hep-th/0507203].
  
\bibitem{Berenstein:2005jq} 
  D.~Berenstein, D.~H.~Correa and S.~E.~Vazquez,
  ``All loop BMN state energies from matrices,''
  JHEP {\bf 0602}, 048 (2006)
  [hep-th/0509015].
  
\bibitem{Aharony:2007rj} 
  O.~Aharony and S.~A.~Hartnoll,
  ``A Phase transition in commuting Gaussian multi-matrix models,''
  arXiv:0706.2861 [hep-th].
  
\bibitem{Berenstein:2008eg} 
  D.~E.~Berenstein, M.~Hanada and S.~A.~Hartnoll,
  ``Multi-matrix models and emergent geometry,''
  JHEP {\bf 0902}, 010 (2009)
  [arXiv:0805.4658 [hep-th]].
  
\bibitem{DiFrancesco:2004qj}
  P.~Di Francesco,
  ``2D quantum gravity, matrix models and graph combinatorics,''
  [math-ph/0406013].

\bibitem{Brezin:1977sv}
  E.~Brezin, C.~Itzykson, G.~Parisi and J.~B.~Zuber,
  ``Planar Diagrams,''
  Commun.\ Math.\ Phys.\  {\bf 59} (1978) 35.
  
\bibitem{Feinberg:1997if}
  J.~Feinberg and A.~Zee,
  ``NonGaussian nonHermitian random matrix theory: Phase transition and addition formalism,''
  Nucl.\ Phys.\ B {\bf 501} (1997) 643
  [cond-mat/9704191].
  
\bibitem{Feinberg:2001vj} 
  J.~Feinberg, R.~Scalettar and A.~Zee,
  ``'Single ring theorem' and the disk annulus phase transition,''
  J.\ Math.\ Phys.\  {\bf 42}, 5718 (2001)
  [cond-mat/0104072].

\end{thebibliography}
\end{document}